\documentclass[superscriptaddress,floatfix,letterpaper,accepted=2021-02-01]{quantumarticle}
\pdfoutput=1

\usepackage{float}
\usepackage[export]{adjustbox}
\usepackage{amsmath,mathtools,dsfont,amssymb}
\usepackage{braket}
\usepackage{enumitem}
\usepackage{amsfonts}
\usepackage{graphicx}
\newcommand{\ids}{\mathds{1}}
\usepackage{hyphenat}
\usepackage{algpseudocode}
\usepackage{algorithmicx}
\usepackage[ruled,vlined]{algorithm2e}
\usepackage{microtype}
\usepackage[caption=false]{subfig}
\usepackage[svgnames]{xcolor}
\usepackage{tabularx}
\usepackage{booktabs}
\usepackage[utf8]{inputenc}
\usepackage{bm}
\usepackage{siunitx}
\usepackage[hang,flushmargin]{footmisc}

\usepackage[numbers,sort&compress]{natbib}
\usepackage{tikz}
\usetikzlibrary{quantikz}

\usepackage{pdfpages}
\makeatletter
\AtBeginDocument{\let\LS@rot\@undefined}
\makeatother
\usepackage{afterpage}

\usepackage{hyperref}
\hypersetup{colorlinks,linkcolor={DarkBlue},citecolor={DarkBlue},urlcolor={DarkBlue}}
\usepackage[capitalize]{cleveref}
\usepackage{url}
\begin{document}

\title{Blueprint for a Scalable Photonic Fault-Tolerant Quantum Computer}
\author{J.~Eli Bourassa}
\affiliation{Xanadu, Toronto, ON, M5G 2C8, Canada}
\affiliation{Department of Physics, University of Toronto, Toronto, Canada}
\listcsgadd{author1affiliations}{\ref{fn:author}}
\orcid{0000-0003-4333-5006}
\author{Rafael N.~Alexander}
\affiliation{Xanadu, Toronto, ON, M5G 2C8, Canada}
\affiliation{Center for Quantum Information and Control, University of New Mexico, Albuquerque, NM 87131, USA}
\affiliation{Centre for Quantum Computation and Communication Technology, School of Science, RMIT University, Melbourne, VIC 3000, Australia}
\listcsgadd{author2affiliations}{\ref{fn:author}}
\orcid{0000-0002-7462-4516}
\author{Michael Vasmer}
\affiliation{Perimeter Institute for Theoretical Physics, Waterloo, ON N2L 2Y5, Canada}
\affiliation{Institute for Quantum Computing, University of Waterloo, Waterloo, ON N2L 3G1, Canada}
\orcid{0000-0002-6711-5924}
\author{Ashlesha Patil}
\affiliation{Xanadu, Toronto, ON, M5G 2C8, Canada}
\affiliation{College of Optical Sciences, University of Arizona, Tucson, Arizona 85719, USA}
\author{Ilan~Tzitrin}
\affiliation{Xanadu, Toronto, ON, M5G 2C8, Canada}
\affiliation{Department of Physics, University of Toronto, Toronto, Canada}
\orcid{0000-0001-5201-3987}
\author{Takaya Matsuura}
\affiliation{Xanadu, Toronto, ON, M5G 2C8, Canada}
\affiliation{Department of Applied Physics, Graduate School of Engineering, The University of Tokyo, 7–3–1 Hongo, Bunkyo-ku, Tokyo 113–8656, Japan}
\orcid{0000-0003-4164-4307}
\author{Daiqin Su}
\affiliation{Xanadu, Toronto, ON, M5G 2C8, Canada}
\author{Ben Q.~Baragiola}
\affiliation{Xanadu, Toronto, ON, M5G 2C8, Canada}
\affiliation{Centre for Quantum Computation and Communication Technology, School of Science, RMIT University, Melbourne, VIC 3000, Australia}
\orcid{0000-0003-3566-2955}
\author{Saikat~Guha}
\affiliation{Xanadu, Toronto, ON, M5G 2C8, Canada}
\affiliation{College of Optical Sciences, University of Arizona, Tucson, Arizona 85719, USA}
\orcid{0000-0002-2581-4380}
\author{Guillaume~Dauphinais}
\affiliation{Xanadu, Toronto, ON, M5G 2C8, Canada}
\author{Krishna K.~Sabapathy}
\orcid{0000-0003-3107-6844}
\affiliation{Xanadu, Toronto, ON, M5G 2C8, Canada}
\author{Nicolas C.~Menicucci}
\orcid{0000-0002-3964-233X}
\affiliation{Xanadu, Toronto, ON, M5G 2C8, Canada}
\affiliation{Centre for Quantum Computation and Communication Technology, School of Science, RMIT University, Melbourne, VIC 3000, Australia}
\author{Ish Dhand}
\thanks{ish@xanadu.ai}
\orcid{0000-0002-7041-2575}
\affiliation{Xanadu, Toronto, ON, M5G 2C8, Canada}

\date{2021-02-02}
\begin{abstract}
Photonics is the platform of choice to build a modular, easy-to-network quantum computer operating at room temperature.
However, no concrete architecture has been presented so far that exploits both the advantages of qubits encoded into states of light and the modern tools for their generation.
Here we propose such a design for a scalable fault-tolerant photonic quantum computer informed by the latest developments in theory and technology. 
Central to our architecture is the generation and manipulation of three-dimensional resource states comprising both bosonic qubits and squeezed vacuum states.
The proposal exploits state-of-the-art procedures for the non-deterministic generation of bosonic qubits combined with the strengths of continuous-variable quantum computation, namely the implementation of Clifford gates using easy-to-generate squeezed states.
Moreover, the architecture is based on two-dimensional integrated photonic chips used to produce a qubit cluster state in one temporal and two spatial dimensions.
By reducing the experimental challenges as compared to existing architectures and by enabling room-temperature quantum computation, our design opens the door to scalable fabrication and operation, which may allow photonics to leap-frog other platforms on the path to a quantum computer with millions of qubits.
\end{abstract}

\maketitle
{\def\thefootnote{*}
    \footnotetext{\textsf{\label{fn:author}These authors contributed equally.}}}

\section{Introduction}\label{Sec:Intro}
On the path to building a scalable fault-tolerant quantum computer, photonic technologies promise important advantages over other approaches. 
These include 
(i) the possibility of room-temperature computation, which allows for full miniaturization, mass manufacturing, the use of inexpensive off-the-shelf components, faster operation, and a more rapid scaling to large numbers of qubits by adopting existing silicon electronics and photonics technology;
(ii) intrinsic compatibility with communication technology, which enables high-fidelity connections between multiple modules without noisy transduction steps required in other platforms; and 
(iii) flexibility in the choice of error-correcting codes, including both the mode-to-qubit encodings and high-dimensional qubit codes using the temporal degrees of freedom of light. 
These advantages motivate serious consideration of architectures for photonic quantum computation.

\afterpage{\clearpage

## Overview of Architecture

Photonics is currently the only platform that enables building room-temperature, modular, and easily-networked quantum computers. The advantages of photonics are augmented by using qubits that are encoded into the state of light using a method proposed by Gottesman, Kitaev and Preskill (GKP). These so-called *GKP qubits* are a leading candidate for optical quantum computation because: (i) an important class of gates, operations, and measurements on these states can be performed with Gaussian resources, which are natively available and easy to implement on integrated photonic devices, and (ii) they are inherently robust to noise and optical losses. Moreover, computation with GKP qubits can be performed at room-temperature, which makes them especially attractive for the scalable fabrication and operation of quantum computers.

Current proposals for quantum computation with optical GKP qubits rely on continuous-variable (CV) cluster states, which are entangled states of many modes of light. These states have been demonstrated experimentally on vast numbers of optical modes. Measuring a CV cluster state can be used to perform Gaussian operations, but non-Gaussian components are required for a fully-fledged, universal computer. This non-Gaussian component is provided by the GKP qubits, whose generation requires a resource beyond those already mentioned. Though the optical properties of integrated photonic platforms are insufficient to provide such a non-Gaussian resource, recent technological developments in photon-counting detectors can save the day.

Specifically, GKP qubits can be produced by Gaussian boson sampling (GBS) devices. These devices displace, squeeze, and interfere light – all Gaussian operations – and then guide it toward photon-counting detectors. When photons in all but one mode of the light are counted, the light in the unmeasured mode emerges in something approaching a GKP qubit, as long as a specific photon-number pattern is observed in the detectors. Although such a process is probabilistic, that is, conditioned on the observation of this pattern, many GBS devices can be run simultaneously to boost the likelihood of making a GKP qubit. But even with this approach, termed *multiplexing*, creating a GKP qubit with near certainty requires very many GBS devices. This requirement hinders existing photonic architectures, which require that GKP qubits be available on-demand.

We propose a scalable architecture for fault-tolerant measurement-based quantum computation that overcomes this severe limitation of GKP qubit production. Our method exploits a *hybrid* resource state comprising GKP qubits at some modes and squeezed states of light at others. Multiplexed GBS devices are still used to generate GKP qubits; however, when these devices fail, the mode is instead guaranteed to be prepared in a squeezed state. This mode becomes entangled with the others, as it would in a CV cluster state. Computation can still be performed on this squeezed-state mode but now the number of GBS devices needed is no longer prohibitive.

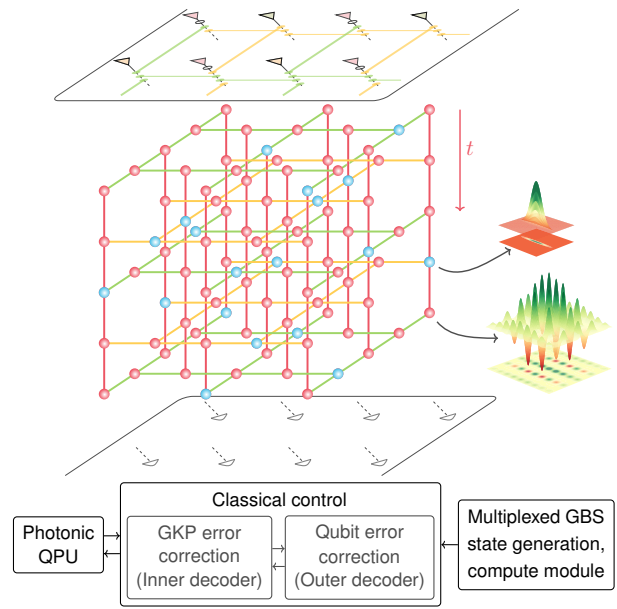

**Photonic quantum computation using hybrid resource states** A planar chip (top) generates a resource state for fault-tolerant quantum computation. The optical modes comprising the lattice are either GKP states of light (red dots) or squeezed light (blue dots); whenever the former is unavailable – its generation is probabilistic – the latter is guaranteed to be there. The light is measured at homodyne detectors (bottom), whose output is carefully decoded. Measurement settings are changed accordingly to perform measurement-based quantum computation.

Introducing Gaussian neighbours to the GKP modes of the cluster state leads to one complication though. When a GKP mode is measured to perform a gate, its intrinsic structure helps reduce the noise in the quantum state through the well-known process of GKP error correction. But when squeezed-state modes are measured, a known amount of random noise is injected into the neighbouring modes, which might degrade the quality of the computation if not accounted for. To tackle this problem, we introduce a novel decoding procedure for the hybrid cluster state. Our *decoder* takes the noisy measurement values and uses the knowledge of the squeezed state locations in order to produce better-informed qubit readout values. Then, usual qubit decoding techniques can be applied to correct any errors that arise in the computation.

Thus our architecture enables scalable fault-tolerant quantum computation with optically-generated GKP states or squeezed states of light. More than that, it uses a room-temperature moderately-sized planar photonic chip, which drastically reduces the difficulties in a scalable fabrication and running of the computer. Our planar architecture also satisfies a crucial requirement towards fault tolerance: ensuring that any noise does not extend beyond a qubit's neighbours. These advantages may allow photonic quantum computation to leap-frog other platforms in the quest to build a scalable fault-tolerant universal quantum computer operating on millions of qubits.

\clearpage}

Current architectures for scalable and universal photonic quantum computing live on two extremes. 
The first type leverages the impressive scalability of continuous-variable (CV) entangled resource states to implement computation on discrete-variable (DV) information (specifically, qubits) encoded in bosonic modes~\cite{Menicucci2014,andersen2015hybrid}. 
While the type of CV resource required for this first approach can be produced deterministically and scalably, these architectures require DV resources also to be generated on-demand and deterministically, which imposes infeasible hardware requirements.
The second type of architectures involves generating entangled resource states made entirely out of the bosonic qubits themselves~\cite{knill2001scheme, Myers2011, fukui2018high, Auger2018, Noh2020}.
Resources that exclusively comprise high-quality bosonic qubits are endowed with a degree of resilience to noise, but are more difficult to make, as either the scalable generation of the qubits or the operations required to combine them into larger states are probabilistic.
In light of this, it is important to devise a scheme that combines the best of both worlds: using CV resources to ease the burden on the preparation of bosonic qubits, but retaining a sufficiently high concentration of bosonic qubits to ensure low-noise operations whenever the CV modes are consumed. 
Here we present such a scheme and analyze how the robustness to error depends on the relative concentration of bosonic qubits in the entangled resource state. 

The first type of architectures aim to use a division of labor between Gaussian and non-Gaussian resources (see \cref{tab:gaussian}). 
The Gaussian resource is provided by easy-to-generate CV cluster states, which are multi-mode Gaussian states~\cite{Menicucci2006}. 
There has been substantial progress in designing and deterministically generating CV cluster states in one~\cite{yokoyama2013ultra, chen2014experimental, yoshikawa2016invited}, two~\cite{Alexander2016, Larsen2019,Larsen2020, Alexander2018,Asavanant2019}, and higher dimensions~\cite{wang2014weaving,wu2020quantum, fukui2020}.
In each of these architectures, the quantum information is encoded in a bosonic qubit introduced by Gottesman, Kitaev and Preskill (GKP)~\cite{Gottesman2001}. 
Clifford circuits---which make up the majority of operations required for a fault-tolerant quantum computer---can be implemented via measurement-based quantum computation (MBQC) on the CV cluster state. 
By circumventing the need for an entangled resource state made entirely out of encoded qubits (as required by the second class of architectures), this approach partially alleviates the burden on the GKP state sources. 
{However, a truly regular supply of GKP encoded states is still required; they provide the necessary non-Gaussianity, implement non-Clifford gates, and correct CV errors. 
Thus far, prior work has required that such qubits can be supplied and coupled to the cluster state deterministically at regular intervals.}

\begin{table*}
\centering
\small
\caption{Examples and implications of Gaussianity and non-Gaussianity in the context of measurement-based quantum computing with GKP qubits. 
Note that only the generation of GKP qubits requires cryogenic temperatures in our architecture. 
Qubit Clifford gates are effected with CV Gaussian transformations; qubit Pauli measurements are performed with CV Gaussian measurements; and non-Clifford gates require ancillary GKP magic states plus Gaussian transformations and measurements.}
\begin{tabular}{ccc}
\toprule
 & \textbf{Gaussian} & \textbf{Non-Gaussian}\tabularnewline
\midrule
\midrule 
\textbf{States} & 
\begin{tabular}{c}
$q$-/$p$-squeezed; CV cluster states
\end{tabular} & \begin{tabular}{c}GKP computational and magic states
\end{tabular}\tabularnewline
\midrule
\textbf{Transformations} & 
\begin{tabular}{c}
squeezing; displacement; linear optics\tabularnewline
\end{tabular} &
\begin{tabular}{c}
None
\end{tabular}
\tabularnewline
\midrule 
\textbf{Measurements} & homodyne & PNRs\tabularnewline
\midrule 
\begin{tabular}{c}
\textbf{Used to}\tabularnewline
\textbf{implement}\tabularnewline
\end{tabular} 
 & Clifford gates & Non-Clifford gates\tabularnewline
\midrule 
\begin{tabular}{c}
\textbf{Experimental}\tabularnewline
\textbf{Characteristics}\tabularnewline
\end{tabular} & 
\begin{tabular}{c}
``Easy''; room temp.; deterministic\tabularnewline
\end{tabular} & 
\begin{tabular}{c}
``Hard''; cryogenic temp.; probabilistic\tabularnewline
\end{tabular}\tabularnewline
\bottomrule
\end{tabular}
\label{tab:gaussian}
\end{table*}

The second type of architectures includes the schemes developed for the cat-basis encoding~\cite{Lund2008, Myers2011}, the GKP encoding~\cite{Fukui2018,Fukui2019}, and the dual-rail encoding~\cite{Rudolph2017}.
These approaches must contend with the non-deterministic generation of individual qubit states, particularly in the former two cases where the states have a complicated structure. 
The latter case has the added challenge of non-deterministic entangling or ``fusion'' gates, which are required to grow a cluster state. 
Each gate is eventually implemented by consuming probabilistically generated photons, which imposes formidable multiplexing requirements for cluster state generation---unlike schemes for generating CV cluster states. 

Any reliance of either type of architecture on deterministic sources of optical GKP qubits is at odds with the current state of theory and technology. 
The numerous procedures for generating optical GKP states that have been proposed tend either to be non-deterministic, as they rely on post-selected measurements directly~\cite{dakna1999,fiurasek2005,Sabapathy2019,Su2019,Quesada2019,Tzitrin2019} or indirectly~\cite{vasconcelos2010all,weigand2018,Eaton_2019}; or require the experimentally challenging conditions of coherent interactions with matter~\cite{motes2017,pirandola2006generating} or extremely strong optical nonlinearity~\cite{pirandola2006generating}.
Recent advances in photon-number-resolving (PNR) detectors~\cite{Lita2008,PhysRevLett.124.013605,PhysRevA.101.031801,Vaidya2020} have substantially improved the viability of the post-selection approach in the near term, with methods based on Gaussian boson sampling (GBS)~\cite{Sabapathy2019,Su2019,Quesada2019,Tzitrin2019} now within reach of state-of-the-art optical devices.
Low-probability sources can be improved with the help of multiplexing at the cost of an increased overhead.
That is, in order to generate states with near-unit probability $1-p_{0}$, the number of required state generation devices scales as $\log(1/p_{0})$, which is prohibitively large as $p_{0} \to 0$. 

In this work, we propose an architecture for measurement-based quantum computing that possesses the advantage of CV-based schemes and yet is compatible with probabilistic GKP qubit sources. 
We consider a \emph{hybrid} CV cluster state where each mode is substituted with a GKP qubit at random and with probability $(1-p_0)$---or said the other way, a qubit cluster state where each node is substituted with a squeezed state with \emph{swap-out probability} $p_0$. 
The precise state we consider is the lattice from the Raussendorf, Harrington, Goyal (RHG) model~\cite{Raussendorf2005,Raussendorf2006,Raussendorf2007}, but our scheme can readily accommodate other error-correcting codes. 
Our use of CV resources affords us an important alternative over existing approaches, wherein a qubit that failed to be produced must be erased from the lattice. 
Instead, we replace the no-show qubit with a squeezed vacuum state: it can still encode logical information (albeit not as well as a GKP state~\cite{pantaleoni2020modular}) but has the distinction of being Gaussian and thus easily producible~\footnote{We cannot replace all modes with squeezed vacua because that would negate the error-correction benefits of the encoded qubits~\cite{bartlett2002efficient}.}. 
This approach -- one of the main innovations in this work -- propels us beyond existing fault tolerance methods, such as those that rely on lattice renormalization to deal with defects~\cite{stace2009thresholds,barrett2010fault,Auger2018,whiteside2014}. 
To characterize the robustness of our architecture as a function of $p_0$, we perform Monte Carlo simulations of our architecture operating as a quantum memory. 
We observe a minimum required squeezing of $\SI{10.5}{\decibel}$ or a maximum tolerable swap-out probability of $p_0 \approx 0.236$; for an experimentally accessible squeezing value of $\SI{15}{\decibel}$~\cite{PhysRevLett.117.110801}, our simulations suggest a swap-out threshold of $p_0 \approx 0.133$, which translates to substantially reduced multiplexing requirements for GKP generation. 
In part these result stem from a tailored decoding procedures that we present and which allow us to perform fault-tolerant computation on our hybrid resource state. 

A key feature of our architecture is that it promises full scalability to a large number of qubits, as required for fault tolerance.  
Working in one temporal and two spatial dimensions ensures that each mode traverses a path of constant optical depth that is independent of the number of qubits in the computer.
This is in contrast to existing schemes for CV computation, where increasing the numbers of qubits requires either longer time delays~\cite{Asavanant2019,Larsen2019}, longer measurement integration times, or more precise spectral resolution~\cite{wang2014weaving,wu2020quantum}.
Such an increase will tend to result in exponentially growing losses so these architectures cannot be scaled indefinitely.
Moreover, the components of our architecture can be arranged as a planar graph: each qubit is connected only to a small and constant number of neighboring qubits and the layout of the computational chip consequently requires no `swaps' or intersecting wave-guides.
This planar structure not only opens the possibility of scalable fabrication but also allows for preserving the local structure of the noise.
An uncorrelated noise structure like the one enabled by our architecture is critical as it allows exploiting the full machinery of fault-tolerance.

Finally, our architecture poses modest experimental requirements as compared to other architectures for photonic quantum computing.
This is because the individual modules involved in our architecture are specialized.
Consider as an example the challenge of low-loss and fast reconfigurable optical switching in cryogenic conditions~\cite{Collins2013}.
For our architecture, the state-generation modules are required to be low-loss, but not reconfigurable; the multiplexing modules pose less severe loss requirements and are more easily made programmable; and the computational modules allow for fairly lossy reconfigurable switches.
Furthermore, the computational module allows for operation at room temperature and without requiring vacuum, thus promising favorable scalability in manufacture via modern lithographic fabrication processes with minimal change. 
Thus, a hybrid resource state for quantum computing along with its accompanying decoder and a scalable and hardware-friendly architecture that computes with this state are the main results of this work.

The paper is structured as follows. 
The inset of the second page provides an overview of the main results, and \cref{Sec:Background} provides the necessary background. 
Following this, the planar architecture is detailed in \cref{Sec:scheme}, and the method to implement quantum error correction, including the specialized decoder for the hybrid lattice, is presented in \cref{Sec:QEC}.
\cref{sec:FTQC} details the fault-tolerant logical-level quantum computation and \cref{Sec:Threshold} presents the fault-tolerance thresholds for our architecture.
We discuss open challenges and technological advantages in \cref{Sec:Open,Sec:Discussions}.

\section{Background on Quantum Computation Using CV Systems}\label{Sec:Background}

In this section, starting with \cref{sec:encodedqubits}, we present the relevant background for our architecture, before which we introduce briefly the field of photonic CV quantum computing,
The physical systems that our architecture computes with -- modes of light -- are infinite dimensional. 
The generalization from a qubit to a qudit computational model, that is, from two-level to $d$-level quantum registers is relatively straightforward. 
But the jump to formally infinite-dimensional systems introduces a few technical complications.

The original CV computational model was proposed in Ref.~\cite{lloyd1999quantum}. 
In analogy to the qubit and qudit stabilizer formalisms~\cite{gottesman1997stabilizer,knill1996non,ketkar2006nonbinary}, CV quantum computation also possesses an efficiently simulable sub-theory, Gaussian computation~\cite{bartlett2002efficient}. 
Unlike in the discrete variable case, however, attempting to encode data in a way that uses the full Hilbert space available to a bosonic mode is physically impossible. 
This is because infinite-dimensional data registers are extremely sensitive to noise. 
Since every mode can be expected to be exposed to (at least) weak noise sources, entangled modal states will be corrupted by high-weight errors that are beyond the capabilities of quantum error correction.  

Nevertheless, consideration of the CV paradigm has been fruitful on at least two fronts:
first, CV cluster states can be generated deterministically, on a large scale, and with constant-depth, local, linear-optical networks; second, the CV state space can house bosonic codes, which are rich families of wave-functions that can be used to encode discrete-variable quantum information with desirable properties such as robustness to decoherence and experimental convenience in the generation of ancill{\ae} and the implementation of logic gates and measurements. 
Using bosonic codes solves the above conundrum: encoded qubits convert high-weight weak noise sources to low-weight qubit-level noise that is compatible with conventional fault-tolerant architectures for quantum computation. 
The next section provides relevant background on bosonic encodings.

\subsection{Qubits Encoded Into Bosonic Modes}
\label{sec:encodedqubits}

Bosonic qubit encodings are two-dimensional subspaces within the infinite-dimensional Hilbert space of a bosonic mode. 
Good choices of this two-dimensional subspace allow for experimentally convenient ways of preparing the encoded qubit states, implementing desired unitary gates, and faithfully performing measurement readout. 
In some cases, the redundancy of the full infinite-dimensional Hilbert space can even be leveraged to detect and correct CV errors -- random Gaussian displacements, rotations, and photon loss, for a few -- without destroying the encoded information. 
Examples of bosonic codes include GKP~\cite{Gottesman2001}, dual-rail~\cite{chuang1995simple,knill2001scheme}, cat~\cite{PhysRevA.59.2631,ralph2003quantum}, hypercat~\cite{PhysRevLett.111.120501,Mirrahimi2014}, binomial~\cite{PhysRevX.6.031006}, and general rotation-symmetric codes~\cite{PhysRevX.10.011058}.

For reasons we will describe, our architecture exploits the GKP encoding. 
For notational convenience we restrict our discussion to square-lattice GKP encoding but the results can be generalized to GKP states on other lattices.
In their ideal form, the GKP qubit states $\ket{0}_{\rm gkp}$ and $\ket{1}_{\rm gkp}$ are defined as Dirac delta combs with a spacing of $2\sqrt{\pi}$ in position space~\cite{Gottesman2001}:
\begin{equation}
    \ket{\mu}_{\rm gkp} = \sum_{n} \ket{(2n+\mu)\sqrt{\pi}}_q,\,\mu=0,1.
\end{equation}
The chief advantage of GKP states is that qubit Clifford operations map to CV Gaussian operations, which can be implemented \textit{deterministically} and \textit{easily} using linear optical elements, homodyne detection, and Gaussian states of light~\cite{Gottesman2001}.
In \cref{sec:optical_elements}, we provide optical circuits for the application of GKP qubit gates in detail.
Briefly, Pauli $X$ and $Z$ gates on GKP qubits correspond to displacements along $q$ and $p$ by $\sqrt{\pi}$, respectively.
The Hadamard gate is simply a $\pi/2$ rotation in phase space, implementable by a phase-shifter.
The qubit phase gate $\sqrt{Z}$ corresponds to phase-space shear, which can be implemented by a single-mode squeezer sandwiched between two phase-shifters.
The qubit $CNOT$ and $CZ$ gates correspond to $CX$ and $CZ$ gates in the CV domain, respectively, which in turn can be broken down into a pair of single-mode squeezers between two beam-splitters.
Deterministic all-optical entangling gates are a distinct advantage of the GKP encoding over dual-rail encoding schemes~\cite{knill2001scheme}.
Moreover, qubit Pauli measurements correspond simply to homodyne measurements, which operate at faster speeds and higher efficiencies than photon-counting detectors~\cite{RevModPhys.81.299, PhysRevLett.117.110801}. 
Finally, non-Clifford operations require a non-Gaussian resource.
Unitary implementations of the $T$ gate can be achieved using a cubic-phase interaction, though this is difficult in practice~\cite{miyata2016implementation}, and does not perform well for finite-energy states~\cite{Hastrup2020}.
As an alternative, $T$ gates can be performed through gate teleportation by preparation of a GKP magic state~\cite{Gottesman2001}, which we also review in \cref{sec:optical_elements}.

Using bosonic codes as the physical qubits for a fault-tolerant quantum computing architecture provides two tiers of protection from noise. 
The first comes from the bosonic code itself. 
GKP qubits possess a degree of intrinsic robustness to those bosonic noise channels that result in small displacements (relative to the $\sqrt{\pi}$ lattice spacing) in phase space. 
This includes weak levels of photon loss, the dominant error mode for quantum communication~\cite{Albert2018,noh2018quantum}. 
Any shift that is less than half the lattice spacing ($\sqrt{\pi}/2$) can be corrected by non-destructively measuring the GKP stabilizers---which are $2\sqrt{\pi}$ shifts in either position or momentum.
This CV error-correction procedure outputs continuous syndrome data, which can be used to undo the displacement with the help of a decoder. 
Noise that leads to larger displacements can result in errors that are undetectable by measuring only the GKP qubit stabilizers. 
In the fault-tolerant regime, these larger displacements are much less likely to occur, and so occasional errors on GKP qubits can be corrected by applying the second layer of protection: a qubit quantum error-correcting code. 
Implementing these codes requires only Clifford gates and Pauli measurements, both of which are easy (Gaussian) for the GKP qubit encoding. 

An essential part of any quantum error correction procedure is the \textit{decoder}, which specifies the recovery operation that has to be applied for given syndrome data.  
The two-layer structure of the error correction described above requires two stages of decoding. 
The first of these stages translates continuous GKP-stabilizer syndrome data into the operations required to return to the GKP code subspace housed in each mode, up to qubit-level errors. 
It can also provide some detailed information about the relative likelihood of different discrete qubit-level errors~\cite{Fukui2018}. 
The second stage maps syndrome data obtained from measuring the higher-level qubit-code stabilizers to a qubit-level recovery operation. 
To avoid confusion, we refer to the former as the \emph{inner decoder}, and the latter as the \emph{outer decoder}. 
Decoders that are tailored for our architecture are described in more detail in \cref{subsec:translator,sec:outer}, respectively. 

Unfortunately, ideal GKP states are non-normalizable states with infinite energy.
Many related methods can be used for defining finite-energy versions of these states~\cite{Tzitrin2019, PhysRevA.102.032408}.
A common approximation is to replace each delta function in the GKP wave-function with a Gaussian of width $\Delta$, in addition to an overall Gaussian envelope of width $1/\Delta$ that damps the peak weighting further from the origin~\cite{Gottesman2001}: 
\begin{align}\label{eq:finiteE_GKP}
\begin{split}
    \ket{\mu^\Delta}_{\rm gkp} \equiv& \frac{1}{N_\mu}\int_{-\infty}^{+\infty}ds\sum_{n}e^{-\Delta^2[(2\sqrt{\pi}+\mu)n]^2/2} \\&\hspace{1.5cm}\times e^{-[s-(2\sqrt{\pi}+\mu)n]^2/2\Delta^2}\ket{s}_q,
\end{split}
\end{align}
where $N_\mu$ is a normalization constant.
While finite-energy effects will be ever-present in any real implementation, the noise they introduce does not preclude GKP states from being useful for error correction and fault-tolerant quantum computation~\cite{PhysRevA.73.012325,Menicucci2014,Albert2018}.

Next, we review a method for preparing approximate GKP states using Gaussian resources and PNR detectors. 
We follow this with a review of how GKP states interface with CV cluster states and qubit quantum error-correcting codes.

\subsection{Generating Bosonic Qubits via GBS} \label{sec:GBScreate}
Recent theoretical and experimental breakthroughs have made post-selected schemes via GBS as the most promising candidates for generating non-Gaussian states.
GBS devices are capable of preparing highly non-Gaussian states -- including GKP qubits -- contingent on the observation of a specific detection pattern in the PNR detectors.
This preparation of non-Gaussian states of light including single bosonic qubits using GBS devices has been developed in Refs.~\cite{Tzitrin2019,Quesada2019,Su2019,Sabapathy2019}, and we summarize the relevant portions briefly here.

\begin{figure}
\includegraphics[width = \columnwidth]{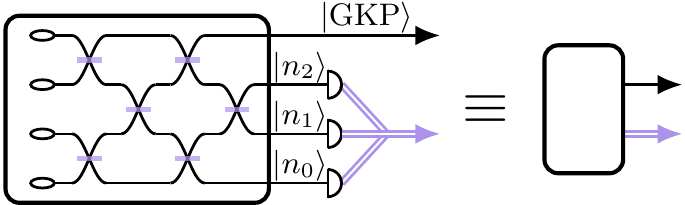}
\caption{GBS devices for state preparation.
(left) A single integrated photonic device implementing GBS-based preparation of non-Gaussian states based on the schemes presented in Refs.~\cite{Tzitrin2019,Quesada2019,Su2019,Sabapathy2019}.
The emitted light from one output port is in a chosen non-Gaussian state subject to obtaining the correct click pattern $\{n_{i}\}$ at the PNR detectors connected to the remaining output ports. 
The double purple lines represent classical logic, which is used to trigger a switch on the emitted port.
(right) A simplified representation of a single GBS device.
\label{Fig:GBS}}
\end{figure}

GBS state preparation consists of sending $N$ displaced squeezed vacuum states into a general interferometer on $N$ modes, followed by PNR detectors on $N-1$ of the modes, as depicted in \cref{Fig:GBS}.
The number of modes, the displacement, squeezing, and inteferometer parameters, as well as the photon number pattern at the PNR detectors, can all be tuned so that the device can herald the desired high-fidelity non-Gaussian output state. 
This procedure exploits the non-Gaussianity of PNR detectors and can generate arbitrary logical single-qubit states for a variety of bosonic encodings, including the GKP and cat encodings. 
As the generation of the desired state requires a particular pattern of photon number detection outcomes to be observed, the generation is non-deterministic but heralded.

As a concrete example, consider that small-scale GBS devices made up of 3-mode interferometers, two PNR detectors registering up to 7 photons, and three momentum-squeezed vacuum states with up to $\SI{12}{\decibel}$  of squeezing have the potential of producing $\ket{0^\Delta}_{\rm gkp}$ GKP states with $\Delta^2=0.1$ ($\Delta_{\text{dB}}=\SI{10}{\decibel}$ , for $\Delta=10^{-\Delta_{\text{dB}}/20}$) with a fidelity of 76\% (92\%, and 96\%) and heralding success probability of 2.1\% (0.4\% and 0.1\%)~\cite{Tzitrin2019}.
We see here a fidelity-probability trade-off for a fixed number of modes.
Comparable results are observed for the preparation of finite-energy GKP magic states.

Although high-fidelity state generation from a single GBS device is non-deterministic, these sources can be multiplexed to obtain higher rates of generation, as we detail in \cref{subsec:MUX} for our architecture.
Additional hardware resources, both in the individual GBS devices and in the multiplexing, can be used to increase the rates and fidelities of the generated states.

\subsection{Measurement-Based Quantum Computing With Canonical CV Cluster States and GKP Qubits} \label{sec:canCVCS}

Modes of light are not well suited to serving as stationary quantum data registers. 
Optical modes can interact with only a few optical elements before they must be measured or else lost. 
Fortunately, this constraint is compatible with the measurement-based model for quantum computing, where each quantum data register is entangled to a constant number of others, and then measured at a detector---the entire computation being specified by which measurements are chosen.
Here we review relevant details and terminology on CV photonic measurement-based quantum computation.

Measurement-based quantum computation in the qubit setting involves preparing an entangled resource state (most commonly, a cluster state~\cite{briegel2001persistent}), and performing a sequence of single-site adaptive measurements~\cite{raussendorf2001one}. 
These cluster states are specified by a graph; for each node a qubit is prepared in the $\ket{+}$ state and for each edge a $CZ$ gate is applied. 
Cluster states are said to be \emph{universal resources} if they enable universal quantum computation when given access to adaptive single-site measurements.

This paradigm can be generalized to the CV degrees of freedom present in a bosonic mode: 
CV cluster states are Gaussian entangled states that enable CV measurement-based quantum computation via local measurements~\cite{Menicucci2006}.
The simplest of these are referred to as \emph{canonical} CV cluster states~\cite{Gu2009}, which are constructed by applying controlled-Z gates $e^{i \hat{q}\otimes\hat{q}}$ to momentum-squeezed vacuum states. 
CV cluster states with any graph can be generated on demand since both the controlled-Z gates, and the preparation of momentum-squeezed vacuum states can be implemented deterministically. 

Ideal CV cluster states cannot be normalized and correspond to unphysical infinite-energy states. In a similar way to the GKP qubit, the approximate nature of physical CV cluster states can be captured by a finite-width Gaussian envelope structure of the state's position space wavefunction:
\begin{align}
\begin{split}
    \ket{C(\mathbf{V}, \epsilon)} &= e^{i \hat{\bm{q}}^{\text{T}} \mathbf{V} \hat{\bm{q}}/2} \left[ \frac{\sqrt{\epsilon}}{\pi ^{1/4}}\int_{-\infty}^{\infty} ds\, e^{-s^{2} \epsilon/2} \ket{s}_q\right]^{\otimes N} \\
    &=\left[\frac{\epsilon}{\sqrt{\pi}}\right]^{\frac{N}{2}}\int_{\mathbb{R}^{N}} d^{N}\bm{s}\, e^{i \bm{s}^{\text{T}} \mathbf{V} \bm{s}/2} e^{-\bm{s}^{\text{T}}  \bm{s}\epsilon/2} \ket{\bm{s}}_{\bm{q}}, \label{eq:CVCS}
\end{split}
\end{align}
where $\mathbf{V}$ is a real symmetric adjacency matrix corresponding to the cluster state's graph and $\epsilon/2$ is the variance of each momentum-squeezed vacuum state in the momentum quadrature. 
The case of $\epsilon\rightarrow 0$ corresponds to the infinite squeezing limit. 

Note that the CV controlled-Z gate $e^{i \hat{q}\otimes\hat{q}}$ is common to both the GKP qubit encoding and canonical CV cluster state generation. 
Therefore it is possible to generate a \emph{hybrid} cluster state with nodes comprising momentum-squeezed states, GKP qubits, and their common $CZ$ gates~\cite{Menicucci2014}. 
As we detail in the results sections, this is one of the key concepts that enables computation with our hybrid resource state.

\subsection{Cluster States and Fault Tolerance}
\label{subsec:mbqc}

That quantum information can be processed reliably in the presence of noise is key achievement of quantum error correction~\cite{shor1995,steane1996a}. 
Given a logical circuit to be implemented, the idea is to redundantly encode the information content of the logical qubits into larger collections of physical qubits and perform computation on these collections.
If physical qubits of sufficient quality are available and if sufficiently precise operations and measurements can be performed on these qubits such that the noise strength remains below the threshold of the specific code used, then the logical quantum circuit can, in principle, be applied with arbitrarily high precision~\cite{aharonov2008,knill1998,Campbell_2017}.
The study of these thresholds and the physical-qubit overheads associated with different error-correcting codes is an active area of research.

In practice, it is often desirable to choose a quantum error-correcting code capable of tolerating a high error rate and not requiring long-ranged connectivity between physical qubits. The \emph{surface code}~\cite{Kitaev2003,bravyi1998quantum} is a commonly used code because it has both these properties, enjoying a high threshold of $\sim 1 \%$~\cite{Raussendorf_2007,PhysRevA.80.052312, Fowler_2012} and only needing nearest-neighbor connectivity of qubits in two spatial dimensions. In addition, it is highly resistant to erasure errors, even up to $50\%$ qubit loss~\cite{stace2009thresholds}. However, optical quantum computing architectures are better suited to measurement-based implementations of quantum error-correcting codes. In the case of the so-called Calderbank-Shor-Steane (CSS) codes~\cite{calderbank1996,steane1996}, this can be implemented using cluster states corresponding to foliated (i.e., layered) lattice sheets that implement CSS code stabilizer measurement gadgets~\cite{Bolt2016}.

\begin{figure}
    \centering
    \includegraphics[width=0.9\columnwidth]{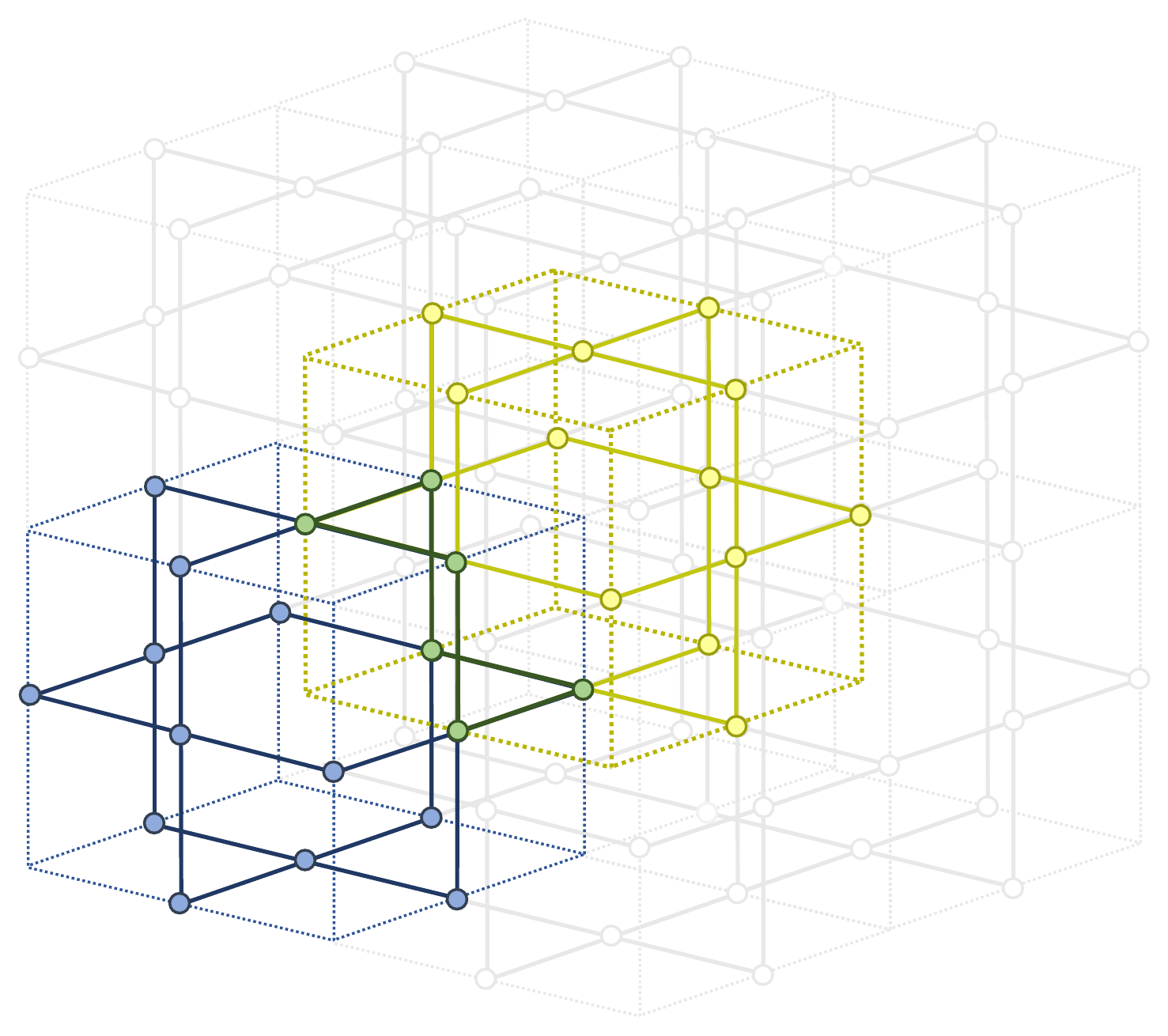}
    \caption{A primal cell of the RHG lattice, i.e., a $2\times2\times2$ stack of unit cells (blue), with the dual cell identified in the middle (yellow). The nodes and edges which overlap are highlighted in green. The faces of the cells are also called primal or dual, and comprise primal or dual boundaries. In surface code terms, smooth (rough) boundaries end on the faces of primal (dual) cubes. All the links are $CZ$ gates.}
    \label{fig:pcell}
\end{figure}

Perhaps the best studied cluster-state based error-correcting code is the RHG lattice. 
In the foliated picture, it can be thought of as alternating so-called primal and dual sheets of 2D cluster states that encode the surface code. 
This special topology is what leads to fault tolerance: error detection and decoding involve multiple mutually entangled layers at a time. 
The RHG lattice serves as a good first candidate to study for our architecture because not only is it universal for MBQC, but it has high fault-tolerant computational error thresholds ($\lesssim 1\%$). 
A full fault-tolerant computation can be performed on the RHG lattice with the help of the RHG scheme~\cite{Raussendorf2005, Raussendorf2006, Raussendorf2007, Fowler2009}. 
A schematic of the RHG lattice is presented in \cref{fig:pcell}.
The details of the full computation on this lattice---state initialization, gate application, measurement, and error correction---will be reviewed and adapted to our architecture in \cref{Sec:QEC,sec:FTQC}.

Other lattices such as non-foliated lattices~\cite{Nickerson2018,Bombin2018,Bombin2018a} or non-CSS codes~\cite{brown2018universal} can also be considered, including possibly new lattice designs tailored to errors in CV quantum computing.

\subsection{Bosonic Codes Concatenated With Topological Codes for Quantum Computation}

The discussion in the previous section deals with cluster states composed of qubits, which can include qubits encoded in bosonic modes.
This two-layer encoding can be seen as a bosonic code \textit{concatenated} with a topological code, a subject of growing interest~\cite{Vuillot2019,Noh2020,Fukui2018,Fukui2019,hanggli2020enhanced,terhal2020towards,yamasaki2020polylog}.
The main motivation here is that CV errors larger than those the inner bosonic code can handle are picked up and corrected by the outer qubit code.

A gate-based model for the concatenated GKP-surface code catering to a superconducting platform was considered in Refs.~\cite{terhal2020towards,Vuillot2019,Noh2020}. 
In these works, each GKP qubit in the surface code first undergoes a round of GKP error correction, after which the stabilizer measurements for the surface code are performed. 
The noise considered in Ref.~\cite{terhal2020towards,Vuillot2019} is a Gaussian classical noise channel applied to all GKP qubits and prior to all homodyne measurements in both the GKP error correction and the parity check measurements. 
Ref.~\cite{Noh2020} considers the same noise model but applied now to the circuit level, and explicitly includes error feedback. 
An improved threshold for the surface-GKP code was obtained in Ref.~\cite{hanggli2020enhanced} by designing bias in the noise.

Measurement-based topological quantum computation using GKP qubits, compatible with a photonic architecture, was considered in Refs.~\cite{Fukui2018,Fukui2019}. 
The approach was to generate a 3D cluster state to implement a measurement-based analogue of the surface code using GKP qubits. 
The cluster state was generated using a post-selection (fusion-based) approach that is non-deterministic by nature. 
This work also introduced an analogue quantum-error-correction scheme, where the real-valued measurement outcome from the homodyne measurement was explicitly used in the decoding procedure. The errors considered in Ref.~\cite{Fukui2018} are finite squeezing effects in the GKP state preparation along with a Gaussian random displacement noise. 
More realistic noise, such as loss in the entangling gates and the homodyne measurements, was also considered in Ref.~\cite{Fukui2019}. 
Furthermore, the GKP state preparation is considered to be deterministic and the final state generated is a connected cluster populated only by GKP states.

\section{An Architecture for Photonic Quantum Computing With Hybrid Resource States} \label{Sec:scheme}

This section describes the different components of our architecture.
The architecture has information encoded in GKP qubits because of the advantages previously discussed.
The encoded information is processed in a measurement-based setting using a hybrid resource state comprising GKP qubits and squeezed states using the components that we now describe.

The architecture comprises four modules:
three modules together generate a computational resource state in one temporal and two spatial dimensions using a planar photonic chip and the final module performs measurement-based quantum computing on this state.
First, the {\it state-preparation} module generates high-quality GKP qubits, albeit with low probability.
The {\it multiplexing} module boosts the qubit generation rates and, in the event of a qubit generation failure, substitutes in a momentum-squeezed vacuum mode.
The {\it computational} module implements the deterministic entangling operations, thereby enabling universal and fault-tolerant measurement-based quantum computation.
The final module, the photonic quantum processing unit or {\it photonic QPU} performs homodyne measurements on the generated resource state in order to perform the required computation.
We note that the first two modules are entirely dedicated to the preparation of single-mode states; entanglement and measurement are relegated to the third and fourth modules.

The resulting quantum state is a hybrid resource, made up of both GKP qubits -- Pauli eigenstates and magic states -- and squeezed-state (or CV) modes.
This structure is compatible with the probabilistic nature of the encoded qubit sources; given access to a hypothetical deterministic sources, a resource state could be constructed entirely out of GKP qubits.   
Though probabilistic sources of qubits can be treated through heralded erasure errors (modelled as the application of a maximal-strength depolarizing channel), we find a better strategy, namely to use readily available momentum-squeezed vacuum states as a substitute for missing GKP states.
We call this replacement a `swap-out'.
Momentum-squeezed states preserve the entanglement structure provided by the $CZ$ gates and do not introduce as much noise as qubit level erasure channels.
We describe the multiplexed state generation in \cref{subsec:MUX}, the computational module in \cref{sec:compmodule}, and the photonic QPU in \cref{Sec:qpu}.

\subsection{Multiplexed GBS Devices for High-Probability GKP Generation}\label{subsec:MUX}

\begin{figure}
\centering
\includegraphics[width = 0.7\columnwidth, valign = c]{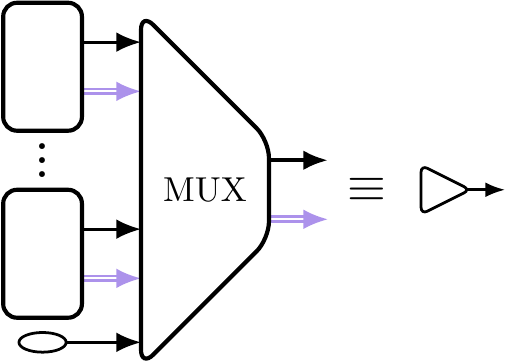}
\caption{\textbf{Multiplexed state generation.} 
Multiplexed GBS devices for increased rate of state preparation.
The multiplexer consists of a binary tree of $2\times2$ switches that either implements an identity or SWAP gate on each optical mode, moving a successfully generated GKP state to the correct output port.
If no GBS device produces a GKP state, we swap the output of the multiplexing device for a deterministically generated momentum-squeezed state (depicted by the ellipse on bottom left).
The right-hand side shows the simplified diagram for the hybrid quantum light source.
Note that the classical information wire is suppressed.
\label{Fig:farm}}
\end{figure}

GBS devices (see Fig. \ref{Fig:GBS}) can be exploited as probabilistic sources of GKP qubits, as reviewed in \cref{sec:GBScreate}.
The success probability of these sources can be boosted at the cost of increased overhead by using multiplexing, i.e., redundantly running multiple sources in parallel.
For a fixed required fidelity, the generation rate for GKP states can be boosted to arbitrary desired probability values $1-p_0$ by using spatial multiplexing, as illustrated in \cref{Fig:farm}.
Specifically, if a single GBS device can prepare a GKP qubit with probability $p_{\text{GBS}}$, we need that 
\begin{equation}
    1 - (1 - p_{\text{GBS}})^{N_\text{GBS}} \geq 1-p_0,
\end{equation}
where $N_\text{GBS}$ is the number of GBS devices.

Multiplexing requires active feed-forwarding of PNR detector outcomes, which can be implemented using $2 \times 2$ ``crossbar" switches.
These switches operate in two modes: a bar that effects the identity and a cross that effects the SWAP gate.
This operation can be realised through variable-transmissivity beam-splitters, or Mach-Zehnder interferometers with variable phase-shifters.
A binary tree of these kinds of switches is sufficient to move a successfully prepared state into the correct output port~\cite{Bonneau2015}.

If $D$ is the depth of this tree, then the total number of switches is given by $\sum_{i = 1}^D 2^{i - 1} = 2^D - 1$.
For a fixed number of GBS devices $N_\text{GBS}$, the depth should be
\begin{equation}
    D =\left \lceil \log_2({N_\text{GBS}} + 1) \right \rceil.
\end{equation}
In the event that no GBS device successfully produces a GKP state, we include an additional switch at the output of the multiplexing device which can swap in a momentum-squeezed state to replace the output of the multiplexed GBS devices, as shown in \cref{Fig:farm}.

While the GBS devices can be operated at high repetition rates by designing sufficiently high-bandwidth squeezers, the maximum repetition rate experienced by state-of-the-art photon number resolving detectors based on transition-edge sensors (TESs) is likely limited to a few MHz.
While other detectors based on superconducting nano-wires are available, these do not meet the number-resolving capabilities required for the suitable operation of the GBS device.
To boost the acceptable operating clock frequency, i.e., the pulse repetition rate, time-to-space demultiplexers can be used to step down the repetition rate seen by each PNR channel from that used to pump the GBS devices {\cite{Lenzini2017}}.
For example, to step down a 1024 MHz pulse train to 8 MHz, each 128 clock cycles on one output are mapped to 128 separate spatial outputs; this requires 7 layers in a binary tree configuration, with 127 active elements.
As the TES readout and data acquisition process takes some time, optical delay lines can be used to buffer the heralded outputs of the GBS device before they reach the space-to-space multiplexers, to provide enough time for the multiplexer routing configuration to be actuated.
Assuming this readout time is limited by the rise time of the TES output voltage traces, a few nanoseconds of buffer delay can be used, corresponding to several metres of low-loss optical fiber or integrated waveguide length.
An alternative to demultiplexing is to interleave pulses arriving from multiple GBS devices, which are operating at the timescale of the TES detectors.
Since the light pulses themselves are short enough, pulses arriving from $k$ different devices can be switched into a single spatial mode but at different arrival times that are separated by the faster clock cycle rates.
This alternative does away with the requirement of fast fed-forward switches at the cost of introducing more squeezers and linear optical elements in GBS devices. 
Either of these alternatives or a combination thereof can be chosen based on the actual hardware considerations.
In a nutshell, the multiplexing module is responsible for the boosting of GKP-generation probabilities and for stepping up the generation rates from the PNR speeds to the computational clock speeds.

\subsection{Generating (2+1)D Computational Resource States} \label{sec:compmodule}

\begin{figure}\centering
\subfloat[\label{Fig:CZa}]{
\includegraphics[width = 0.9\columnwidth]{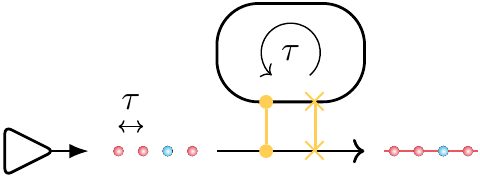}}\\
\subfloat[\label{Fig:CZb}]{
\includegraphics[width = 0.7\columnwidth]{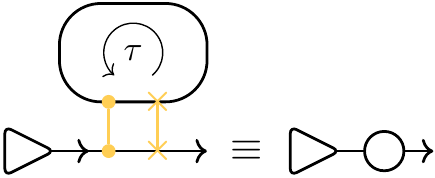}}
\caption{
\textbf{Generating 1D qubit cluster in the time domain}. 
(a) On the left, a `GBS factory' comprising multiplexed GBS devices is used to generate the sequence of pulses, where each pulse contains either a GKP $\ket{+}$ state or a momentum-squeezed state.
Each input interacts with the previous input (which is in the loop) via a $CZ$ gate, enters the loop mode via the swap, interacts with the next mode, and then is swapped into the output mode by the same swap. 
(b) Simplified diagram for 1D time-domain cluster state source.
}
\end{figure}

\begin{figure*}
\centering
\subfloat[\label{Fig:NewChipd}]{\includegraphics[valign=c,width = 0.6\columnwidth]{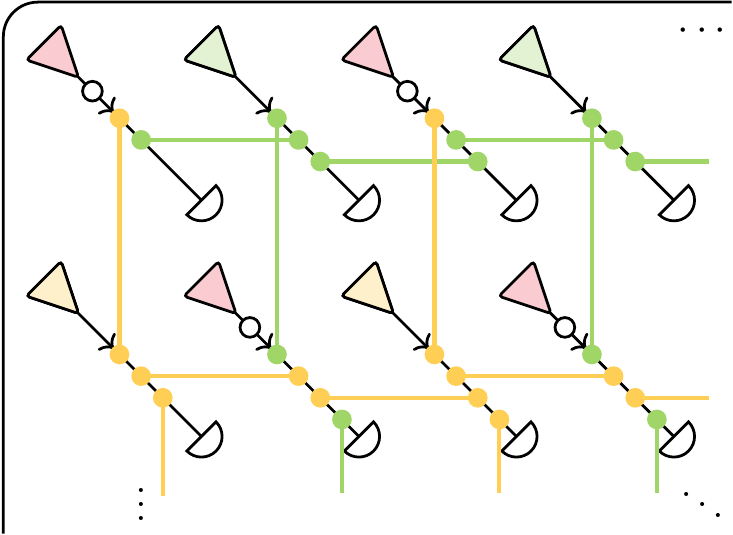}}
\hspace{0.5cm}
\subfloat[\label{Fig:NewChipa}]{\includegraphics[valign=c,width = 0.3\columnwidth]{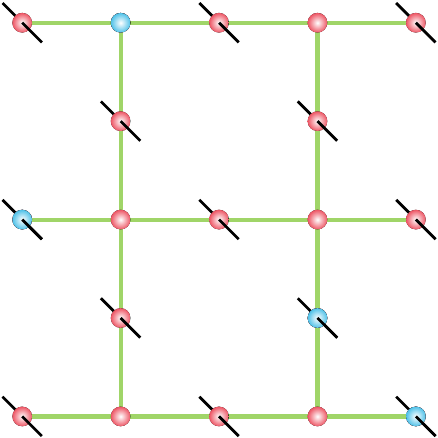}\hspace{0.2cm}\includegraphics[valign=c,width = 0.3\columnwidth]{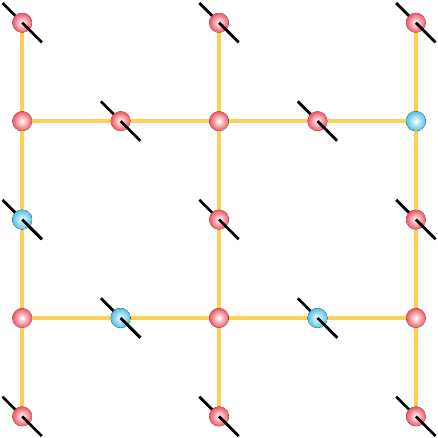}}
\hspace{0.5cm}
\subfloat[\label{Fig:NewChipRHG}]{\includegraphics[valign=c,width = 0.6\columnwidth]{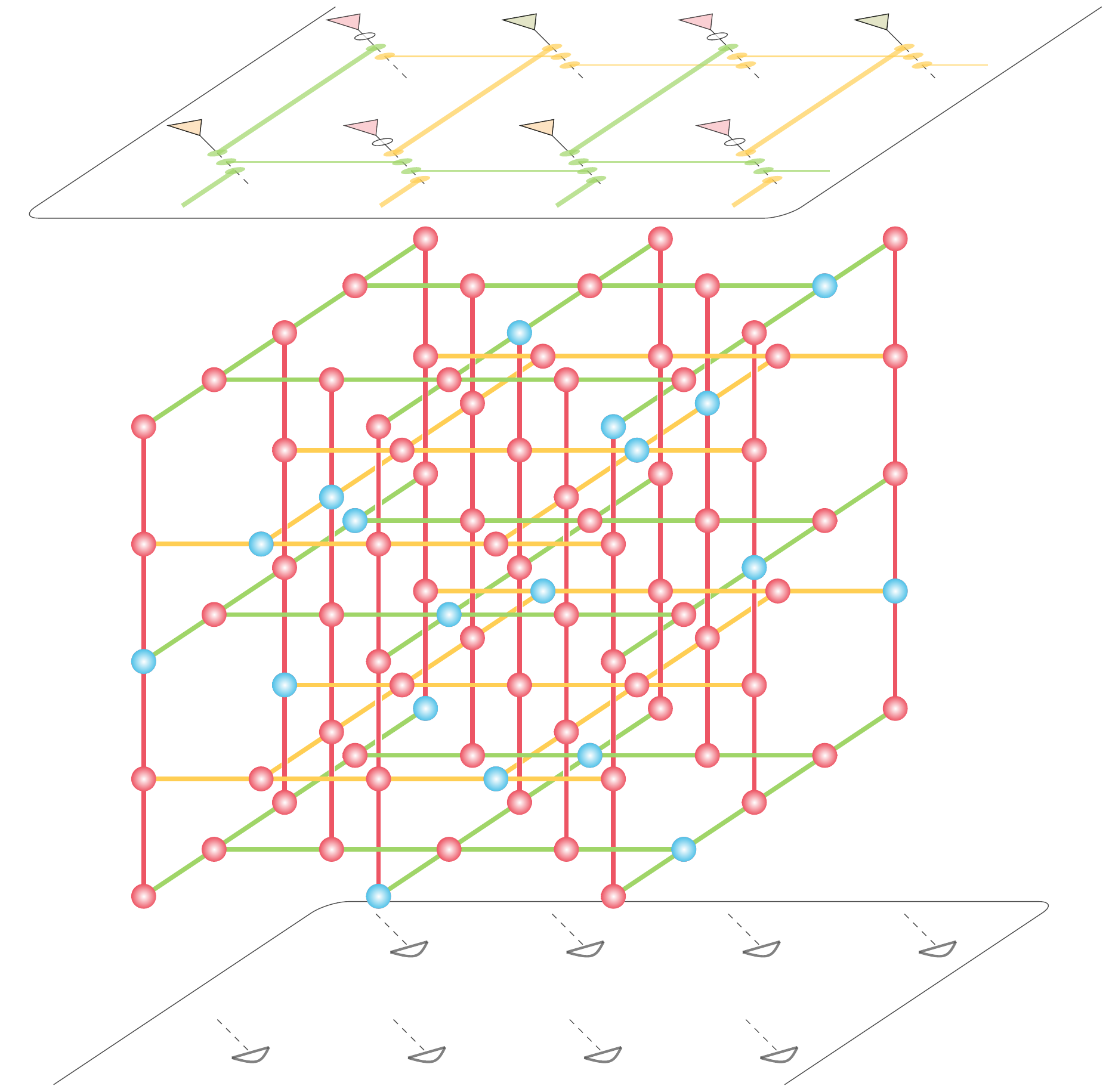}}
\caption{\label{Fig:Raussendorf-Gen}
(a) Chip layout to generate the hybrid RHG lattice. 
Light from three kinds of sources is incident on the chip. 
The red triangles with circles are sources of hybrid 1D cluster states in the time domain, with a sequence of entangled qubits being emitted at a time delay of $\tau$. 
The remaining yellow and green triangles are simply qubit sources, but these sources fire only at a time interval of $T = 2\tau$, with yellow sources firing only at the $(2n-1)\tau$ times and the green sources firing at the $(2n)\tau$ times. 
The lines on the chip represent $CZ$ gates. The yellow $CZ$ gates are turned on only at odd times and the green ones at even times. 
Together, these qubits and gates generate the different layers of the RHG lattice and the connections between them.
(b) A representation of two layers of the RHG lattice. 
Recall that the dots represent individual computational qubits and the connections between them show the entanglement. 
Here the two sub-figures represent the even and odd layers of the RHG lattice. 
(c) The hybrid RHG lattice generated by the chip of (a).
}
\end{figure*}

With suitable boosting, the state generation module outputs $\ket{+^\Delta}_{\rm{gkp}}$ states [\cref{eq:finiteE_GKP}] with probability $1-p_0$ and momentum-squeezed vacuum states with probability $p_0$.
The outputs of the multiplexed state-generation module are fed into the computational module, which we now describe.

The first step in the generation of the resource states is to create one-dimensional hybrid cluster states that extend in the temporal direction. 
Recall from \cref{sec:encodedqubits,sec:canCVCS} that both GKP qubit cluster states and CV cluster states require $CZ=e^{i \hat{q}\otimes \hat{q}}$ gates. 
Given a single physical $CZ$ gate (implemented via beam-splitters and squeezers as shown in \cref{fig:Gauss_gates}), a linear cluster state can be generated in the time domain using optical delay lines~\cite{Menicucci2011,Dhand2018,Lubasch2018}.
\cref{Fig:CZa} depicts the setup for the generation of this 1D cluster state. 
The circuit receives as input the states generated using GBS state preparation.
The first mode is swapped into the optical delay line, whose length is set equal to the distance between subsequent optical pulses.
This mode returns to the interferometer and interacts with the next mode at the $CZ$ gate.
At the final step of operation, the cycling light is kicked out of the delay line by the same physical swap gate.
This interaction repeats for each of the incoming modes.
In this way, a one-dimensional cluster state is generated.

In terms of on-chip implementations, this cluster-state generation involves a pair of fast actively switchable beam-splitters, controllable phase shifters, a delay line, and inline squeezers. 
The last of these requirements can be eliminated, in principle, by moving to a macro-node approach (see \cref{Sec:Open}).
The delay line is set to one clock period, and is required to be phase stable; therefore, integrated implementations of this module are preferable.
We also note that the clock speeds are ultimately limited by the speeds of the final detections; in our case, these are homodyne detections, which can be faster than other photonic and non-photonic platforms for quantum computation. 

Next, additional $CZ$ gates are implemented in the two spatial dimensions to generate the 3D structure of the RHG lattice.
Consider a 2D spatial array of 1D time-domain cluster state sources, interspersed by additional state-preparation modules and connected in the spatial domain by a nearest-neighbor array of optical $CZ$ gates, as shown in \cref{Fig:NewChipd}. 
These extra state-preparation modules are broken into two sets, indicated by the green and yellow coloring in \cref{Fig:NewChipd}. 
Half emit states at even clock cycles, and the other half emit at odd. 
The $CZ$ gates are also divided into two sets---indicated by green and yellow coloring in \cref{Fig:NewChipd}---and are applied during even and odd clock cycles, respectively. 
Thus, the additional spatial connectivity of the lattice for even and odd clock cycles is as shown in \cref{Fig:NewChipa}. 
The resulting cluster state has a lattice structure, as shown in \cref{Fig:NewChipRHG} in (2+1)-dimensions. 
After traversing through the $CZ$ gates, all modes are sent to homodyne detectors.

\subsection{Measurement-Based Quantum Computation With a Photonic QPU}\label{Sec:qpu}

The actual computation is performed in the photonic quantum processing unit (QPU), which includes an array of homodyne detector cells and fast classical control.
By merely changing the phases of the local oscillator, the QPU can correct the errors that are detected by the decoders (\cref{Sec:QEC}) and perform the logical computation (\cref{sec:FTQC}).

The homodyne cells implement quadrature measurements on each of the modes at each clock cycle, with the measured quadrature angle controlled by the local-oscillator (LO) phase~\footnote{In principle, the cells need not be arranged as a 2D plane, but this is desirable in order to keep information ``local”, i.e., the quadrature values read out from a given node will likely be used to inform the next quadrature angles for measurement on spatially nearby cells.}.
In commonly employed implementations, each homodyne detector cell consists of a vertical coupler, an LO channel, a fast phase shifter, a 50/50 beam-splitter, and a pair of photodiodes.
This functionality can all be accomplished on a planar silicon photonic chip, e.g. using Silicon-Germanium for the photodiodes and standard silicon-photonic phase modulator.
We note that modulator loss is not important here, since the phase-shifter acts on classical LO light.
The photo-currents are subtracted in order to implement balanced detection and suppress LO noise.
Each cell of the homodyne layer sends its corresponding photocurrent difference output to an electronic quadrature discriminator layer, which amplifies (via an integrated transimpedance amplifier) and digitizes (via an analog-to-digital converter) the signal, extracting a value for the quadrature measurement.
These readout values are sent as inputs to the classical QPU controller.

The QPU controller is a fully classical digital-electronic system responsible for calculating each set of quadrature angles to be implemented on the subsequent clock cycle.
This calculation takes as its inputs the quadrature measurement readout values from the photonic QPU on the previous clock cycle, the input state record from the state generation module, and the program instructions (encoding the user’s compiled quantum program).
After the subsequent clock cycle’s quadrature settings are calculated, the information is passed to the photonic QPU to actuate the LO phase-shifters before the arrival of the pulses in the subsequent clock cycle.
The QPU controller also records the results of the computation by storing the quadrature readout values in (classical) memory, to be decoded and passed back to the user.
Though fully classical, the performance requirements of the QPU controller are likely to be substantial, which motivates the development of dedicated digital electronic application-specific integrated circuits operating at very high clock speeds.

\section{Error Correction for a Quantum Memory \label{Sec:QEC}}

Having laid out the details of our architecture in the previous sections, we move on to describing its operation for quantum computation on logical qubits. 
The simplest logical computation is the identity logical operation or the \textit{quantum memory}. 
Here we elucidate the steps required to implement quantum error correction for a quantum memory. 

\begin{algorithm}
\begin{enumerate}[leftmargin=*,align=left]
\item \emph{Initialization}. Prepare a resource state on $N$ quantum modes corresponding to the nodes of the RHG lattice. With probability $1-p_0$ and $p_0$, the state of each node is either a noisy GKP state or a finitely-squeezed momentum eigenstate, respectively. Both node states are characterized by a noise variance parameter $\delta$.
\item \emph{Measurement}. Obtain a list of real-valued outcomes corresponding to $p$-homodyne measurements on all the modes. 
\item \emph{Inner decoder.} Map the real-valued homodyne outcomes to binary qubit measurement outcomes using local and global information via \Cref{alg:translator}.
\item \emph{Outer decoder}. Apply qubit decoding techniques for the RHG lattice such as those in \cref{alg:decoder} to obtain a recovery operation which has a corresponding CV implementation. 
\item \emph{Error correction.} Perform CV feed-forward operations based on the outcomes obtained and processed in steps 2 and 3. 
These, combined with the qubit recovery operation obtained in step 4, return the complete CV recovery operation, which can be tracked in software.
\end{enumerate}
\caption{Quantum error correction procedure for a quantum memory}
\label{alg:ec}
\end{algorithm}

At a high level, quantum error correction in the architecture consists of performing homodyne measurements on a subset of nodes of the RHG lattice, followed by processing of the measurement data to output a recovery operation to be applied on the remaining active nodes of the lattice. 
In our case, the data processing procedure consists of two decoders, the first of which is an inner (CV) decoder that converts the real-valued homodyne measurements into qubit outcomes and probabilities of $Z$-type qubit-level errors.
This information, in turn, is fed into an outer (qubit-level) decoder, which returns an outer recovery operation. 
As described below, our outer recovery operation can exploit analog information from the inner decoder, resulting in suitable inner recovery operation to be applied on the physical modes of the system. 

Thus, the full error correction procedure is specified by the choice of inner decoder (applied to the GKP code) and the outer qubit code (applied to the RHG lattice).
We first introduce a noise model for our hybrid lattice in the next subsection.
The inner and the outer decoders are tailored to both the noise model and the hybrid GKP/squeezed-state structure of our architecture. 
The step-wise procedure for implementing the quantum error correction procedure on a quantum memory is overviewed in \cref{alg:ec}.

\subsection{Error Model}\label{subsec:noise}
In order to motivate our choice of inner decoder and check its efficacy, we first construct and analyze a simple noise model for our hybrid RHG lattice. 
This section summarizes the noise model and main conclusions that we draw from it, with full details available in \cref{sec:noise_model}.

A reasonable model, which is standard in the CV literature, for capturing part of the noise effect of finite-energy GKP states is obtained by the application of a Gaussian noise channel~\cite{hall1994gaussian}
\begin{align}
\label{eq:noise}
    {\cal N}_{\bm{Y}}(\hat{\rho}) = \int_{\mathbb{R}^{2}} \frac{d^2\bm{\xi}}{\pi \sqrt{\det{\bm{Y}}}} \exp\left[- \frac{1}{2}\bm{\xi}^T \bm{Y}^{-1} \bm{\xi}\right] \hat{\mathcal{D}}(\bm{\xi}) \hat{\rho} \hat{\mathcal{D}}(\bm{\xi})^{\dag}
\end{align} 
to the ideal GKP states with noise of variance $\frac{\delta}{2}$~\cite{Menicucci2014,Noh2020} in both quadratures, with $\delta=\Delta^2$ from \cref{eq:finiteE_GKP}. 
While this noise model does not capture the peak-damping envelope of \cref{eq:finiteE_GKP}, it captures the finite width added to each delta-function in phase space. 
In our case, we find that the same noise model framework can be used to model the replacement of $\ket{+}_{\rm{gkp}}$ states with $p$-squeezed states, setting $\epsilon = \Delta^{2} = \delta$ from \cref{eq:CVCS}. 
In particular, we notice that adding Gaussian noise of variance $\delta/2$ ($\frac{1}{2\delta}$) in $p$ ($q$) quadrature makes the $\ket{+}_{\rm{gkp}}$ mimic the Wigner function of a mixture of $p$-squeezed states. 
In the context of \cref{eq:noise}, the noise matrices for GKP and $p$-squeezed states are given by:
\begin{align}
\label{eq:Ymat}
    \bm{Y}_{\rm gkp} = \frac{1}{2}\begin{pmatrix} \delta & 0 \\ 0 & \delta \end{pmatrix}, ~~ \bm{Y}_{\rm p} = \frac{1}{2}\begin{pmatrix} \delta^{-1} & 0 \\ 0 & \delta \end{pmatrix}.
\end{align}
{More concretely, the $\ket{+}_{\rm{gkp}}$ Wigner function has rows of positive peaks periodically arranged in phase space along even integer multiples of $\sqrt{\pi}$ in the $p$ quadrature, and alternating positive and negative peaks for odd multiples (see Fig. 1a of \cite{PhysRevLett.123.200502}).
The broad distribution in $q$ from $\bm{Y}_{\rm p}$ causes the rows of positive and negative peaks to cancel, and the rows of positive-only peaks to add, washing away the Wigner negativity and yielding a distribution mimicking a mixture of $p$ squeezed states spaced by even multiples of $\sqrt{\pi}$ in $p$.
While this is not a true $p$-squeezed state, we do not expect it to provide an underestimation of the error probability of the quantum memory, especially since a mixture of states (as opposed to a pure $p$-squeezed state) would only add more noise and hence make the decoding problem more difficult.}

Given these two types of initial states, both modelled as GKP states having undergone independent and different Gaussian noise channels, we then model the encoding into the RHG lattice, which simply consists of repeated applications of $CZ$ gates. 
Propagating the initial state noise through the $CZ$ gates results in a \textit{correlated} Gaussian noise channel, where the correlations depend on the locations of $p$-squeezed states and on the lattice-dependent pattern of $CZ$ gates applied to the nodes. 
We assume that the dominant source of noise is the noise in the input states. 
Additional noise sources include photon loss and noise introduced in $CZ$ gates, which we leave to analyze or improve in future work.

From our model, we can formally write down the distribution of $p$-homodyne data.
Since all the modes are measured in the $p$-quadrature when the computer is operating as a quantum memory, we can use this model for the distribution to inform our choice of inner decoder. 
In the case of no initial-state noise, sampling from the distribution of $p$-homodyne outcomes would simply correspond to sampling a lattice point $\bm{n}\sqrt{\pi}$ in $p$-space, where $\bm{n}$ is dictated by the qubit state of the RHG lattice. 
However, under the correlated Gaussian noise channel in our model, we find that each lattice point in $p$-space is converted into a correlated Gaussian distribution centered at the same point with covariance matrix $\widetilde{\bm{\Sigma}}_p$. 
Here, $\widetilde{\bm{\Sigma}}_p$ is the momentum part of the covariance matrix for the Gaussian peaks of the Wigner function in the phase space for the state of our hybrid lattice, as we show in \cref{sec:noise_model}. 
$\widetilde{\bm{\Sigma}}_p$ contains the aforementioned correlations and can be used to our advantage in the inner decoder as we show in the next section.

\begin{algorithm}
\textbf{Input}: Vector $\bm{p} = (p_0, ..., p_4)$ of homodyne measurement outcomes, with $p_i \in \mathbb{R}$. Mode $0$ is the $p$-squeezed state, rest are GKPs.
\begin{enumerate}[leftmargin=*,align=left]
    \item Apply the following change-of-basis $\bm{T}\bm{p}=\bm{p}'$ where:
    \begin{equation*}
    \bm{T} = \begin{pmatrix}
        1 &  0 & 0 & 0 & 0\\
        0 & 1 &  1 & 1 & 1\\
        0 & 1 &  1 & -1 & -1\\
        0 & 1 &  -1 & 1 & -1\\
        0 & 1 &  -1 & -1 & 1\\
    \end{pmatrix}.
    \end{equation*}
    We note the column vectors of this transformation are eigenvectors of $\widetilde{\bm{\Sigma}}_p$ in this case.
    \item Bin the first component of $\bm{p}'$ to the nearest integer multiple of $\sqrt{\pi}$ to return $n'_0\sqrt{\pi}$, since the $p$ quadrature outcome of mode 0 is uncorrelated from the others.
    \item Of the last three components of $\bm{p}'$, find the component $i$ that is closest to an integer multiple $n'_i$ of $\sqrt{\pi}$. 
    Round $\bm{p}'_i$ to $n'_i\sqrt{\pi}$. 
    We only choose the last three components since we do not trust the second component which corresponds to homodyne results along $(0,1,1,1,1)$ which has excessive noise of order $\frac{1}{2\delta}$.
    \item If $n'_i$ is even (odd), round the remaining two components other than $\bm{p}'_0$, $\bm{p}'_1$ and $\bm{p}'_i$ to the nearest even (odd) integer multiples of $\sqrt{\pi}$ for each component. This yields $\sqrt{\pi}\bm{v}'=\sqrt{\pi}(n'_0,\bm{p}'_1/\sqrt{\pi},n'_2,n'_3,n'_4)$, because on applying the change of basis $\bm{T}$ to an integer vector, the last four components of the new vector should either all be even or all odd.
    \item If $(n'_2+n'_3+n'_4)\bmod 4 = 0,1,2,3$, then round $\bm{p}'_1$ to the nearest $n'_1\sqrt{\pi}$ with the constraint that $n'_1\bmod4 = 0,3,2,1$. This yields $\sqrt{\pi}\bm{n}'=\sqrt{\pi}(n'_0,n'_1,n'_2,n'_3,n'_4)$. Again, this is because on applying the change of basis $\bm{T}$ to an integer vector, the second component and the last three components respect this rule, so this guess should respect it too.
    \item Undo the change of basis on the integer-valued vector $\bm{T}^{-1}\bm{n}'=\bm{n}$.
    \item Take $\bm{n}\bmod{2}=\bm{s}$ to be the five-component binary string output.
\end{enumerate}
\textbf{Output}: 5-qubit measurement values $\bm{s}$.
\caption{Inner decoder applied to 5-modes: a $p$-squeezed state surrounded by 4 GKP states}
\label{alg:translator_eg}
\end{algorithm}

\begin{algorithm}
\textbf{Input}: Vector $\bm{p} = (p_0, ..., p_N)$ of homodyne measurement outcomes, with $p_i \in \mathbb{R}$, and the noise model.
\begin{enumerate}[leftmargin=*,align=left]
    \item  Identify directions that are noisy and those that are not using the noise matrix.
    \item Perform a suitable change of basis to the homodyne data to obtain CV results for joint quadratures, a smaller number of which have reduced noise. In particular, an integer-valued transformation would allow for certain consistency checks (e.g. parity) when making a guess for the $p$-space lattice point $\bm{n}$.
    \item Apply binning along the new directions to round results to nearest ideal peak position, taking into account self-consistency of the results.
    \item Undo the change of basis to return a candidate lattice point $\bm{n}\sqrt{\pi}$.
    \item Obtain a binary string by taking $\bm{n} \bmod 2$.
\end{enumerate}
\textbf{Output}: Interpreted qubit measurement outcome
\caption{Inner decoder}
\label{alg:translator}
\end{algorithm}

\subsection{Inner Decoder}\label{subsec:translator}
As described above, an inner decoder $\mathcal{T}$ is a function that takes real-valued homodyne data and outputs binary data interpreted as qubit measurement outcomes, i.e., 
\begin{align}
    \mathcal{T} : {\mathbb R}^n \to \{0,1\}^n.
\end{align}
These qubit outcomes can then be combined into stabilizer measurement outcomes and used in the subsequent decoding procedure of the outer code~\cite{Raussendorf2007}. 
Additionally, we use our model for noise and the inner decoder strategy to calculate (marginal or correlated) probabilities of qubit error in our readout, which in turn can then be used to inform our outer decoder strategy that we outline in the following subsection.
The standard map from homodyne measurement outcomes to qubit measurement outcomes is a binning function derived from the translational symmetry of the original GKP state, i.e., the perfect periodicity in the $q$ and $p$ directions. 
The $\ket{+}_{\rm gkp}$ and $\ket{-}_{\rm gkp}$ states are each $2\sqrt{\pi}$-periodic in momentum but shifted relative to each other by $\sqrt{\pi}$. 
Therefore we can place the homodyne outcomes into bins of width $\sqrt{\pi}$ that are centred at integer multiples of $\sqrt{\pi}$, associating with $\ket{+}_{\rm gkp}$ ($\ket{-}_{\rm gkp}$) the outcomes that fell in bins centered about even (odd) integer multiples of $\sqrt{\pi}$. 
We refer to this procedure as ``standard binning".
While this binning procedure uses the original symmetry of the GKP states, it does not account for the correlations in the covariance matrix introduced by the $CZ$ gates and the presence of $p$-squeezed states, as described in the error model.

As a key proof-of-concept improvement to illustrate the importance of taking correlations into account, consider the example of a momentum-squeezed state at the centre of a primal face of the RHG lattice, which we denote as node 0, surrounded by four neighboring GKP states on nodes 1--4. 
For simplicity, in this example we assume that all the continuous-variable $CZ$ gates are the same, but this trivially generalizes if the signs of the $CZ$ gates change. 
The joint quadrature $p_0 + \sum_{j=1}^4 q_j$ has a large variance on the order of $\frac{1}{2\delta}$. 
Without using the correlations, the na\"ive inner decoder described above would result in a high-strength dephasing channel on the four neighboring GKP qubits, since the marginal distributions along $p_{j}$ would be broadened by $\frac{1}{2\delta}$ and standard binning does not leverage correlations between nodes. 
On the other hand, by taking correlations into account, the high covariance along the joint quadrature will result in either the identity gate, or a correlated four-body ring of Z operators on the neighboring qubits, which acts trivially on the code space.

More explicitly, consider the binning strategy that makes use of the correlations between optical modes. 
The momentum part of the noise matrix resulting from the application of the $CZ$ gates is
\begin{align}
     \widetilde{\bm{\Sigma}}_p &= \frac{1}{2}\begin{pmatrix}
       5\delta & 0 & 0 & 0 & 0 \\
        0 & \delta+\delta^{-1} & \delta^{-1} & \delta^{-1} & \delta^{-1}\\
        0 &\delta^{-1} &  \delta+\delta^{-1} & \delta^{-1} & \delta^{-1}\\
        0 & \delta^{-1} & \delta^{-1} &  \delta+\delta^{-1} & \delta^{-1}\\
        0 & \delta^{-1} & \delta^{-1} & \delta^{-1} &  \delta+\delta^{-1}\\
        \end{pmatrix},\nonumber\\
        &= \frac{1}{2}\left[5\delta \oplus \left(\delta 1\!\!1_4 + \delta^{-1} \ket{\bm{\nu}}\bra{\bm{\nu}}\right)\right], \, \bm{\nu} = (1,1,1,1)^T,
\end{align}
where we label the modes $0,1,\ldots,4$ with the momentum state corresponding to mode $0$. 
We see that the noise matrix is non-diagonal, i.e., the CV noise is correlated, but it has a specific structure that can be exploited. 
Two immediate observations are that mode 0 is uncorrelated from the other modes, meaning we can simply apply standard binning to it; and that there is correlated noise along the direction $(0,1,1,1,1)$ in $p$-space. 
\cref{alg:translator_eg} presents a strategy for dealing with this correlated noise, taking into account consistency checks that our guesses for modes 1--4 must respect. 

In general, the problem of finding a better inner decoder for our hybrid architecture is to find a decoder that takes into account the location of  GKP and $p$-squeezed states, and  knowledge of the structured $CZ$ gates that have been applied to form the cluster state.

The distribution of $p$-homodyne outcomes consists of a periodic arrangement of Gaussian distributions all with covariance $\widetilde{\bm{\Sigma}}_p$ on $N$ modes, each Gaussian centred at a point $\bm{n}\sqrt{\pi}$ where $\bm{n}$ are integer valued vectors from a set that corresponds to the ideal state of the qubits.
Suppose we obtain the values $\bm{p}$ after the homodyne measurements. 
If we assume $\bm{p}$ could have resulted from a Gaussian distribution centered at any of the lattice points $\bm{n}$, then the so-called \emph{responsibility} \cite{bishop2006pattern} of a given lattice point for the result $\bm{p}$ is given by:
    \begin{equation}
        r(\bm{n}) = \exp\left[-\frac{1}{2}(\bm{n}\sqrt{\pi}-\bm{p})^T \widetilde{\bm{\Sigma}}_p^{-1}(\bm{n}\sqrt{\pi}-\bm{p})\right].
    \end{equation}
The responsibility is directly related to the Gaussian distributions at each lattice point and provides a relative way of ordering which lattice points were most likely to have generated $\bm{p}$. 
Specifically, the lattice point which was most likely to have produced the point $\bm{p}$ is:
    \begin{equation}
        \bm{n}_{\text{IQP}} = \arg\min_{\bm{n}\in\mathbb{Z}^N} (\bm{n}\sqrt{\pi}-\bm{p})^T\widetilde{\bm{\Sigma}}_p^{-1}(\bm{n}\sqrt{\pi}-\bm{p}),
    \end{equation}
where we have chosen the subscript IQP to indicate that this is an integer quadratic program, i.e., a minimization of a quadratic function over an integer domain. 
As mentioned above and for simplicity, we are using the standard approximation that all peaks in the GKP state have equal weight~\cite{Menicucci2014}.
However, one could also include an envelope that weights peaks differently, in which case this information could also be included in the calculation of the responsibility. In general, integer quadratic programs are NP-hard~\cite{Buchheim2015, Park2017}, so we will require a heuristic strategy that is computationally tractable. 
Our approach for a generalized version of \cref{alg:translator_eg} is summarized in \cref{alg:translator}, with the case of more complicated configurations of $p$-squeezed states left to future study.

\begin{algorithm}
\textbf{Input}: Qubit measurement outcome from \cref{alg:translator}
\begin{enumerate}[leftmargin=*,align=left]
    \item \emph{Syndrome identification}. Construct relevant stabilizer measurement outcomes from the input qubit outcomes. 
    \item \emph{Matching graph construction}. Construct a complete graph using:
        \begin{itemize}
            \item Vertices: One vertex for each unsatisfied stabilizer (include additional vertices if needed for specific boundary conditions.)
            \item Edges: Connect every pair of vertices
            \item Weights: The edges are assigned weights reflecting probability of the most likely error that could have given rise to the pairs of unsatisfied stabilizers. The choice of weights can be informed by the CV noise model.
        \end{itemize}
    \item \emph{Matching algorithm}. Find a minimum-weight perfect matching by running Edmonds' algorithm~\cite{Edmonds1967} on the matching graph from the previous step. 
    \item \textit{Qubit recovery operator.} The recovery operator is given by this rule: for each pair $(u,v)$ in the matching, flip the binary outcomes of qubits in the most likely error that could have given rise to $u$ and $v$. 
\end{enumerate}
\textbf{Output}: Qubit recovery operator
\caption{Outer decoder using MWPM}
\label{alg:decoder}
\end{algorithm}

\subsection{Outer Decoder and Error Correction}
\label{sec:outer}
After obtaining and binning the outcomes of the homodyne measurement, error correction is performed for the outer qubit code. 
The details of the error correction problem we solve are summarized in \cref{alg:decoder} for a particular, standard, choice of decoding algorithm: minimum-weight perfect matching (MWPM)~\cite{Dennis_2002,Raussendorf2006,Wang_2011}. 
Note that there are many other decoding algorithms that could be used such as those presented in Refs.~\cite{delfosse2017unionfind,bravyi2014,duclos-cianci2010,duclos-cianci2014,bravyi2013,anwar2014,wootton2012,hutter2014,panteleev2019,roffe2020,fowler2014,torlai2017,varsamopoulos2018,tuckett2018,Herold2015,herold2017a, Harrington2004, Dauphinais_2017}.

A few comments are in order. 
The weights of the matching graph edges in \cref{alg:decoder} are derived from the homodyne measurement outcomes, as well as the positions of the $p$-squeezed states in the lattice. 
An example of such weights is presented in \cref{Sec:Threshold}.
Furthermore, using the homodyne measurement outcomes to calculate matching graph weights has been explored in the context of the toric code~\cite{Noh2020,Vuillot2019}, but the knowledge of the locations of the $p$-squeezed states gives us additional information that can be used to improve the performance of the decoder. 
We discuss this point in more detail in \cref{Sec:Threshold}.

As mentioned earlier, due to the measurement-based computation model, feed-forward operations based on the outcomes obtained from the homodyne measurements and the inner decoder are combined with the qubit-recovery operation obtained from the outer decoder.
Together, these inform the complete CV recovery operation that needs to be applied to the active computation layers.
In practice, the combined recovery operation need not actually be applied on the qubits; instead, we would keep track of the recovery operations in classical control programming by updating the Pauli frame~\cite{Knill2005,Fowler_2012}.

\section{Fault-Tolerant Universal Quantum Computation} \label{sec:FTQC}

\begin{figure}[t]
\centering
\subfloat{\includegraphics[width=0.95\columnwidth]{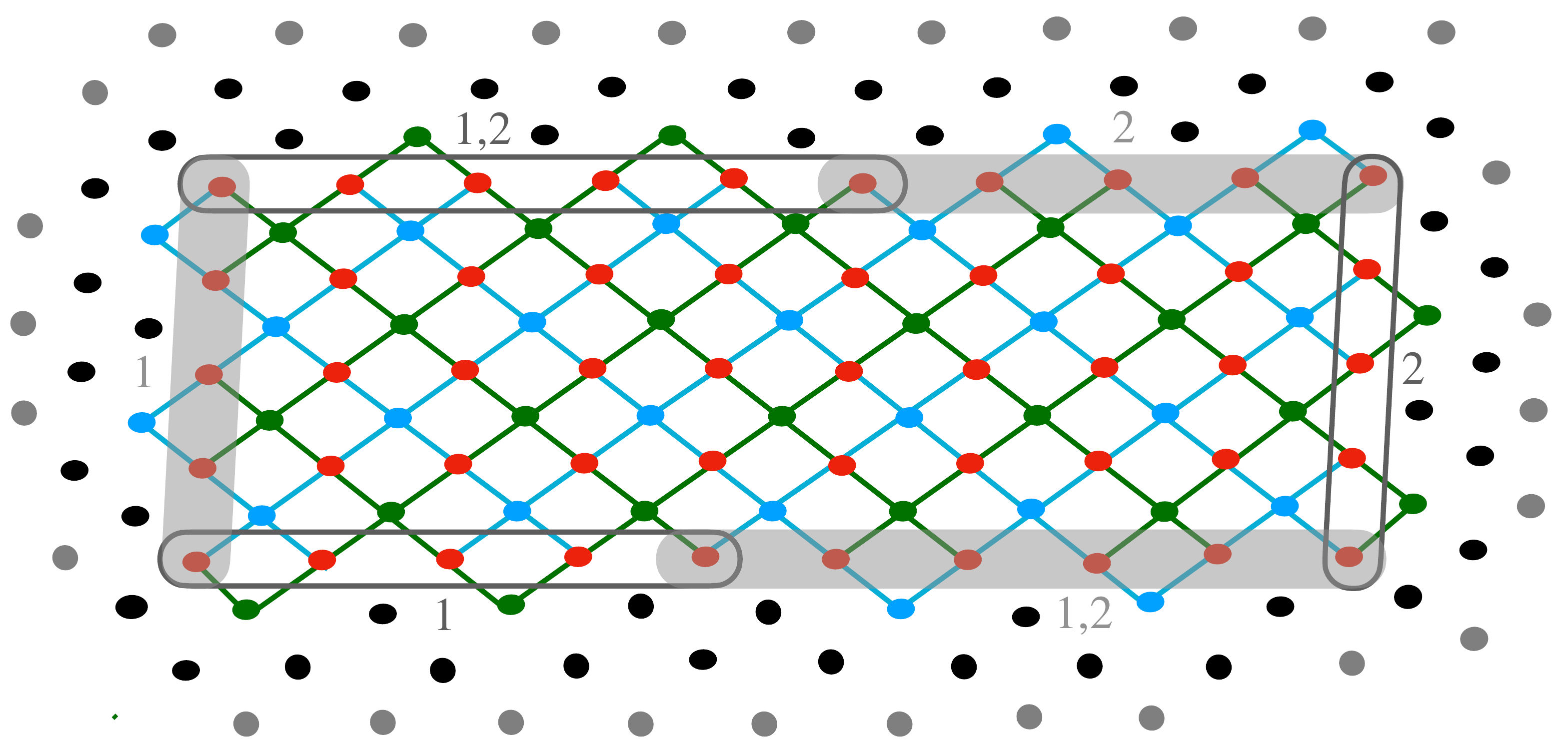}}

\subfloat{\includegraphics[width=0.95\columnwidth]{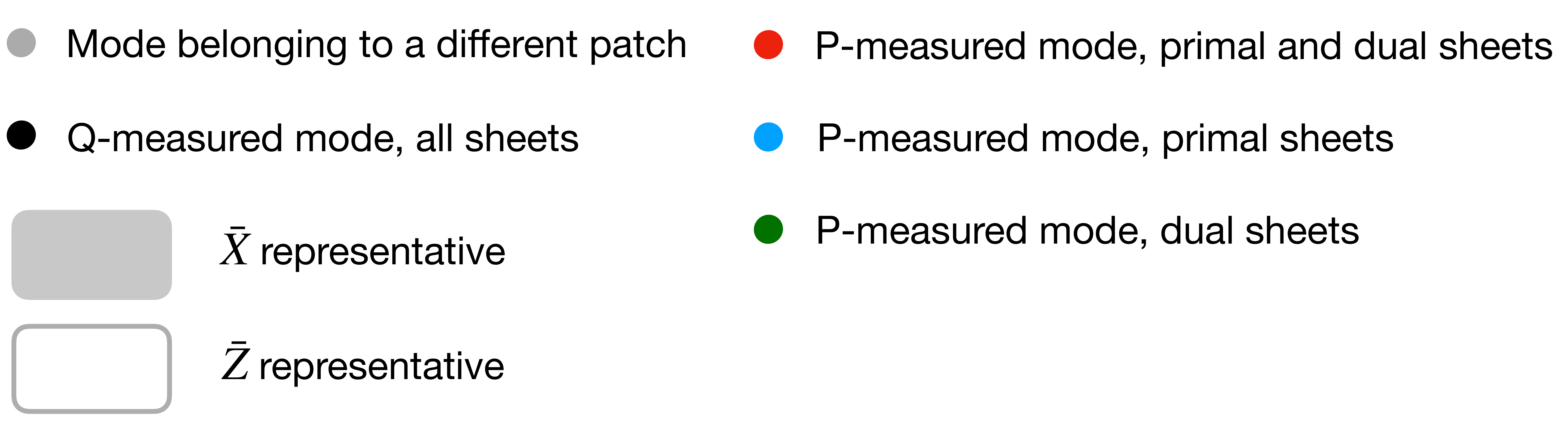}}

\caption{Top view of the measurement pattern defining a patch encoding two logical qubits with distance 5. The black nodes, which are measured in the $q$ basis to disentangle them from the cluster, define a patch consisting of the red, green and blue nodes. The grey nodes either belong to other patches or are measured in the $q$ basis as well.
The red, blue, and green nodes are eventually all measured in the $p$ basis. The red nodes are present in both primal and dual sheets of the foliated surface code, while the blue (green) ones, are only present in the primal (dual) sheet. The red nodes represent data qubits, while the blue and green ones are ancillas.
Representatives of the Pauli logical operators $\bar X$ ($\bar Z$) are highlighted by shaded (lighter unshaded) grey boxes. They consist of strings of Pauli $X$'s or $Z$'s applied to the circled data qubits. The numbers next to the boxes identify the affected logical qubit. For simplicity, the Pauli operators shown above are applied on a primal sheet. For dual sheets, they would be conjugated by Hadamard gates.}

\label{Fig:patch_definition}
\end{figure}

The hybrid architecture from \cref{Sec:scheme} allows for the generation of a (2+1)D cluster state suitable for performing scalable quantum computation. 
For completeness, we present a possible scheme for the fault-tolerant implementation of logical algorithms on such a state. 
The scheme we choose and review here is based on lattice surgery for the surface code~\cite{Horsman_2012, PhysRevX.7.021029, Litinski_2018, Litinski2019}, particularly its measurement-based version~\cite{Herr_2018, brown2018universal}.
Note that while the schemes considered here have lower overheads~\cite{Litinski_2018} than other surface-code-based approaches, they are heuristics, since finding an optimal implementation is NP-hard~\cite{Herr_2017}. 
This section overviews the fundamental components needed to perform fault-tolerant quantum information processing in our architecture.
We use as examples codes of distance 5, the distance being the weight of the smallest representative of a logical operator.
Inner (GKP) code states will be denoted by a subscript, whereas outer (RHG) code states will have a bar. The reader may refer to \cref{fig:pcell} for clarity on the terms primal, dual, rough, and smooth. We use the \emph{data} and \emph{ancilla} qubit nomenclature from the foliated picture~\cite{Bolt2016} where useful; see \cref{Fig:patch_definition} for more details.
While much of this section is meant as a review for a photonics audience and might be familiar to an expert in fault tolerance, it does highlight specific points about how these logical operations could be implemented in our architecture.

\paragraph*{Logical Qubits.}
In lattice surgery, quantum information is encoded by way of \emph{patches}. After the chip produces a large RHG lattice, homodyne measurements in the $q$ quadrature disentangle certain nodes from the lattice, effectively creating holes. 
We refer to these $q$-measured regions as \emph{gaps}, and we define a patch as a continuous part of the lattice completely surrounded in the spatial dimensions by a gap~\footnote{The gap region is referred as the vacuum~\cite{Raussendorf2006}, but we forgo this terminology to avoid confusion with the more common notion of vacuum. Note also that the patches we consider here do not contain defects, that is, internal gaps.}.
\cref{Fig:patch_definition} shows an example of a patch and how two logical qubits are encoded with it. A patch can be deformed to move around the logical information and minimize overheads. 
Patch deformations can be achieved by changing the $q$ homodyne measurements to $p$ in the gap surrounding the patch (without connecting it to other patches), followed by a sufficient number of rounds of error correction. 

\paragraph*{State Initialization.}

The state of a single logical qubit can be initialized in either $\vert \bar{+} \rangle$ or $\vert \bar{0} \rangle$ by measuring the ancillary qubits of the first temporal layer in the $p$ or $q$ quadrature, respectively, while the rest of the measurements are all performed in the $p$ quadrature. The initialization can be performed fault-tolerantly by measuring a number of layers that scales linearly in the code distance and then performing error correction as described in \cref{Sec:QEC}. Alternatively, the error correction can be performed during a subsequent patch deformation~\cite{Litinski2019}; the latter is more resource-efficient, and is thus the approach we favour. 

\paragraph*{Logical $\bar{Z}$ and $\bar{X}$ Operators and Measurements.} 

Logical $\bar{Z}$ ($\bar{X}$) gates in the code are effected through chains of physical $Z$ ($X$) operations connecting the appropriate borders, that is, $\sqrt{\pi}$ displacements along the $p$ ($q$) quadrature, in a primal sheet. The $Z$ and $X$ physical operations are reversed when applied on a dual sheet. While we do not apply logical operators in that manner in this work, they are helpful to understand multi-qubit Pauli measurements.

Destructive logical Pauli measurements can be effected through homodyne measurements on the active layer of data qubits of the patch. 
Measuring a primal sheet in the $q$ ($p$) quadrature will result in a logical $\bar{Z}$ ($\bar{X}$) measurement; for dual sheets the measurement basis is swapped. The actual measurement outcome depends on the parity of the results along the corresponding chain from the previous paragraph. This procedure can be viewed as the inverse of state initialization. 
For non-destructive logical Pauli measurements, a set of ancilla qubits can be coupled to a single data qubit each, followed by homodyne measurements in the $p$ basis. 
Once again, the clock cycle at which the data and ancilla qubits are coupled determines whether $\bar{Z}$ or $\bar{X}$ is measured. 
Performing error correction is required to reliably infer the logical measurement outcome.

\begin{figure}
\centering
\subfloat[]{\includegraphics[width=.95\columnwidth]{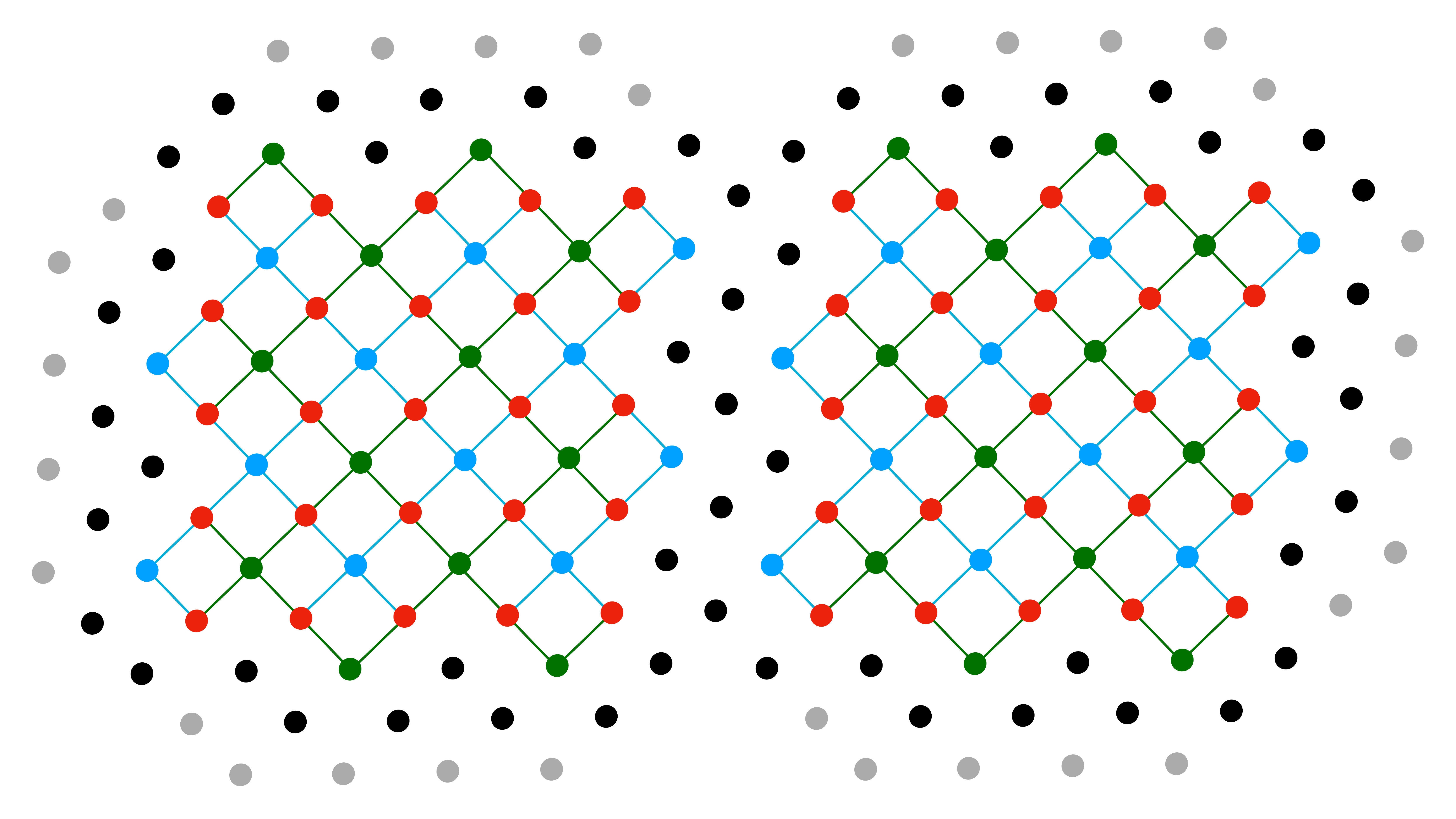}}\\
\subfloat[]{\label{subfig:merged}\includegraphics[width=.95\columnwidth]{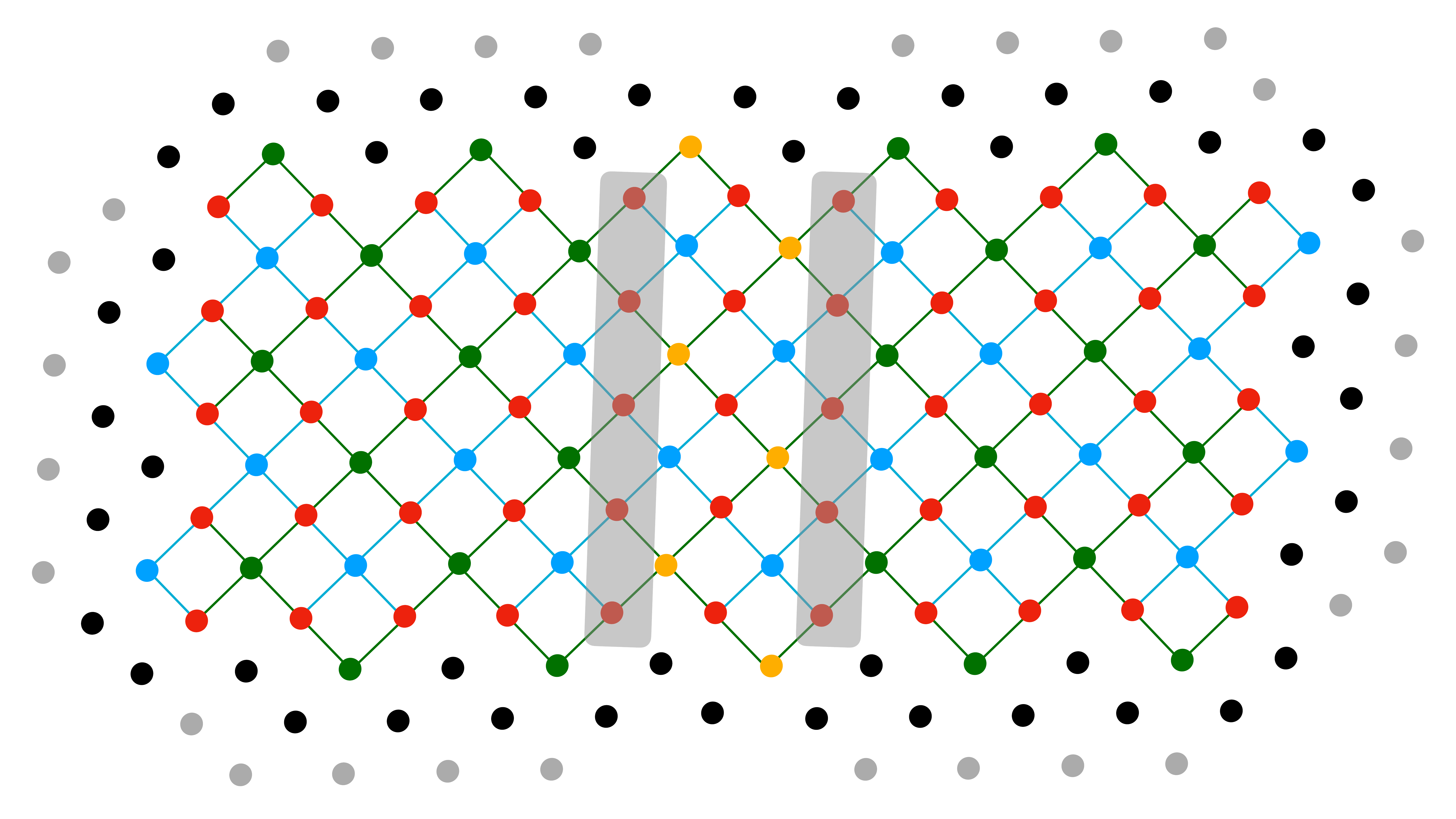}}
\caption{Two neighboring patches before merging (or after splitting) in (a); and the patches after merging (or before splitting) in (b). The legend used is the same as in \cref{Fig:patch_definition}. Measurements of the nodes in between two patches in (a) are switched from the $q$ quadrature to $p$, resulting in the measurement of the new surface code stabilizers. The merged patch encodes a single logical qubit. After five rounds of error correction, corresponding to the code distance, the measurement outcomes of the logical operator $X_1 \otimes X_2$ can be inferred, as indicated by the shaded boxes shown in (b), which corresponds to the product of the new stabilizers, shown in yellow. A single logical qubit remains encoded in the patch after the merging process. The splitting of the patch in (b) is achieved by measuring the data nodes in between the two patches in the $p$ quadrature and the ancillas in the $q$ quadrature for a number of rounds of error correction sufficient enough for allowing fault-tolerance, after which the data nodes can also be measured in the $q$ quadrature. After the splitting process, the logical state of the system transforms according to $\alpha \vert \bar{0} \rangle + \beta \vert \bar{1} \rangle \rightarrow \alpha \vert \bar{0} \bar{0} \rangle + \beta \vert \bar{1} \bar{1} \rangle$.}
\label{Fig:patch_merging}
\end{figure}

\paragraph*{Multi-Qubit Operations.}
Merging and splitting different patches allows one to perform logical entangling gates and multi-qubit Pauli measurements. 
Merging (splitting) is achieved by changing the measurement pattern from $q$ to $p$ ($p$ to $q$) in the gap between patches, as illustrated in \cref{Fig:patch_merging}.
As in the case of deformation, error correction must be performed after the merging for a fault-tolerant implementation~\footnote{For the case of patch merging, it is sometimes useful to introduce twist defects in the process~\cite{Bombin_2010, PhysRevX.7.021029}. 
This allows for the measurement of a multi-qubit logical Pauli operator involving a logical $Y$, alleviating the cost of distilling the eigenstates of the Pauli-$Y$ operator $\vert y \rangle$ to implement logical $P$ gates with the surface code~\cite{Dennis_2002, Fowler_2012}.}.

\begin{figure}
\centering
\subfloat[]
{\label{fig:meas_before} \includegraphics[width=.95\columnwidth]{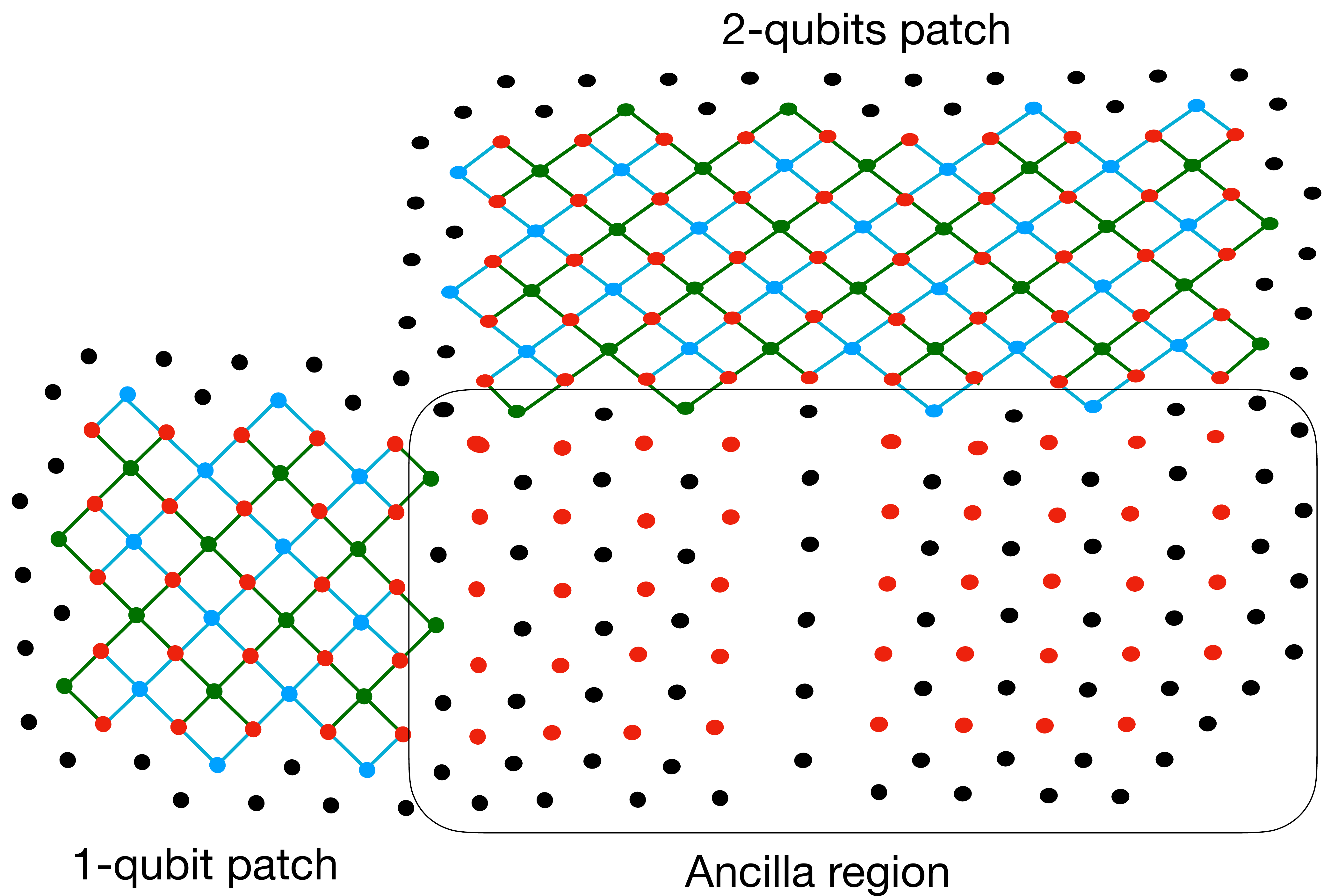}}

\subfloat[]{\label{fig:meas_after}\includegraphics[width=.95\columnwidth]{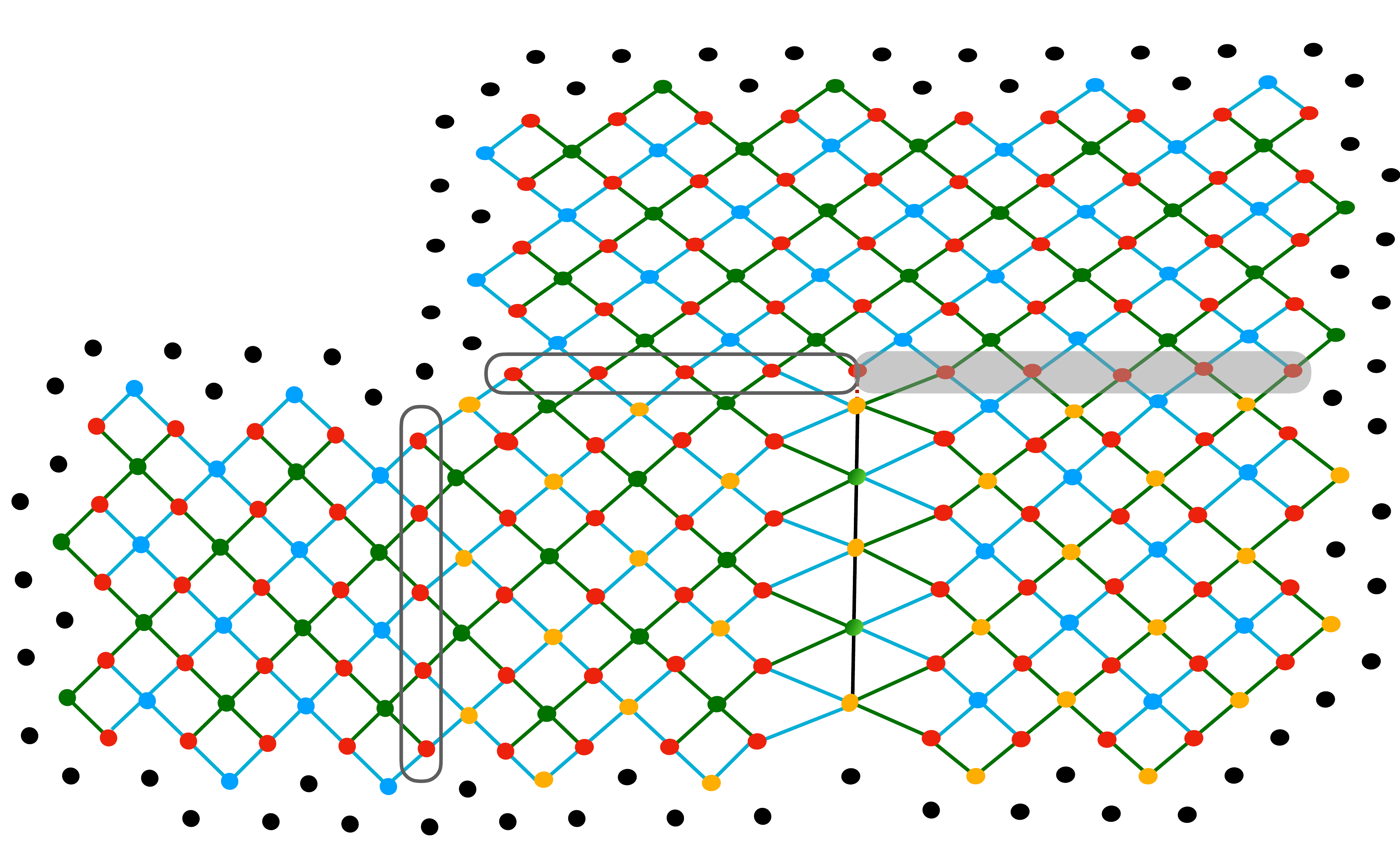}}
\caption{{A fault-tolerant measurement of the operator $\bar{Z}_1 \otimes \bar{Y}_2 \otimes \bar{X}_3$. See Fig.~\ref{Fig:patch_definition} for the legend. In (a), an ancilla region is prepared by initializing the data nodes into a tensor product of $\vert + \rangle$ states (on a primal sheet) by measuring the ancilla nodes in the $q$ quadrature. A dislocation region, within the ancillas, is aligned with the meeting point of the rough and smooth boundaries of the two-qubit patch. This specific geometry is chosen so that the product of stabilizers, identified by yellow nodes, gives the desired product of logical operators, represented by the grey boxes in (b). Note that a $\bar Y_2$ representative has support on the qubits in both the shaded and unshaded horizontal boxes. After five rounds of error correction (5 being the distance of the code), the measurement outcomes can be inferred from the measurement results.}}
\label{fig:multi_pauli_meas}
\end{figure}

An adjacent ancillary patch can be used to measure the tensor product of a Pauli operator $\mathcal P$ associated with logical qubits living on different patches. 
This ancilla patch does not encode logical information; it simply consists of a tensor product of physical $\vert + \rangle$ states. In order to perform the measurement, the ancilla patch is merged via the relevant boundaries of the patches encoding the logical qubits. Measuring the additional stabilizers along the concerned boundaries associated with $\mathcal P$ in the ancilla patch and performing error correction allows one to fault-tolerantly measure $\mathcal P$. 
A specific example of this process is illustrated in \cref{fig:multi_pauli_meas}. 

\paragraph*{Magic State Injection and Distillation.}

\begin{figure*}
\centering
\subfloat[]{\includegraphics[width=.3\textwidth]{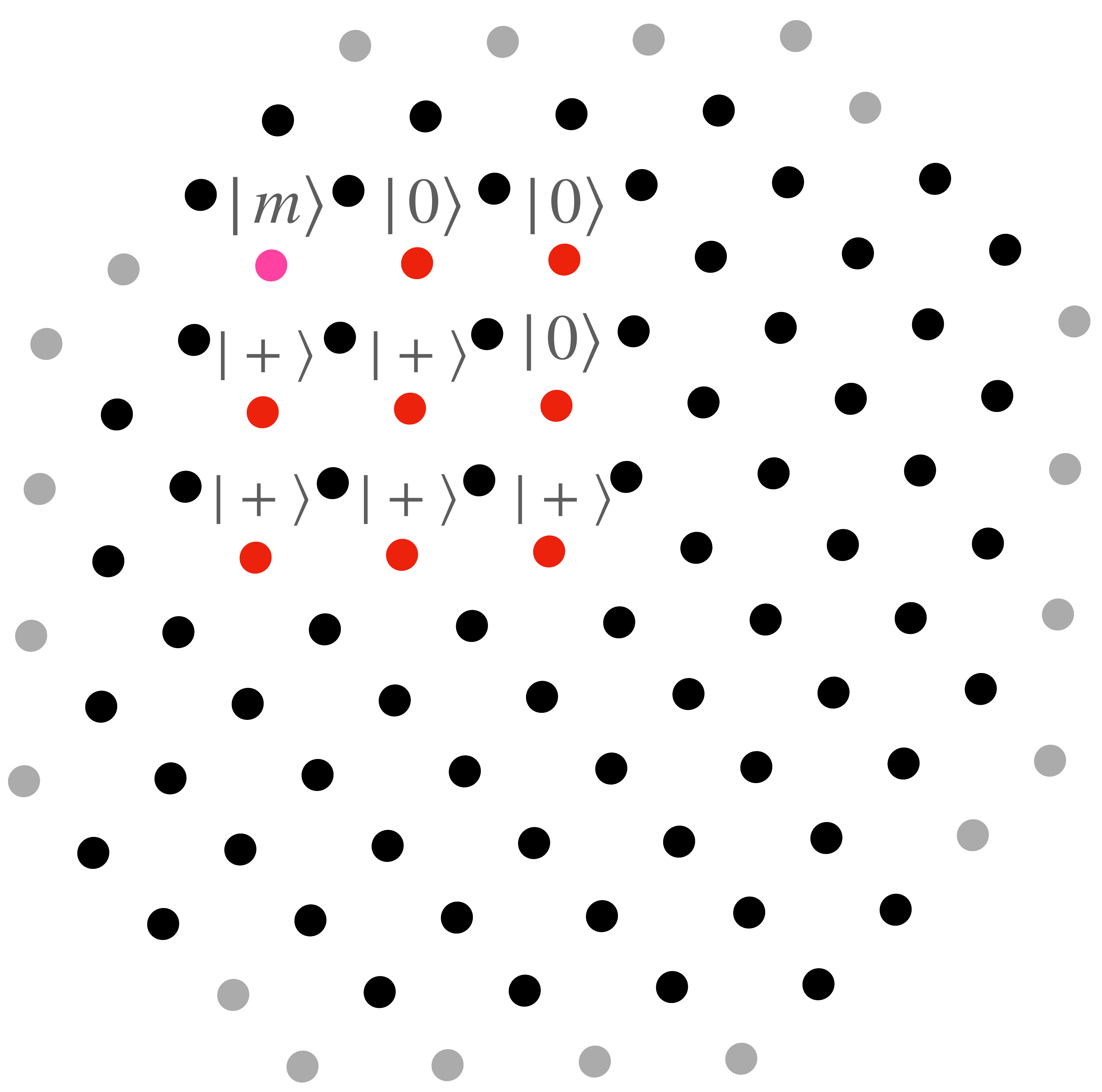}}
\hspace{.2cm}
\subfloat[]{\includegraphics[width=.3\textwidth]{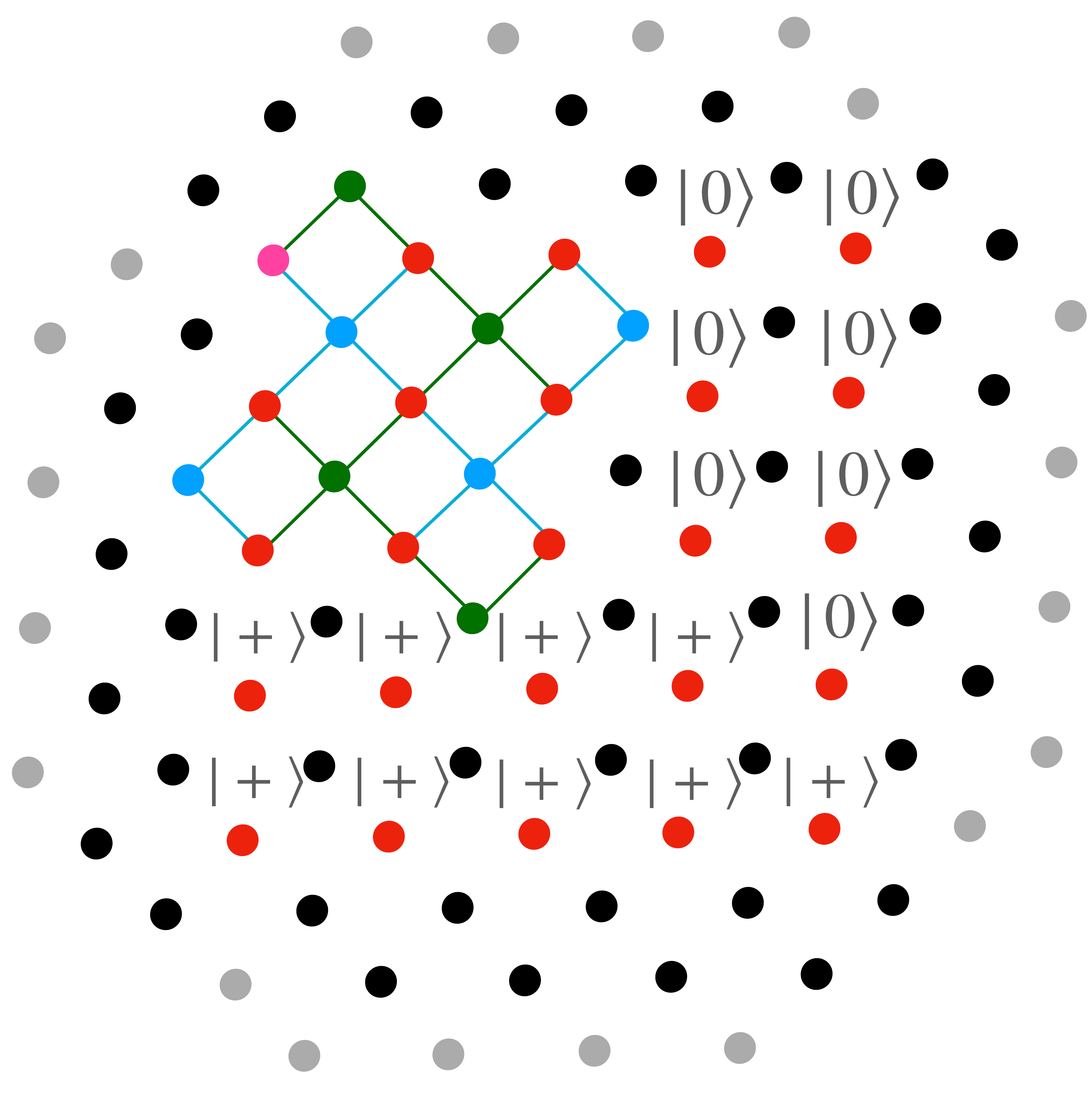}}
\hspace{.2cm}
\subfloat[]{\includegraphics[width=.3\textwidth]{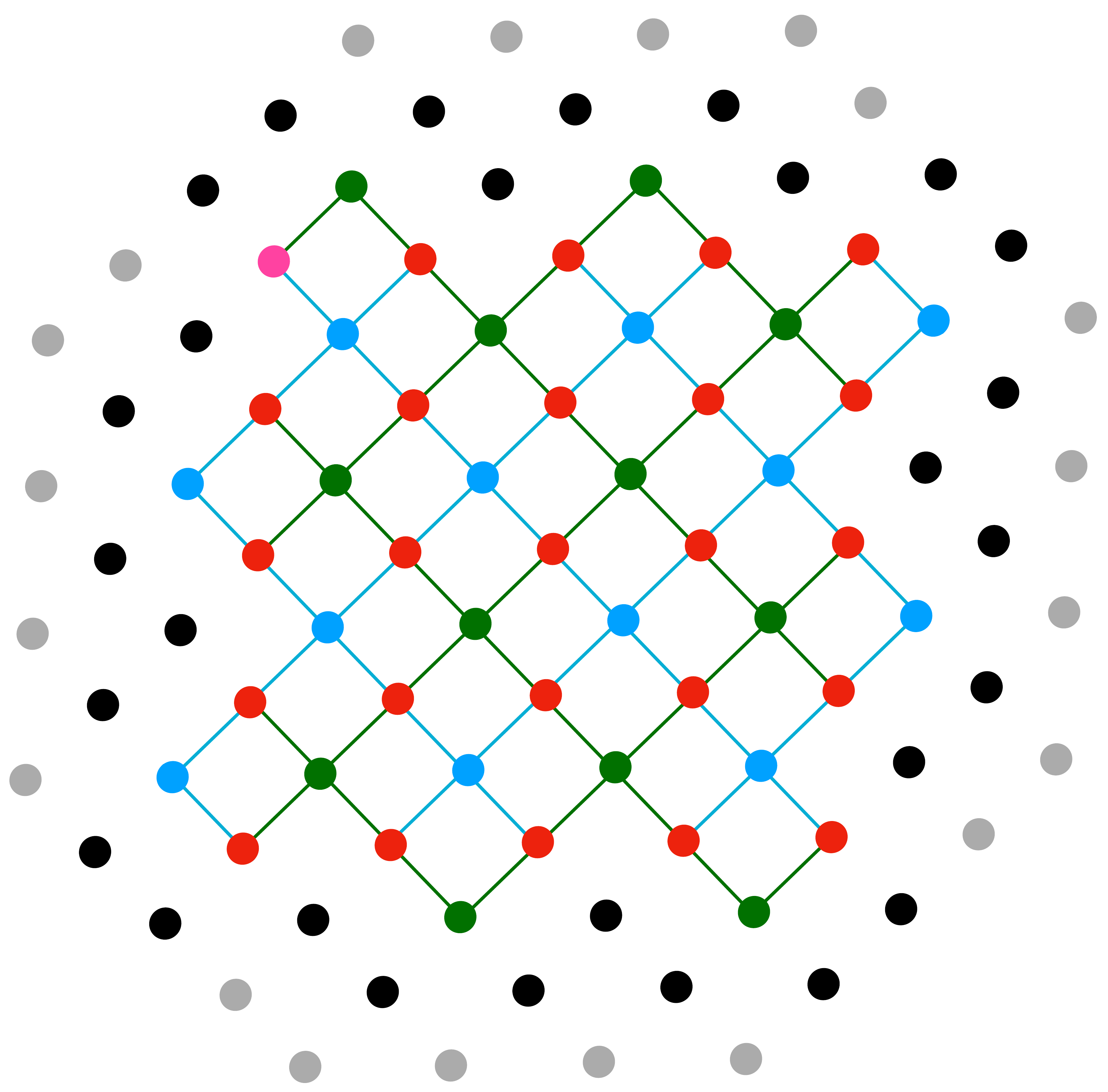}}
\caption{The three main steps for state injection of a physical magic state $\vert m \rangle$ into a logical magic state $\vert \bar{m} \rangle$ of distance $d_2$. A physical magic state and enough data nodes to prepare a distance $d_1 < d_2$ code are initialized according to the pattern shown in (a). The stabilizers are then measured for two rounds, shown in (b). If an error is detected, the process is restarted. If no errors are detected, the patch is deformed with data states appropriately chosen until reaching a larger code with distance $d_2$, illustrated in (c), and followed by error correction. To ensure a high probability of success, provided that $d_2$ is large enough, the first stage of the process can be performed in parallel, keeping a single successful instance.
}
\label{Fig:injection}
\end{figure*}

Homodyne detection on GKP qubits alone does not give us access to the non-Pauli measurements used in the original proposal~\cite{Raussendorf2006}. 
It is therefore necessary for us to inject \emph{physical} magic states $\vert m \rangle = \Ket{0}_{\rm gkp} + e^{i\frac{\pi}{4}} \Ket{1}_{\rm gkp}$ created by the GKP factories into the RHG lattice. 
This process -- illustrated in \cref{Fig:injection} -- requires some classical post-selection (which can be performed in parallel to ensure a high probability of success) and a subsequent $d$ rounds of error correction~\cite{Li_2015}. 
In our case, the error correction is performed after the patch merging step of the multi-qubit Pauli measurement, as described in the previous paragraph.

One can distill multiple copies of noisy magic states into a single high-fidelity state. 
This process is necessary in our architecture in order to obtain an encoded magic state with low enough noise to be able to implement logical gates. Several different magic state distillation procedures exist~\cite{PhysRevA.71.022316, Bravyi_2012, Haah2018codesprotocols, Haah2017magicstate, PhysRevA.87.042305, Hastings_2018}. 
The details of the noise afflicting the hardware as well as the logical algorithm specified by the user will inform the preferred choice.

\paragraph*{Running a Logical Quantum Computation.}

Before running the algorithm on the hardware, it ought to be compiled in such a way that all logical Clifford operations are commuted through to the end of the circuit and absorbed into the logical measurements. This turns the single logical Pauli measurements into a sequence of multi-qubit Pauli measurements.

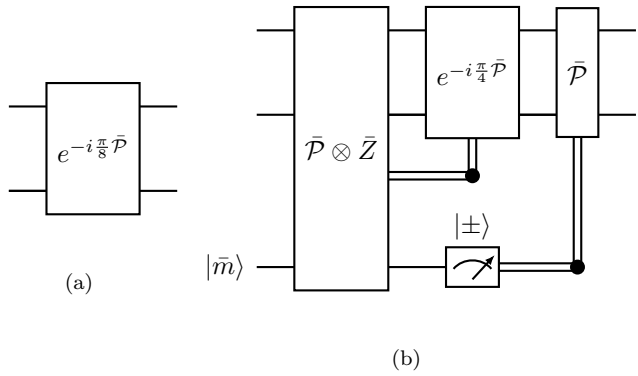
\begin{figure}
\centering
\subfloat[Non-Clifford gate][]{
\begin{quantikz}
& \gate[wires=2]{e^{-i \frac{\pi}{8} \bar{\mathcal P}}} & \qw \\ 
& \qw & \qw \\
\end{quantikz}
}
\subfloat[Non-Clifford gate with magic state][]{
\begin{quantikz}
& \gate[4,nwires={3}]{\bar{\mathcal P} \otimes \bar Z} & \gate[2]{e^{-i \frac{\pi}{4}\bar{\mathcal P}}} & \gate[2]{\bar{\mathcal P}} & \qw \\
& & \qw & \qw & \qw \\
& & \cwbend{-1} \\
\lstick{$\ket{\bar{m}}$} & & \meter{$\ket{\pm}$} & \cwbend{-2} \\
\end{quantikz}
}
\caption{In the compilation method we use, non-Clifford gates as in (a) can be implemented by consuming a logical magic state $\vert \bar{m} \rangle$, as in (b). The first gate in circuit (b) represents the measurement of $\bar{\mathcal P} \otimes \bar Z$, with $\bar{\mathcal P}$ a Pauli operator. Since both the gates $e^{-i\frac{\pi}{4} \bar{\mathcal P}}$ and $\bar{\mathcal P}$ belong to the Clifford group, they can be commuted to the end of the circuit and absorbed in the multi-qubit Pauli measurements. These circuits (drawn using Quantikz~\cite{Kay2019}) can be straightforwardly generalized to an arbitrary number of qubits.
}
\label{fig:non_cliff_gate}
\end{figure}

 The remaining non-Clifford multi-qubit rotations are performed by consuming a (logical) magic state~\cite{Litinski_2018, PhysRevB.97.205404}, as shown in \cref{fig:non_cliff_gate}. Since logical T-gates are not native to the surface code, a high-quality logical magic state must be injected into the computation. Magic state distillation~\cite{PhysRevA.71.022316} can be used to produce a high quality encoded magic state starting from several lower quality ones. As the distillation is resource intensive, several algorithms for the minimization of the number of non-Clifford operations can be used during compilation~\cite{10.5555/2535649.2535653,PhysRevLett.110.190502, gosset2013algorithm, Heyfron_2018, 6516700, PhysRevA.87.042302, 6899791}.
Running a compiled logical quantum algorithm essentially decomposes into two stages. First, the non-Clifford logical rotations are implemented. Each one of these consumes a distilled logical magic state and requires the measurement of a multi-qubit logical Pauli operator and of single-qubit logical $\bar{X}$ operators. Depending on the measurement outcomes, some additional Pauli operations may need to be implemented. These can be commuted to the end of the circuit, in a procedure known as modifying the Pauli frame~\cite{Knill2005}.
Second, once all the non-Clifford operations have been performed, the multi-qubit logical Pauli measurements are performed.

\paragraph*{Compatibility With Hybrid Architecture.}

Our proposed architecture is well-suited to accommodate modifications to these general steps. The different layouts for the magic state factories, the data and the ancilla blocks, and the boundary conditions of the data patches can be easily modified by an appropriate selection of the homodyne measurement quadratures. Note that as long as the fraction of swap-outs is below a critical value (which depends on the squeezing $\Delta$), the general structure of our computing scheme carries through in the presence of swap-outs: both the GKP states in the lattice and $p$-squeezed states encode $\Ket{+}$ and act appropriately under the physical operations we discussed~\cite{pantaleoni2020modular}. However, the swapped-out nodes add correlated noise to their neighborhood, an effect we deal with in the error-correction procedure, which we address in the following section.

\section{Threshold Estimation for a Quantum Memory}\label{Sec:Threshold}

The operation of the architecture as a fault-tolerant quantum computer is underpinned by the concept that the logical error rate of an encoded computation can be arbitrarily lowered by increasing the size of the code.
This concept is based on the idea of fault-tolerance thresholds.
The existence of such thresholds for qubit-based architectures has been a subject of extensive research for over twenty years~\cite{aharonov2008,knill1998,preskill1997,gottesman1997stabilizer,kitaev_1997,Dennis_2002,aliferis2006,divincenzo2007,aliferis2008,aliferis2008a,kovalev2013,gottesman2014,fawzi2018} but the existence of thresholds for CV-based architectures~\cite{Menicucci2014} is less well understood.
Furthermore, the question of whether hybrid architectures remain fault-tolerant with \emph{probabilistic} sources of GKP qubits is not obvious.
Here we provide numerical evidence that our architecture does indeed have a threshold in the presence of errors arising from finite squeezing and for a range of swap-out probabilities. 
As we detail in this section, in order to calculate the threshold, we simulate the hybrid architecture operating as a quantum memory and run a complete error-correction procedure~\cite{terhal15}. 
We detail the various steps involved in the simulation of the thresholds in \cref{alg:ft}. 

\begin{algorithm}
\begin{enumerate}[leftmargin=*,align=left]
\item \textit{Parameters.} Choose lattice size $d$, swap-out probability $p_0$, and noise variance $\delta$
\item \textit{Simulated homodyne measurements.} Generate homodyne measurement outcomes consistent with the noise matrix as given in \cref{eq:pvalue} through a suitable sampling method.
\item \textit{Inner decoder.} Apply the inner decoder of \cref{alg:translator} on the homodyne data to obtain qubit measurement outcomes. 
\item \textit{Outer decoder.} Apply the outer decoder to the qubit outcomes to obtain a recovery operation using \cref{alg:decoder}
\item \emph{Error correction.} Apply the recovery operation. 
\item \textit{Success check.} Note the success/failure of the error correction procedure.
\item \textit{Error rate.} Repeat steps 2-6 sufficient number of times to obtain an error rate.
\item {\textit{Thresholds}}. Repeat steps 1-7 for different lattice sizes and noise parameters $(p_0,\delta)$.
\end{enumerate}
\caption{Procedure for obtaining thresholds for a quantum memory}
\label{alg:ft}
\end{algorithm}

We now briefly review the numerical procedure for estimating the error threshold of a quantum memory.
Consider a family of codes of growing size, parameterized by $d$. 
In the case of the RHG lattice, $d$ is the code distance (the weight of the minimal weight non-trivial logical operator) and the number of qubits is $n = O( d^3 )$.   
Another parameter is the noise channel, which in our case is described by two numerical parameters: the noise variance $\delta$ and the swap-out probability $p_0$. 
To estimate the error threshold, we run many trials of Monte Carlo simulations to determine the logical error rates as a function of our physical noise parameters.
This is done for different lattice sizes $d$. 
In each trial, we generate homodyne measurement outcomes according to the noise parameters, then we run our error correction procedure (inner decoder followed by outer decoder as described in \cref{Sec:QEC}), and finally check if error correction has been successful. 
Let us assume that we fix $p_0$ and vary $\delta$. 
Then if a threshold, $\delta_c$, exists, we expect to see the following behaviour. 
For $\delta > \delta_c$, increasing the size of the code (increasing $d$) increases the logical error rate. 
But for $\delta < \delta_c$, increasing the size of the code exponentially decreases the logical error rate. 
We note that the largest code sizes we consider involve $n \approx 5000$ qubits.
Simulation of such a large number of qubits is possible due to the fact that we use a classical noise channel to model approximate GKPs and the circuits we simulate belong to the Clifford group, which makes them efficiently classically simulable~\cite{PhysRevA.70.052328}.

\begin{figure}
    \centering
    \subfloat[]{\includegraphics[width=.45\columnwidth]{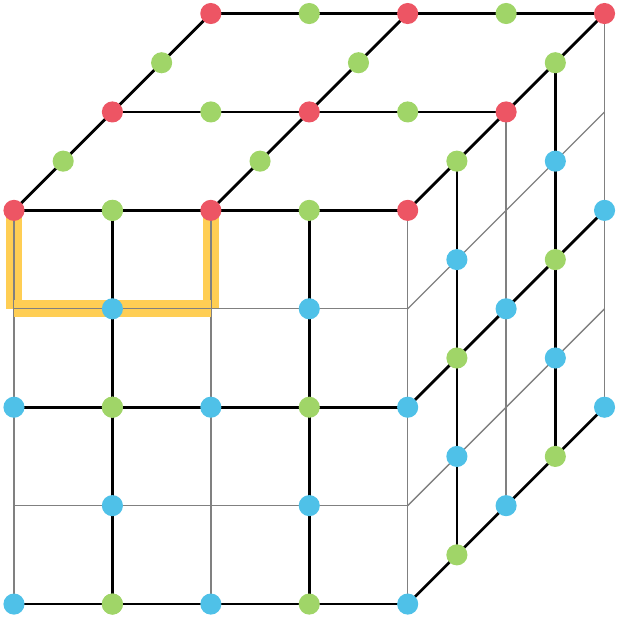}}
    \hfill
    \subfloat[]{\includegraphics[width=.45\columnwidth]{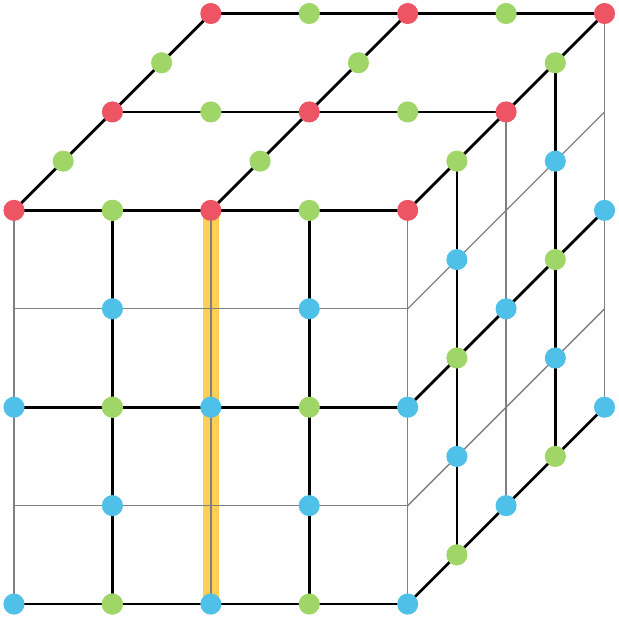}}
    \caption{
    $X$ correlation surface in the RHG lattice. The blue circles are the primal qubits, the green circles are the dual qubits, and the pink circles are the primal qubits in the $X$ correlation surface. The yellow highlighted edges represent $Z$ operators, i.e.\ primal qubits on yellow highlighted edges have a Pauli $Z$ applied to them. In a) we show a logical identity operator that commutes with all the stabilizers and the correlation surface, whereas in b) we show a non-trivial logical operator that commutes with all the stabilizers but does not commute with the correlation surface.
    }
    \label{fig:corr-surf}
\end{figure}

While the other steps of \cref{alg:ft} are relatively straightforward, we explain the success-check step of \cref{alg:ft}. 
After applying the recovery operation, all the cluster state stabilizers are guaranteed to be satisfied. 
Therefore, error correction is successful if the product of the qubit error and the recovery operator is a stabilizer (logical identity operator) and error correction fails if the product of the qubit error and the recovery operator is a non-trivial logical operator. 
Such operators anti-commute with at least one of the correlation surfaces of the cluster state~\cite{Raussendorf2005}. 
\cref{fig:corr-surf} shows the $X$ correlation surface (the $Z$ correlation surface is analogous). 
To summarize, if the product of the qubit error and the recovery operator anti-commutes with either correlation surface, then error correction has failed. 

The remainder of this section is structured as follows. In \cref{subsec:FT_sim}, we describe our simulations in detail and compare the performance of different inner and outer decoding strategies. Then, in \cref{subsec:thresh_res}, we present the threshold simulation results for our architecture operating as a quantum memory. 

\subsection{Simulation Details}\label{subsec:FT_sim}

\begin{figure*}
    \centering
    \subfloat[]{\includegraphics[width=.32\linewidth]{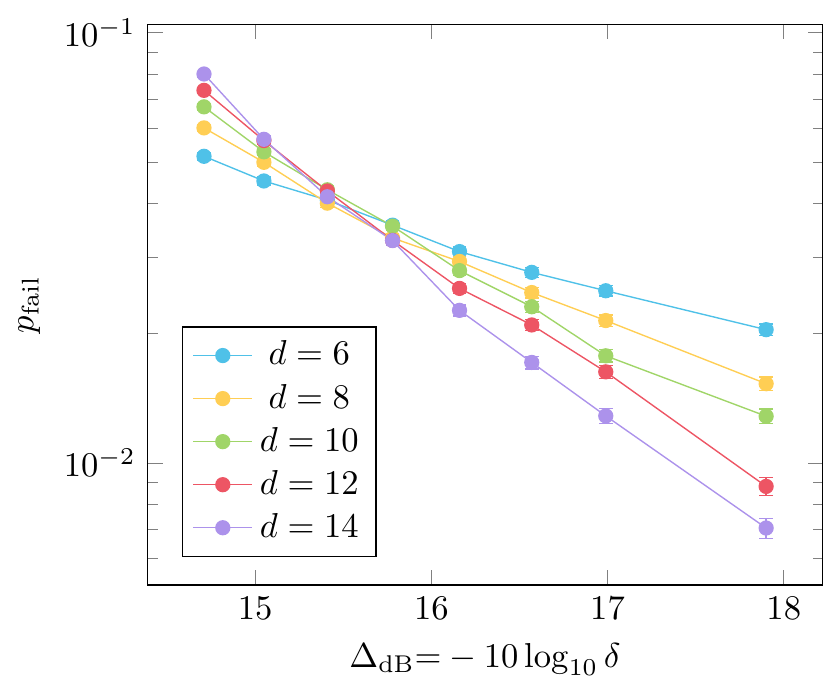}}
    \hfill
    \hfill
    \subfloat[]{\includegraphics[width=.32\linewidth]{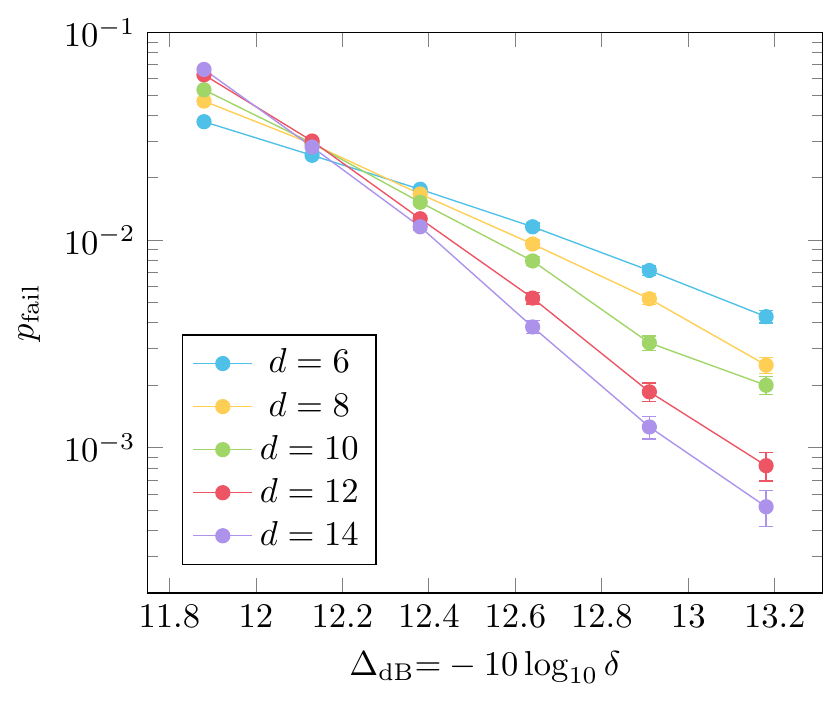}}
    \hfill
    \subfloat[]{\includegraphics[width=.32\linewidth]{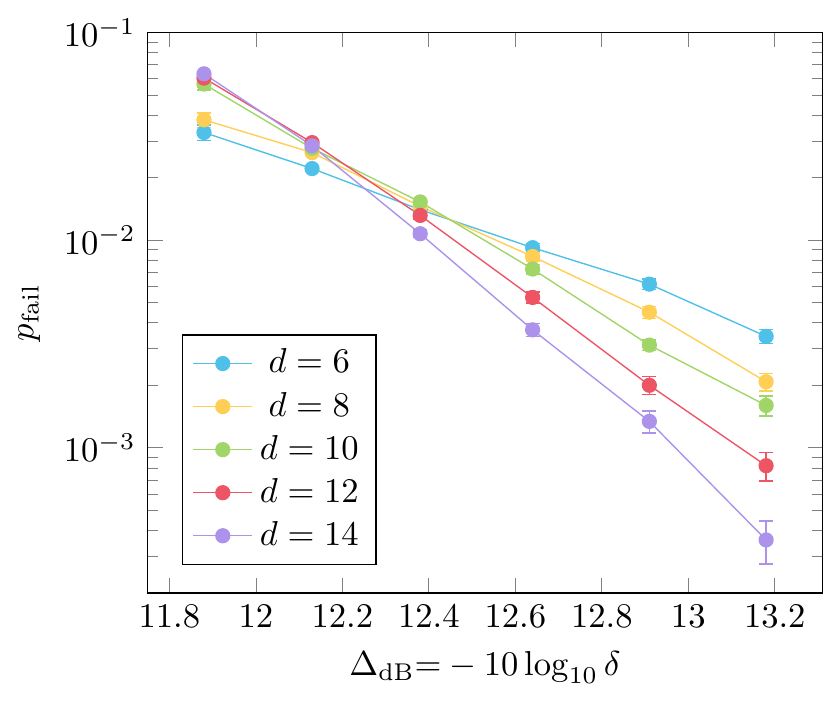}}
    \caption{
        Performance comparison for the various inner decoders considered for $p_0 = 0.06$. For (a) and (b), standard binning is used as the inner decoder for every node, and the weights assigned to the edges of the matching graph are either (a) all equal, or (b) assigned following \cref{eq:error_prob}. In (c), \cref{alg:translator_eg} is first used on GKP nodes connected to isolated momentum states, standard binning is used for the remainders, and weights described by \cref{eq:error_prob} are used in the matching graph.
    }
    \label{fig:inner_comparison}
\end{figure*}

Here we provide some details on the simulations performed to find the thresholds. 
First, we note that we only simulate error correction of the primal lattice nodes, as the error correction problem for the dual lattice nodes is the same and each problem can be solved independently. 
We consider RHG lattices with size parameterized by $d$, where the left and right boundaries are equivalent to distance $d$ surface codes, and there are $d$ layers of nodes in between these two boundaries (see \cref{fig:corr-surf}). 

We now return to the calculation of the matching graph weights (Step 2 in \cref{alg:decoder}). 
The first step is to construct a decoding graph based on the RHG lattice. 
For the sake of brevity, we will describe this construction for an RHG lattice with periodic boundary conditions (see~\cite{Wang2010} for the case of lattices with boundaries). 
We refer to the elements of the decoding graph as vertices and arcs, to avoid confusion with the nodes and edges of the RHG lattice.
The decoding graph has a vertex for each six-body $X$ stabilizer acting on the primal qubits of the RHG lattice. These stabilizers are formed from products of cluster state stabilizers surrounding a dual cell.
Vertices are connected by arcs if their corresponding stabilizers share a qubit. 
As each qubit is in the support of two such stabilizers, the arcs of the decoding graph are in a one-to-one correspondence with the primal qubits of the RHG lattice, and hence with a subset of the modes of the cluster state.
We assign weights to the arcs of the decoding graph as follows. 
Consider the mode $q$ corresponding to an arc in the decoding graph. 
Let $m$ be the number of swapped-out modes neighboring $q$ and let $z$ be the outcome of the homodyne measurement of $q$. We assign to this mode a heuristic error probability as follows:
\begin{equation}
    w(z, m, \tilde \delta) = 
    \begin{cases} 
        2/5 & \mbox{if } m = 4, \\
        1/3 & \mbox{if } m = 3, \\
        1/4 & \mbox{if } m = 2, \\
        \frac{\sum_{n \in \mathbb Z} \exp [-(z-(2n+1)\sqrt{\pi})^2/\tilde \delta]}
        {\sum_{n \in \mathbb Z} \exp [-(z-n\sqrt{\pi})^2/\tilde \delta]} & \mbox{if } m \leq 1.
    \end{cases} 
    \label{eq:error_prob}
\end{equation}
If a mode has one swapped-out neighbor, then there are no errors due to swap-outs as the net effect of a single swap-out after applying the $CZ$ gates is a stabilizer. 
In this case, the error probability is the probability of incorrectly binning the state~\cite{Noh2020}, using the standard binning function and assuming a classical noise channel with parameter $\tilde \delta$, which we derive from $\tilde \Sigma_p$. 
For $m\geq 2$, we derive the weights in \cref{eq:error_prob} from simulations which we detail in \cref{sec:decoder_weights}. 
The weight of the corresponding arc in the decoding graph is then $-\log w(z, m, \tilde \delta)$~\cite{Wang_2011}. 

Given the decoding graph, we construct the matching graph weights as follows. 
For each pair of vertices in the matching graph, we compute the total weight of the minimum weight path between the corresponding vertices in the decoding graph using Dijkstra's algorithm~\cite{Dijkstra1959}. 

Many variants are possible for the inner decoder introduced in \cref{subsec:translator}. In this work, we considered two simple ones. The first is performing standard binning of the homodyne outcomes, irrespective of the presence of momentum-squeezed state in its vicinity. Second, for those momentum-squeezed states which are isolated from others, in the sense that no connected node is also connected to another squeezed state, a variant of \cref{alg:translator_eg} is used. The modifications are required because of the variable number of neighbors and signs present in the physical application of the $CZ$ gates. We emphasize that, as mentioned in \cref{subsec:translator}, more complex strategies can be devised and are likely to improve the overall decoding performance.

\begin{figure}
    \centering
    \includegraphics[width=\columnwidth]{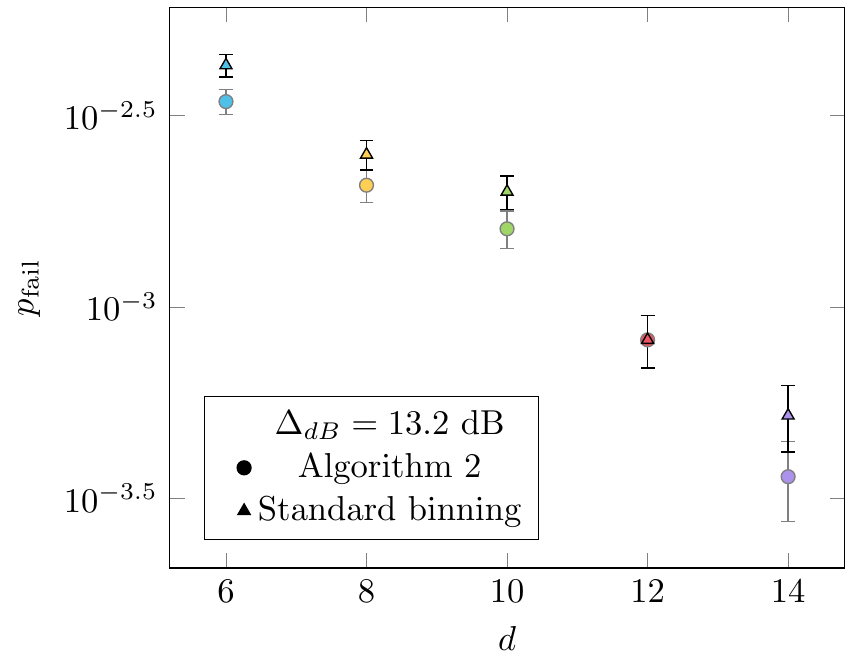}
    \caption{{A comparison of the failure probability as a function of code distance $d$, for $\Delta_{\text{dB}}=13.2$ dB using two different inner decoders. At this point below threshold, we see that \cref{alg:translator_eg} provides equal or lower error rates than standard binning.}}
    \label{fig:translator_vs_sb}
\end{figure}

Simulation results for both possibilities are shown in \cref{fig:inner_comparison}, for $p_0 = 0.06$. \cref{fig:inner_comparison} (a) shows results for na\"ive binning using uniform weights in the matching graphs, while (b) uses weights as described in \cref{eq:error_prob}. In (c), \cref{alg:translator_eg} is used, and weights are given by \cref{eq:error_prob}{; we chose $p_0 = 0.06$ as a representative example to test-drive \cref{alg:translator_eg}, as it is best suited to cases of isolated swap-outs, which are common for this value of $p_0$.} Incorporating the analog information into building the matching graph clearly improves the performance, the threshold decreasing from $\sim \SI{15.5}{\decibel}$ to $\sim \SI{12.2}{\decibel}$, with both variants of the inner decoder. 
We note that modifying the inner decoder to leverage \cref{alg:translator_eg} did not result in any significant differences for the thresholds themselves but the failure rates below threshold {are equal or lower using~\cref{alg:translator_eg}, as we show in \cref{fig:translator_vs_sb}.}
Quantifying and understanding the origin of this effect is left for future work.

\subsection{Threshold Results \label{subsec:thresh_res}}

\begin{figure}
    \centering
    \includegraphics[width=\columnwidth]{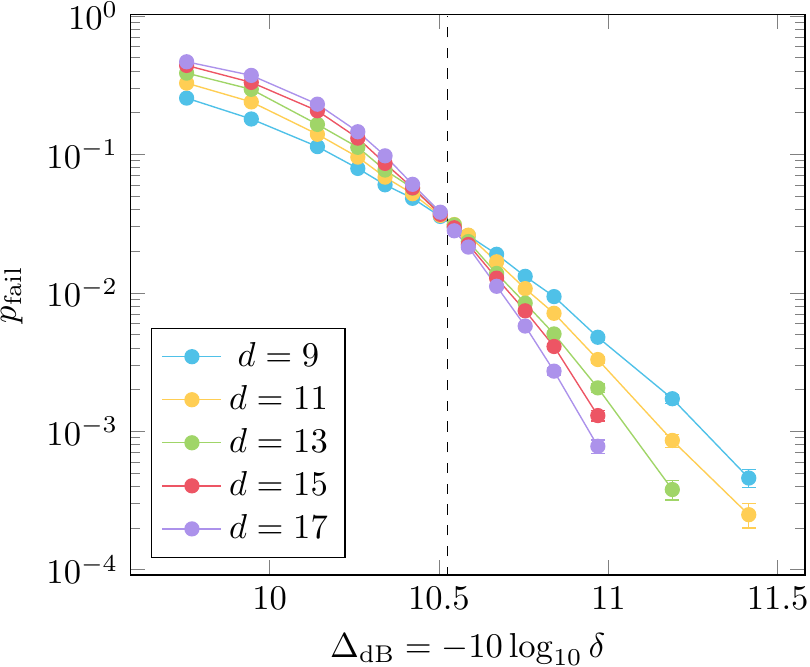}
    \caption{
        Error threshold of the RHG-GKP code with approximate GKP states using standard binning and matching graph weights derived from \cref{eq:error_prob}. 
        We estimate the logical error rate $p_{\rm fail}$ using Monte Carlo simulations for different lattice sizes $d$ and noise variance $\delta$. 
        The error threshold is the point where the curve for different values of $d$ intersect. Each data point in the average of $\eta \geq 10^4$ trials and has at least 25 failure events. 
        The error bars show the standard error of the mean $\sqrt{p_{\rm fail} (1 - p_{\rm fail}) / \eta}$. 
        We use the fitting procedure described in~\cite{Harrington2004} to systematically obtain our threshold estimate (dashed vertical line).
    }
    \label{fig:raussendorf-gkp}
\end{figure}

Now we are ready to present the thresholds of our hybrid architecture.
Our first result is the error threshold of the RHG-GKP code with approximate GKP states, which we model as ideal states suffering a random displacement with noise variance $\delta$, as discussed in \cref{subsec:noise}.
In our noise model, this corresponds to the limit of no swap-outs, i.e.,\ $p_0 = 0$. 
Similar simulations have been carried out in previous works for the toric-GKP code~\cite{Vuillot2019} and the surface-GKP code~\cite{Noh2020,hanggli2020enhanced}. We use standard binning and matching graph weights derived from \cref{eq:error_prob}.
We observe an error threshold of $\Delta_{\rm dB} = -10\log_{10}(\delta) \approx 10.5~\rm dB$, which is comparable with results for similar noise models in the aforementioned works. 
The data are shown in \cref{fig:raussendorf-gkp}.

\begin{figure}
    \centering
    \includegraphics[width=\columnwidth]{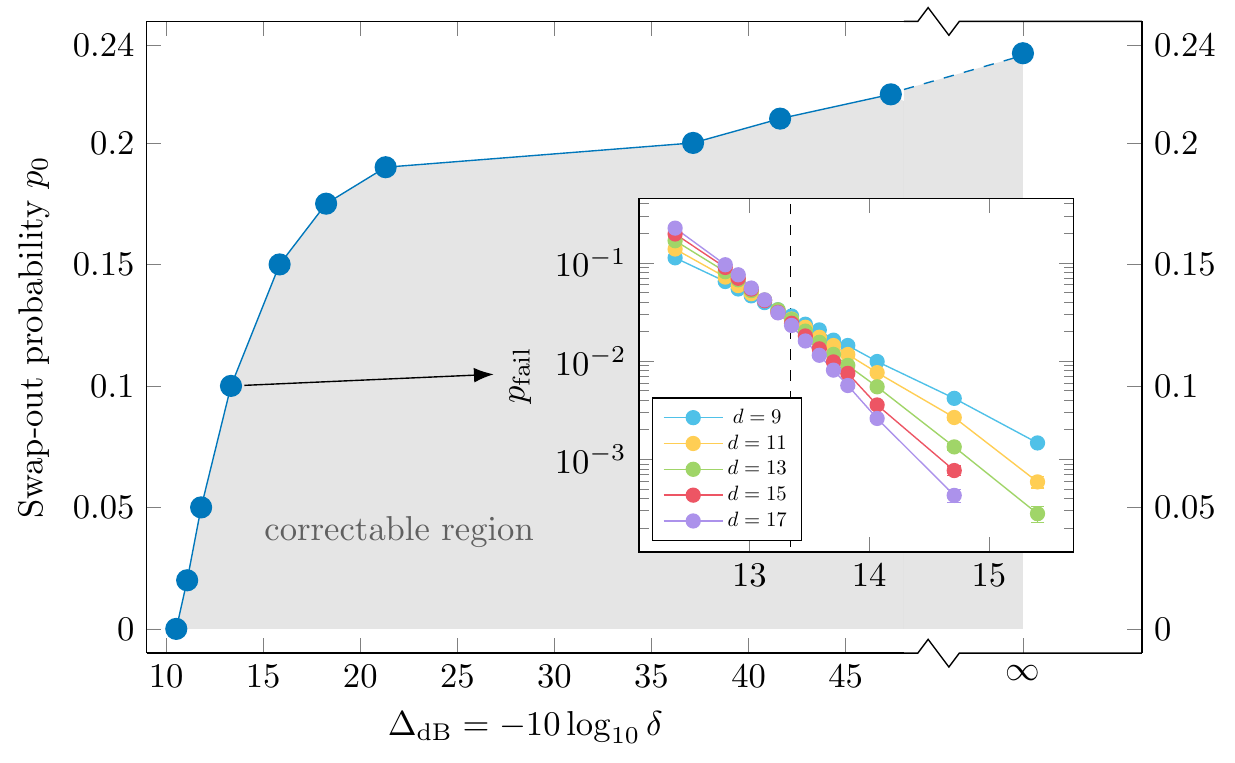}
    \caption{
        Error threshold estimates for a quantum memory using standard binning and matching graph weights derived from \cref{eq:error_prob}. Each blue point represents a threshold plot, where we fix either $p_0$ or $\delta$ and use Monte Carlo simulations to find the threshold value of the other noise parameter. The inset shows the threshold plot for $p_0 = 0.1$, with the corresponding threshold $\Delta_{\rm dB} \approx 13.3$ shown as a dashed line. Pairs of parameters below the blue line lie in the correctable region of parameter space where error correction works.     }
    \label{fig:thresholds}
\end{figure}

As described above, the full noise model we use involves two noise parameters, the noise variance $\delta$ and the swap-out probability $p_0$; the error threshold is a line in $(\delta, p_0)$ parameter space rather than a single point. 
To estimate this error threshold, we run Monte Carlo simulations as described in \cref{alg:ft} for different values of $\delta$, $p_0$ and $d$ (the lattice size). 
For a particular value of $p_0$, we can extract the corresponding threshold $\delta$ value by plotting the logical error probabilities, $p_{\rm fail}$, for a range of values of $\delta$ and $d$. 
The error threshold is then the point where curves for different $d$ intersect. 
Equivalently, we can instead fix a value of $\delta$ and vary $p_0$ and $d$.
In the inner decoder we use standard binning, and we use matching graph weights derived from \cref{eq:error_prob} in the outer decoder.
\cref{fig:thresholds} shows the below-threshold region in $(\delta, p_0)$ parameter space, alongside an example threshold plot for $p_0=0.1$. 
We find a high tolerance to swap-outs, with a maximum swap-out threshold of $p_0 \approx 0.236$ (for $\delta = 0$). 
For $p_0 = 0$, the noise variance error threshold is $\approx \SI{10.5}{\decibel}$, where the dB value is given by $-10\log_{10}\delta$.
As expected, an increase in the swap-out probability leads to an increase in the squeezing thresholds.
For an experimentally accessible~\cite{PhysRevLett.117.110801} squeezing value of $\SI{15}{\decibel}$ , our simulations suggest a swap-out threshold of $p_0 \approx 0.133$.
We note that the noise variance ($\delta$) tolerance of our decoder is markedly better for $p_0 \lesssim 0.19$ than for values nearer the swap-out threshold. 
Understanding this behaviour is an open problem, with one possible reason being that the inner and outer decoders we are using for the current simulations might be sub-optimal for this regime. 
Therefore, to investigate this phenomenon further, we should compare our decoding strategy with e.g.\ maximum-likelihood decoding, in order to ascertain whether the sharp decrease in performance is a fundamental property or an artifact of our decoding strategy. 
We leave this analysis for future work.

Previous works~\cite{barrett2010fault,whiteside2014,Auger2018} have studied the error threshold of the RHG cluster state model when qubits are erased with some probability. 
This is a natural noise model in optics and bears some resemblance to our model, as one assumes that the locations of the erasures are known. 
The relationship between the erasure threshold and the $Z$ error threshold was found to be approximately linear~\cite{barrett2010fault} and there is a fundamental erasure threshold of $0.249$, which is set by percolation theory~\cite{Lorenz_1998}. 
It is difficult to directly compare our results with those of~\cite{barrett2010fault} because of the differences in the noise models. 
However, our swap-out threshold is close to the percolation theory erasure threshold, and it is natural to ask whether we can increase the swap-out threshold beyond the erasure threshold by further optimizing our decoder. 
There are many ways we could improve our current decoder (see \cref{Sec:Open}), so, unless there is a fundamental limit due to percolation of swap-outs, we are hopeful that we can surpass the erasure threshold. 
In addition, the question as to whether our decoder has an advantage over the equivalent erasure decoder for finite values of $\Delta_{\rm dB}$ remains open and could be a subject of future work.

\section{Open Problems}
\label{Sec:Open}
Thus far, we have discussed the theoretical advances made in this architecture.
As we detail in the next section, this architecture can provide important advantages over other architectures and platforms.  
However, there are many ways in which the current architecture can be improved.
Here we list some open questions and suggested routes for further improvement of the architecture.

The open problems belong broadly in three categories: hardware-focused improvements to the architecture, better encoding, and better decoding strategies.
From the hardware perspective, one of the important open problems is to devise a passive implementation of the hybrid architecture.
In the current architecture, the $CZ$ gates on the computational module are active transformations, i.e., they require in-line squeezing and displacements.
If such active transformations can be replaced by passive transformations, then this could further simplify the computational module and thus reduce the experimental requirements.
Possible paths to explore in this direction could be to obtain a hybrid version of macronode-based architectures reviewed in~\cite{Larsen2020} or perhaps to use techniques demonstrated in~\cite{Walshe2020}.

Another challenge is that of reliable state preparation.
In particular, a key experimental goal highlighted by this work is to gain access to { high-quality GKP qubits whose teeth are squeezed at or exceeding the 15 dB level. This motivates the improvement of existing methods for state generation with GBS devices, which previously considered states with up to 10 dB of per-peak squeezing~\cite{Sabapathy2019,Su2018,Quesada2019,Tzitrin2019}. Reaching 15 dB states with the techniques from  Refs.~\cite{Sabapathy2019,Su2018,Quesada2019,Tzitrin2019} will involve going to higher-order truncations of the Fock basis, which is computationally more demanding. This motivates the development of new numerical simulation techniques, or the application of breeding or distillation protocols to lower quality states to reach the 15 dB level. Progress in these directions is critical to accurate  resource estimates for photonic quantum computing.}
Though the present architecture only couples GBS devices to optical switches, we leave open the possibility of using any surplus GKP qubits to check and improve the quality of outputs, for example, by using them as flag qubits~\cite{shi2019fault}. 
Overhead, particularly the required number of bosonic qubit states, can be reduced by using fore-knowledge about the quantum computation. For example, we expect there to be little (if any) benefit to having bosonic qubits in place of squeezed states at $q$-measured nodes described in \cref{sec:FTQC}.

From the hardware perspective, yet another important task is to develop more realistic yet tractable noise models.
Our main focus here was on a Gaussian noise model for state preparation---in tandem with the  non-deterministic nature of the bosonic qubit sources. 
Further analysis is required to incorporate the non-Gaussian features of approximated states output by the GBS devices, to model other noise sources arising, for example, from the $CZ$ gates and the homodyne detectors, and transmission losses.
We anticipate that this task might be simplified in moving to a macro-node based approach, wherein the noise from $CZ$ gates does not enter into the picture at all.

The second category of open problems deals with better strategies for encoding logical qubits in our architecture.
This includes strategies for choosing the best qubit encodings, i.e., about the choice of the inner encoding.
The current architecture relies heavily on features of the GKP encoding, including the fact that GKP qubits can be entangled with squeezed-light modes via the same optical elements as those that entangle them with other GKP qubits, thus opening the possibility of performing swap-outs.
The choice of GKP encoding is also motivated by its near-optimal loss tolerance.
That said, it might be possible to use other encodings if they could provide these advantages and overcome some of the shortcomings of GKP qubits (the primary being the low probabilities of state generation or equivalently large multiplexing requirements).
In particular, other bosonic encodings might be considered if they are compatible with swap-outs.
A more general question that has yet to be explored is the viability of employing other hybrid resource states comprised of different types of bosonic encoded qubits.

While the current work focuses on the RHG lattice as a paradigmatic example of an outer qubit code, significant benefit can be expected by moving to other outer encodings.
In particular, as we learn more about the structure of noise in realistic GKP-based computation, this opens the possibility of devising better outer encodings that are tailored for the specific noise structure.
Furthermore, noise-tailoring along the lines of~\cite{hanggli2020enhanced} may provide substantial enhancement to the thresholds obtained in our hybrid architecture.

The final set of open problems that we discuss are related to better methods for decoding in the current architecture.
One question is about the possibility of obtaining a further advantage from accounting for real-valued homodyne outcomes.
Although our inner decoder is exploiting the structure of the CV noise, there is still more information that could in principle be exploited, for example, at the level of the outer decoder.
Is may be possible to use ideas from analog quantum error correction~\cite{Fukui2019} and maximum-likelihood decoding~\cite{Vuillot2019} to further reduce our squeezing thresholds.

The question of optimal methods for decoding is also closely related to more fundamental questions related to swap-out-based resource states. 
Specifically, what is the fundamental swap-out threshold of our architecture? 
An alternative to swap-outs (i.e., replacing the GKP-node no-shows with squeezed light modes) is to treat the no-shows directly as erasure errors, but for these the threshold is set by percolation theory to be around 24.9\%.
Is the swap-out threshold higher than the erasure threshold set by percolation theory~\cite{barrett2010fault,Lorenz_1998} and in which regions of $(\delta,p_0)$ parameter space is it beneficial to have swap-outs rather than erasures? 
With the current decoders, we obtained around 23.6\% swap-out threshold, which likely can be improved substantially as numerous upgrades can be made to our inner and outer decoders.
We expect that development of an inner decoder that can treat more complicated arrangements of $p$-squeezed states in the lattice will provide fewer errors in the readout of qubit outcomes. 
Furthermore, we expect improvements to the outer decoder by using a more sophisticated method for assigning weights that takes the structure of the inner decoder into account.

A direction that becomes especially relevant to our photonics-based approach is that of developing fast decoders.
As we detail in the next section, the clock speeds of our architecture could be as fast as GHz.
However, to exploit these fast time scales, we need to develop efficient methods of decoding real-valued homodyne signals that need to be processed in order to change the homodyne local-oscillator phase as required by the logical computation.
We are confident that solving these open problems will further augment the advantages offered by this architecture and bring us closer to a scalable fault-tolerant quantum computer.

\section{Summary and Technological Advantages}\label{Sec:Discussions}

We have proposed a concrete and scalable architecture for quantum computing with light.
By using a hybrid resource state that can be generated and manipulated using near-future photonic technology, our architecture synthesizes modern techniques in scalable entangled resource state generation and bosonic codes.
This ``best of both worlds'' hybrid approach comes with a novel error structure that arises from the Gaussian model of state imperfections and the use of probabilistic bosonic qubit sources. 
Numerical results show that such errors can be handled by our tailored two-tier decoder that makes use of continuous- and discrete-variable syndrome data. 
We find that fault-tolerant quantum computation is possible in the regime where the swap-out probability -- the likelihood that that any given bosonic qubit source failed and the input was swapped with a squeezed state -- is smaller than $\approx 23.6\%$. 
For an experimentally accessible squeezing value of $\SI{15}{\decibel}$, our numerical results suggest that the maximum tolerable swap-out probability is $\approx 13.3\%$. 
This level of squeezing has been attained in free-space implementations~\cite{PhysRevLett.117.110801}, while the current state-of-the-art level of squeezing demonstrated in integrated sources is 8 dB~\cite{zhang2020single}; it remains an open experimental problem to match the level of squeezing in integrated sources to the free-space record.
In the remainder of this section, we discuss some technological aspects of the architecture. 

Our architecture provides several important technological advantages over competing architectures on photonic and other platforms. 
These include its modular nature, minimal cryogenic requirements, and fast clock speeds.
The first point deals with modularity: the various aspects of computation -- namely state preparation, multiplexing, cluster state generation and measurement -- impose different hardware requirements.
The distinctions between these requirements allow for a modular design in which different chips are given different tasks.
For instance, encoded bosonic states can be generated using non-reconfigurable circuits with low-loss interconnects leading to PNR detectors or perhaps even low-loss connections to PNR detectors that are integrated on the same chip.
The stitching of the cluster can also be performed on a non-reconfigurable chip.
The measurements on the generated cluster require reconfigurable homodyne detection fed-forward from measurements on other homodyne detectors.

Our architecture also poses minimal cryogenic requirements.
The state generation chips require low-loss non-reconfigurable circuits, which motivates on-chip PNR detectors with entire chips placed in a cryogenic environment until room temperature PNR technology is available~\cite{Collins2020}.
For this purpose, a static integrated platform will suffice.
The entire remaining architecture can operate at room temperature. 
Specifically, the switching network, required in the state-generation part to boost the production rate of GKP states, can be maintained at room temperature, thus exploiting any delays introduced in extracting the light out of the cryostat.
The cluster manipulation requires reconfigurable homodyne detection and delay lines to enable feed-forward, all at room temperature.
The generation and manipulation of the cluster can thus be performed in a scalable an integrated manner.

The cryogenic requirements of our architecture are modular: we do not need the extensive connectivity seen in other photonic architectures and other platforms entirely.
While other platforms require custom-built `jumbo' cryostats~\cite{Cho2020}, our architecture imposes no such requirements as it can make use of small, commercially available cryogenic technology.
This can provide significant advantage in the cost and reliability of our quantum computing architecture.

The final technological advantage of the architecture is that it allows for the fastest clock speeds among existing quantum computing architectures, which could enable very low-loss delay lines on the chip.
Unless a slower process is present in the final generation procedure, the timescale of the cluster generation and manipulation is ultimately set by the timescales of homodyne detection.
This is a positive feature, as homodyne detection can be much faster than PNR detectors used in the multiplexing procedure or threshold detectors used in other photonic encodings.
Faster time scales mean that the cluster-generation delay lines are shorter and thus incur lower losses.
We expect that these massive technological advantages will make photonics the leading platform for building a fault-tolerant photonic quantum computer.

\begin{acknowledgements}
J.E.B. is supported through an Ontario Graduate Scholarship, the Lachlan Gilchrist Fellowship, and by Mitacs through the Mitacs Accelerate program grant. 
I.T. is supported through by an Ontario Graduate Scholarship and a Mitacs Accelerate program grant. 
We thank Z.~Vernon for many suggestions in the design of the architecture, M.~Collins, D.~Mahler, B.~Morrison and D.~Phillips for valuable discussions about hardware aspects of the architecture, J.~M.~Arrazola, L.~G.~Helt, F.~Laudenbach, N.~Quesada, M.~Schuld and C.~Weedbrook for a careful proofreading, and P.~Dhara and J.~M.~Arrazola for their assistance with the figures.
Research at Perimeter Institute is supported in part by the Government of Canada through the Department of Innovation, Science and Economic Development Canada and by the Province of Ontario through the Ministry of Colleges and Universities. 
This research was enabled in part by support provided by Compute Ontario (\url{www.computeontario.ca}) and Compute Canada (\url{www.computecanada.ca}).
\end{acknowledgements}

\newpage
\appendix

\section{Noise Model for a Hybrid RHG Lattice Operating as a Memory}\label{sec:noise_model}
In this Appendix, we provide the full details of the error model summarized in \cref{subsec:noise}, that we in turn used to justify our choice of inner decoder.

\subsection{Noisy Initial States}
\subsubsection{Additive Gaussian Noise Channel}
The cluster state that the hardware generates will be populated by two kinds of states as mentioned in the main text: the  $\ket{+}_{\rm gkp}$ state and the momentum-squeezed state. Since the computer is operating in memory mode, we do not need to consider magic states as one of the possible states prepared. The position wavefunction of the ideal GKP state is ~\cite{Gottesman2001}
\begin{align}
    \ket{+}_{\rm gkp} = \sum_{n=-\infty}^{\infty} |n \sqrt{\pi}\rangle_q,
\end{align}
where $|\cdot \rangle_{q}$ corresponds to a position eigenstate. 
To model the state initialization error, we apply the single-mode additive Gaussian noise channel given by~\cite{hall1994gaussian} 
\begin{align}
\label{eq:noise_app}
    {\cal N}_{\bm{Y}}(\hat{\rho}) = \int_{\mathbb{R}^{2}} \frac{d^2\bm{\xi}}{\pi \sqrt{\det{\bm{Y}}}} \exp\left[- \frac{1}{2}\bm{\xi}^T \bm{Y}^{-1} \bm{\xi}\right] \hat{\mathcal{D}}(\bm{\xi}) \hat{\rho} \hat{\mathcal{D}}^{\dag}(\bm{\xi}), 
\end{align}
where $\bm{Y}\geq 0$ is the noise matrix, applied independently on each mode depending on the state that populates it. The Weyl-Heisenberg displacement operator is defined as ${\cal \hat{D}}(\bm{\xi})= \exp[i \bm{\xi}^T \bm{\Omega} \hat{\bm{r}}]$, where $\bm{\xi}=(\xi_q,\xi_p)^T \in \mathbb{R}^2$ for a single mode, $\bm{\Omega}$ is the anti-symmetric symplectic metric, and $\hat{\bm{r}} = (\hat{q},\hat{p})^T$. 
For the GKP states and the momentum states, the corresponding noise matrices are chosen as 
\begin{align}
\label{eq:Ymat_app}
    \bm{Y}_{\rm gkp} = \frac{1}{2}\begin{pmatrix} \delta & 0 \\ 0 & \delta \end{pmatrix}, ~~ \bm{Y}_{\rm p} = \frac{1}{2}\begin{pmatrix} \delta^{-1} & 0 \\ 0 & \delta \end{pmatrix}.  
\end{align}

In other words, we  start with ideal  $\ket{+}_{\rm gkp}$ states and either apply the noise channel ${\cal N}_{\bm{Y}_{\rm gkp}}$ or ${\cal N}_{\bm{Y}_{\rm p}}$ with probabilities $1-p_0$ and $p_0$, respectively. Note that we know what the state is at each site, so we know which noise channel was applied. For GKP states, although this noise model does not capture the damping of peaks due to finite energy seen in \cref{eq:finiteE_GKP}, it captures the broadening of peaks that results from finite-energy effects~\cite{Menicucci2014,Noh2020}. Similarly, this method approximates the realistic momentum state well in the position basis but gives it a periodic structure in momentum space. These points can be viewed transparently through the Wigner picture. In both cases, the application of a noise channel renders the output states mixed. 

\subsubsection{The Wigner Picture}
The Wigner function for ideal GKP states consists of a linear combination of two-dimensional  $\delta$-functions in phase space~\cite{Gottesman2001}:
\begin{align}\label{eq:ideal_wigner}
        \begin{split}
        W_{\ket{+}_{\rm gkp}}(\bm{r}) &= \frac{\sqrt{\pi}}{2}\sum_{s,t=-\infty}^{\infty} (-1)^{st}\delta(\bm{r}-\bm{\mu}_{s,t}),\\
        \bm{\mu}_{s,t}^T &= \frac{\sqrt{\pi}}{2}(s,2t). 
        \end{split}
\end{align}
Note that the lattice spacing of the Dirac-delta peaks in the momentum direction is twice that of the position direction in the Wigner picture. For a clear diagram of the phase space unit cell distribution, see Fig. 1a of~\cite{PhysRevLett.123.200502}. Treating each $\delta$-function as a Gaussian of infinitely small width in phase space, we see that the effect of the noise channel is to replace the $\delta$-functions with  Gaussian distributions with covariance $\bm{Y}_{\rm gkp}$, by using \cref{eq:noise_app}. Thus, the linear combination of $\delta$-functions is mapped to a linear combination of Gaussian functions centred at the same points in phase space and with the same weights in the linear combination of \cref{eq:ideal_wigner}.

With regard to the momentum states in the RHG lattice, consider the same noise model of \cref{eq:noise_app}, now instead with covariance 
\begin{align}
    \bm{Y}_{\rm p}(\epsilon,\delta) = \frac{1}{2}\begin{pmatrix} \epsilon^{-1} & 0 \\ 0 & \delta \end{pmatrix}.  
\end{align}
Returning again to the Wigner function picture for the $\ket{+}_{\rm gkp}$ state, this noise replaces the $\delta$-functions with Gaussians of covariance $\bm{Y}_{\rm p}(\epsilon,\delta)$. In the limit that $\epsilon \rightarrow 0$, we see that for odd values of $t$, these Gaussians in phase space cancel each other out, while for even values of $t$, the Gaussians add together. This is due to the phase factor $(-1)^{st}$ in \cref{eq:ideal_wigner}. The resulting phase space distribution is a periodic \textit{mixture} of $p$-quadrature eigenstates, each separated by $2\sqrt{\pi}$, and each with a Gaussian noise of variance $\delta/2$ applied in the $p$-quadrature. That is, the noise channel $\bm{Y}_{\rm p}(\epsilon,\delta)$ turns $\ket{+}_{\rm gkp}$ into a classical mixture of noisy $p$-squeezed states. Since we are examining the regime where $\delta$ is small, we set $\delta=\epsilon$, which leads to $\epsilon^{-1}$ being large as needed for the states to approach $p$-squeezed states. {{While it is the case that this noise model for momentum squeezed states returns a Wigner function which still, in principle, has a periodic structure, we do not expect it to positively bias the decoding procedure. The periodic structure in position space is essentially washed away by the broadness of the envelope of order $\delta^{-1}$, while the discrete $2\sqrt{\pi}$ translational symmetry in momentum introduces a mixture of momentum eigenstates, and hence more noise than a pure momentum squeezed state.}}

The initialization step for all $N$ nodes can therefore be written compactly in one equation as:
\begin{align}
    \hat{\rho}_0 &= {\cal N}_{\bm{\Sigma}_0}(\ket{+}_{\rm gkp}\bra{+}^{\otimes N})\nonumber\\
    &= \int_{\mathbb{R}^{2N}} \frac{d^{2N}\bm{\xi}}{\pi^N \sqrt{\det{\bm{\Sigma}_0}}} \exp\left[-\frac{1}{2} \bm{\xi}^T \bm{\Sigma}_0^{-1} \bm{\xi}\right] \nonumber\\
    &~~~~~~\times \hat{\mathcal{D}}(\bm{\xi}) (\ket{+}\bra{+}_{\rm gkp}^{\otimes N}) \hat{\mathcal{D}}^{\dag}(\bm{\xi}),
\end{align}
where  $\bm{\xi}=(\bm{\xi}_{\bm{q}},\bm{\xi}_{\bm{p}})^T = (\xi_{q_1},\cdots,\xi_{q_N},\xi_{p_1},\cdots,\xi_{p_N})^T \in \mathbb{R}^{2N}$, and $\hat{\bm{r}} = (\hat{q}_1,...,\hat{q}_N,\hat{p}_1,...,\hat{p}_N)^T$, corresponding to $N$-modes. Here, $\bm{\Sigma}_0$ is a direct sum of matrices, where the $i^{th}$ matrix in the direct sum is either of the form $\bm{Y}_{\rm gkp}$ or $\bm{Y}_{\rm p}$ depending on whether the $i^{th}$ mode is a GKP or a $p$-squeezed state. In other words,
   \begin{equation}
    \bm{\Sigma}_0 = \begin{pmatrix} \bm{\Sigma}_x & \bm{0} \\ \bm{0} & \frac{\delta}{2}\bm{\ids} \end{pmatrix},
    \end{equation}
    where $\bm{\Sigma}_x$ is a diagonal matrix with elements $\frac{1}{2\delta}$ or $\frac{\delta}{2}$ depending on if the mode is $p$-squeezed or GKP, respectively.

There are several reasons to model the state preparation error with the noise channel described in \cref{eq:noise_app,eq:Ymat_app}. For one, there are many physical gates that use a measurement-based squeezing operation~\cite{filip2005,ukai2011,yoshikawa2008} that naturally leads to  imperfections modeled as the classical noise channel. Furthermore, this type of noise is closely related to pure loss---following a pure loss channel by an amplifier of the inverse strength leads to a classical noise channel~\cite{caruso2006one,saba2010,sabapathy2011robustness,garcia2012majorization}. In settings where loss can be treated this way, such as in measurement imperfections, this relationship would play an important role. 

The classical noise channel is easily described in the Heisenberg picture, so we use this representation in our simulations. Let us consider the quadrature operators 
$\hat{r}$ of the $N$-modes. The noise channel on each mode can be described as 
\begin{align}
\hat{\bm{r}} \to \hat{\bm{r}} + \bm{\xi},      
\end{align}
where $\bm{\xi}$ is a vector of random variables drawn from the corresponding normal distribution $\bm{\Sigma}_0$ associated with the state initialization errors. 

A final note is that we assume that the $CZ$ gates and the measurement procedure are noiseless. Imperfections in both these modules are likely to reduce the error threshold. Inefficiencies in the measurement outcomes can be modeled as a lossy channel and can be converted into a classical noise channel by virtually applying an amplifier that would affect the measurement readout. Similarly, classical noise channels in the $CZ$ can also be tracked due to the Gaussian nature of the noise. However, for simplicity of the presentation, we leave the analysis of this case and possibly more complicated noise models to future work.

\subsection{Propagation of Noise in the Cluster State Preparation}

For each node, with probability $p_0$ prepare a momentum-squeezed state, and with probability $(1-p_0)$ prepare a $\ket{+}_{\rm gkp}$. Next in our model, we apply $CZ$ gates perfectly according to the structure of the cluster state, i.e., apply $CZ$ gates to each pair of qubits connected by an edge. We invert some of the $CZ$ gates to match the CV toric code convention~\cite{Demarie2014}. Since we are operating in memory mode, no further gates are applied before the $p$-homodyne measurements.

The symplectic transformation for a $CZ$ gate defined as $\exp[i\hat{q_1} \hat{q_2}]$ in the $(q_1,q_2,p_1,p_2)$ basis ordering is given by
\begin{equation}
    \bm{S}_{CZ} = \begin{pmatrix} 
    1 & 0 & 0 & 0 \\ 
    0 & 1 & 0 & 0 \\
    0 & 1 & 1 & 0 \\
    1 & 0 & 0 & 1\end{pmatrix} = 
  \begin{pmatrix} \ids & \bm{0} \\ \bm{A} & \ids \end{pmatrix}, 
\end{equation}
where $\bm{A}$ is a $2 \times 2$ adjacency matrix. This motivates the symplectic matrix that links all $N$ optical modes into an RHG lattice as
\begin{equation}
    \label{eq:srl}
    \bm{S}_{RL} = \begin{pmatrix} \ids & \bm{0} \\ \bm{A}_{RL} & \ids \end{pmatrix},
\end{equation}
where $\bm{A}_{RL}$ is the $N\times N$ matrix with 1 at position $(i,j)$ if two nodes are entangled with a $CZ $ gate and 0 otherwise, with a suitable {\sf parity} function dictated by the toric code convention. $\bm{A}_{RL}$ corresponds to the links depicted in \cref{fig:pcell,Fig:Raussendorf-Gen}. 

It is also instructive to look at the effect of the cluster state preparation on the noise matrix. Under the action of all the $CZ$ gates, the full noise matrix evolves under the symplectic transformation to
    \begin{align}
       \bm{\Sigma}_0 \rightarrow \widetilde{\bm{\Sigma}}_0 &= \bm{S}_{RL}\bm{\Sigma}_0\bm{S}_{RL}^T \\
       &=  \begin{pmatrix} \bm{\Sigma}_x & \bm{\Sigma}_x \bm{A}_{RL}^T \\ \bm{A}_{RL}\bm{\Sigma}_x & \frac{\delta}{2}\ids + \bm{A}_{RL}\Sigma_x \bm{A}_{RL}^T\end{pmatrix}.
    \end{align}
Since we are mainly concerned with the momentum homodyne measurement values, it turns out that the momentum component of the covariance matrix is useful to write down for subsequent sections. To achieve this, we trace out the position degrees of freedom of the covariance matrix of the noise channel to obtain
\begin{align}
    \widetilde{\bm{\Sigma}}_p = \frac{\delta}{2}\ids + \bm{A}_{RL}\bm{\Sigma}_x \bm{A}_{RL}^T. 
    \label{eq:noisemat}
\end{align}

\subsection{Probability Distribution in Momentum Space}
So far we have only focused on the noise model and the correlated noise matrix that one obtains once all the $CZ$ gates have been applied to the initial states in each mode. We now detail the connection between the noise matrix obtained in \cref{eq:noise_app} and the homodyne distribution. 

Let us define the unitary corresponding to the symplectic transformation in \cref{eq:srl} as $\hat{U}_{RL}$. Since the preparation of the RHG cluster state and the noise channel on the initial states are both Gaussian, we can conjugate $\hat{U}_{RL}$ through the noise matrix to obtain
\begin{align}
    \hat{\rho}_{RL} &= \hat{U}_{RL} \left[{\cal N}_{\bm{\Sigma}_0}(\ket{+}\bra{+}_{\rm gkp} ^{\otimes N})\right] \hat{U}_{RL}^{\dagger} \nonumber \\
    &= {\cal N}_{\widetilde{\bm{\Sigma}}_0}\left[ (\hat{U}_{RL}\ket{+}\bra{+}_{\rm gkp}^{\otimes N} \hat{U}_{RL}^{\dagger})\right].
\end{align}
This corresponds to taking the ideal state of the RHG lattice had all the GKP states been initialized perfectly without noise  and then applying a correlated multimode Gaussian noise channel with covariance $\widetilde{\bm{\Sigma}}_0$.

To understand the probability distribution produced by conjugating the unitary through the Gaussian noise channel, we first show that the probability distribution in $p$-quadrature of a state under a Gaussian noise channel is given by the convolution of the noiseless probability distribution with the marginal Gaussian distribution of the noise channel along the $p$-quadrature.

Consider a Gaussian random displacement channel applied to a state:
\begin{equation}
        \mathcal{N}_{\bm{\Sigma}}(\hat{\rho})= \int_{\mathbb{R}^{2N}} \frac{d^{2N}\bm{\xi}}{\pi^N \sqrt{\det{\bm{\Sigma}}}} \exp\left[-\frac{1}{2}\bm{\xi}^T \bm{\Sigma}^{-1} \bm{\xi}\right] \hat{\mathcal{D}}(\bm{\xi}) \hat{\rho} \hat{\mathcal{D}}^{\dag}(\bm{\xi}).
\end{equation}
Next we note we can break the displacement into displacements $\hat{X}(\cdot)$ and $\hat{Z}(\cdot)$ along $\bm{q}$ and $\bm{p}$ in phase space, respectively, along with a phase factor:
\begin{align}
      \hat{\mathcal{D}}(\bm{\xi})\ket{\bm{p}} &= e^{i\bm{\xi}_{\bm{x}}\cdot\bm{\xi}_{\bm{p}}/2}\hat{X}(\bm{\xi}_{\bm{x}})\hat{Z}(\bm{\xi}_{\bm{p}})\ket{\bm{p}}\\  
      &=e^{-i\bm{\xi}_{\bm{x}}\cdot\bm{\xi}_{\bm{p}}/2}e^{-i\bm{\xi}_{\bm{x}}\cdot\bm{p}}\ket{\bm{p}+\bm{\xi}_{\bm{p}}}.
\end{align}
Thus:
\begin{align}
    \begin{split}
    &\bra{\bm{p}}\hat{\mathcal{D}}(\bm{\xi})\ket{\bm{p}'}\bra{\bm{p}''} \hat{\mathcal{D}}^{\dag}(\bm{\xi})\ket{\bm{p}} \\
      =& \delta(\bm{p}-\bm{p}'-\bm{\xi}_{\bm{p}})\delta(\bm{p}-\bm{p}''-\bm{\xi}_{\bm{p}})e^{i\bm{\xi}_{\bm{x}}\cdot(\bm{p}''-\bm{p}')}
    \end{split}
\end{align}

Putting these equations together, we find:
\begin{align}
    \begin{split}
    &\bra{\bm{p}}\mathcal{N}_{\bm{\Sigma}}(\hat{\rho})\ket{\bm{p}} \\
    &= \int_{\mathbb{R}^{2N}} \frac{d^{2N}\bm{\xi}}{\pi^N \sqrt{\det{\bm{\Sigma}}}} \exp\left[- \frac{1}{2}\bm{\xi}^T \bm{\Sigma}^{-1} \bm{\xi}\right] \rho(\bm{p}-\bm{\xi}_{\bm{p}},\bm{p}-\bm{\xi}_{\bm{p}})\\
    &= \int_{\mathbb{R}^{N}} \frac{d^{N}\bm{\xi_p}}{\sqrt{\pi^N \det{\bm{\Sigma}_p}}} \exp\left[- \frac{1}{2}\bm{\xi}_{\bm{p}}^T \bm{\Sigma}_p^{-1} \bm{\xi}_{\bm{p}}\right] \rho(\bm{p}-\bm{\xi}_{\bm{p}},\bm{p}-\bm{\xi}_{\bm{p}}),
    \end{split}
\end{align}
where in the last step we performed the Gaussian integral over $\bm{\xi}_{\bm{x}}$.

Therefore, returning to the probability distribution in $p$-space of our hybrid lattice, we find:
\begin{align}
\label{eq:pvalue}
    &\bra{\bm{p}}{\cal N}_{\widetilde{\bm{\Sigma}}_0}(\hat{U}_{RL}\ket{+}\bra{+}_{\rm gkp}^{\otimes N}  \hat{U}_{RL}^{\dagger} )\ket{\bm{p}}\nonumber\\
    &= \int_{\mathbb{R}^{N}} \frac{d^{N}\bm{\xi_p}}{\sqrt{\pi^N\det{\widetilde{\bm{\Sigma}}_p}}} \exp\left[- \frac{1}{2}\bm{\xi_p}^T \widetilde{\bm{\Sigma}}_p^{-1} \bm{\xi_p}\right] |\psi_{RL}(\bm{p}-\bm{\xi_p})|^2,     
\end{align}
where $\psi_{RL}(\bm{p})$ is the wavefunction in $p$-space of the ideal RHG cluster state and $\widetilde{\bm{\Sigma}}_p$ was defined in \cref{eq:noisemat}. We know that $|\psi_{RL}(\bm{p})|^2$ consists of a lattice in $p$-space, where each point of the lattice is located at $\bm{n}\sqrt{\pi}$, where $\bm{n}$ is an integer-valued $N$-component vector chosen from a set dictated by the ideal qubit state of the RHG lattice. The addition of the Gaussian noise channel broadens each of these lattice points into Gaussian functions with covariance $\widetilde{\bm{\Sigma}}_p$. Therefore, we see that we can interpret homodyne momentum outcomes as being sampled from the noise matrix $\widetilde{\bm{\Sigma}}_p$ using \cref{eq:pvalue}. Given the measurement outcomes, we then apply a classical decoder to these values to yield us the net recovery operation. 

\begin{figure*}
    \centering
    \includegraphics[width = 0.80\linewidth]{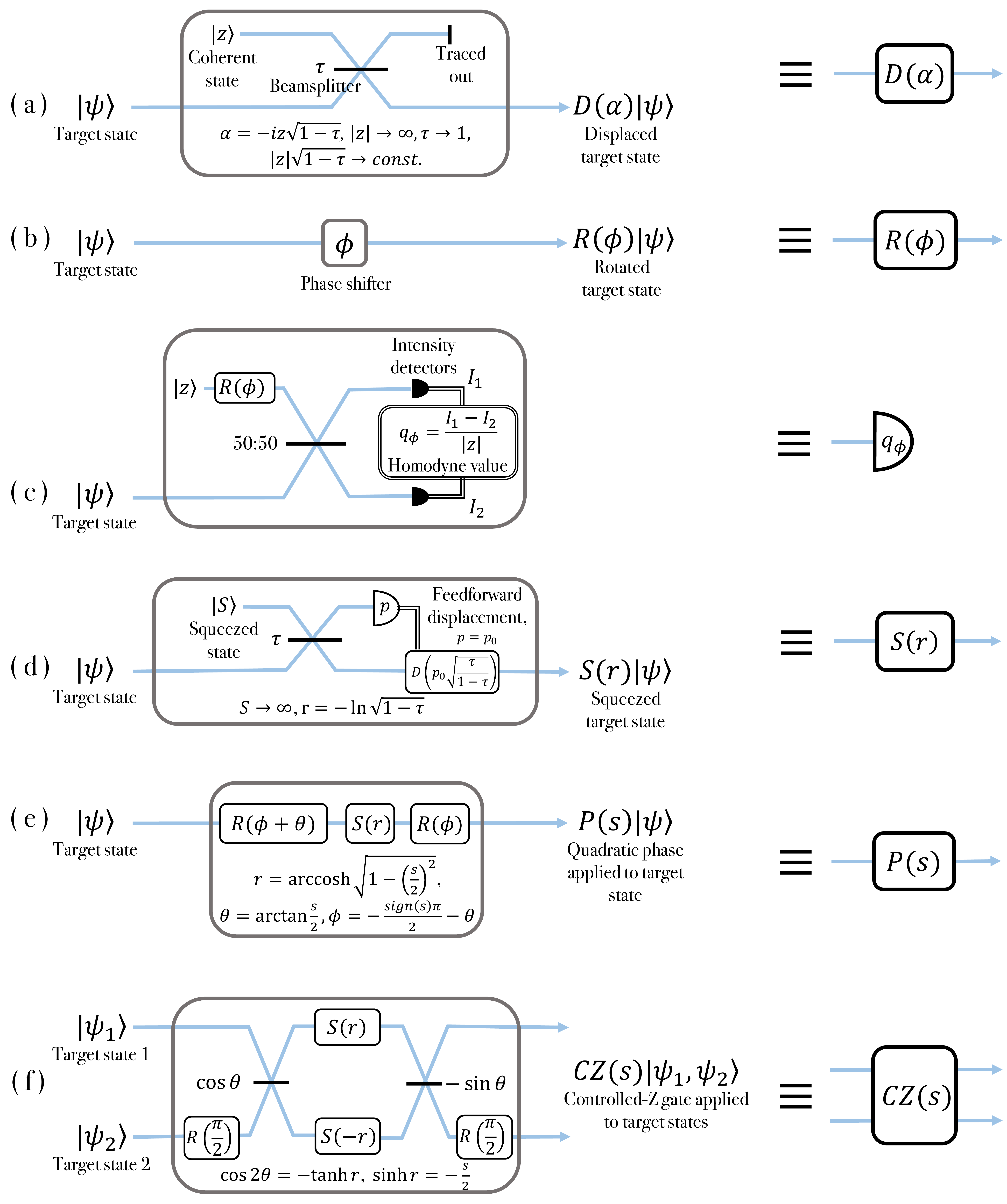}
    \caption{A review of optical implementations of the gates and measurements required for Clifford operations in the GKP encoding, including limits required to achieve ideal, perfect CV gate application. (a) A general displacement module~\cite{Paris1996}. Displacement by $\sqrt{\pi}$ in $q$ ($p$) corresponds to a GKP qubit Pauli $X$ ($Z$) gate. (b) Rotation module as performed by e.g. an optical thermoelectric heating element. $\phi=\pi/2$ corresponds to the CV Fourier transform as well as the GKP-qubit Hadamard gate.  (c) Homodyne measurement module. Changing the rotation $\phi$ changes the axis in phase space along which the measurement is performed. $\phi=0$ ($\pi/2$) corresponds to $q$ ($p$) homodyne measurement, which is the GKP qubit Pauli $Z$ ($X$) measurement. (d) Measurement-based squeezing module~\cite{filip2005}. On-demand, in-line squeezing is in general required for implementing CV quadratic phase and Controlled-X/phase gates, and a measurement-based approach allows for offline preparation of squeezed resource state. (e) Quadratic phase gate module~\cite{Killoran2019}. $s=\pm 1$ corresponds to the GKP qubit phase gate. (f) CV $CZ$ gate module. $s=\pm 1$ corresponds to the GKP qubit $CZ$ gate. Application of $\pi/2$ rotations on the second mode before and after the $CZ$ gate implements a CV $CX$ gate~\cite{Killoran2019} with Target state 1 becoming the control and Target state 2 becoming the target, and thus a GKP qubit $CNOT$ gate.}
    \label{fig:Gauss_gates}
\end{figure*}

\section{Optical Components for GKP Qubit Operations}\label{sec:optical_elements}

A primary advantage of the GKP qubit encoding is the fact that Clifford gates and measurements correspond to CV Gaussian gates and measurements. 
In \cref{fig:Gauss_gates}, we provide optical circuits for the application of GKP Clifford gates and measurements in an optical setting. 
These circuits present how the gates would be implemented in a circuit-based setting. 
In contrast, the actual gates on our physical qubits are implemented in a measurement-based manner, and hence their implementation would only involve performing homodyne measurements on the computational resource state. 
Thus, this section is included for completeness of the background material and to demonstrate the accessibility of Gaussian resources in optics, rather than as the actual implementation of how the gates would be performed in our architecture.

\begin{figure}
    \centering
    \includegraphics[width = \columnwidth]{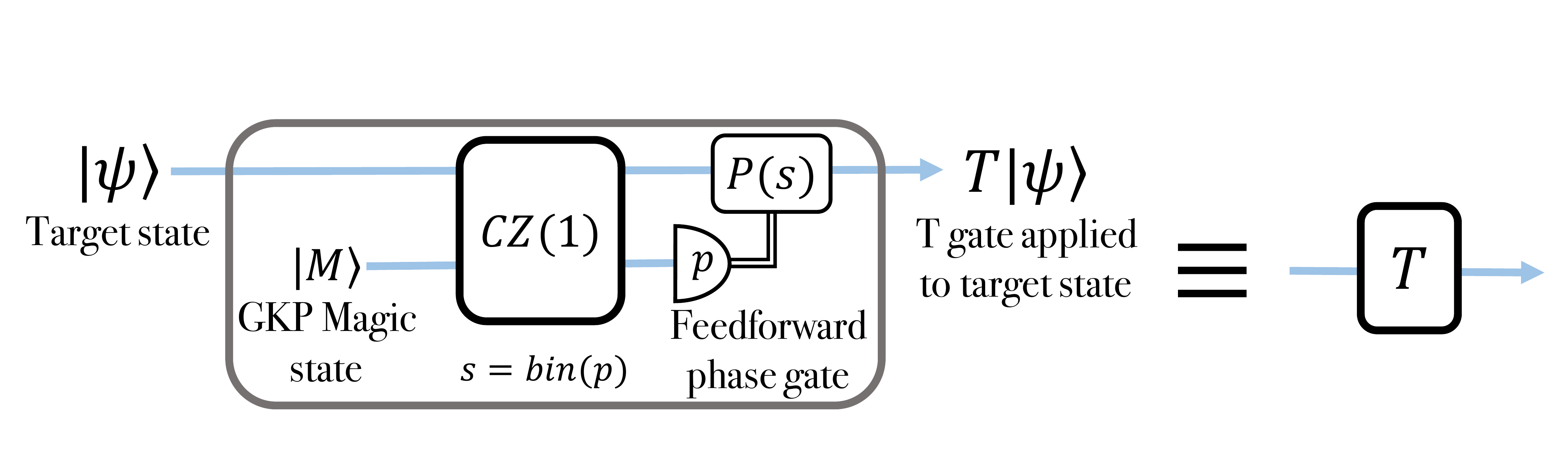}
    \caption{Optical implementation of the GKP qubit $T$ gate up to global phase, following the method from~\cite{Gottesman2001}. Here, in the ideal limit, $\ket{M}=e^{-i\pi/8}\ket{+}_{\rm gkp}+e^{i\pi/8}\ket{-}_{\rm gkp}$, and the feedforward phase gate is applied if the ancillary mode detects $\ket{-}_{\rm gkp}$ via a qubit $X$ measurement (CV $p$ homodyne).}
    \label{fig:Tgate}
\end{figure}

For the non-Clifford $T$ gate, non-Gaussian CV gates are required. In \cref{fig:Tgate}, we provide an optical circuit for $T$ gate application via gate teleportation using a GKP magic state as a resource. In~\cite{Tzitrin2019}, it was observed that magic states and Pauli basis states are comparably resource-intensive to produce using GBS state preparation.

\section{Heuristic Weights for the Outer Decoder}\label{sec:decoder_weights}
In the outer decoder algorithm detailed in \cref{sec:outer} and applied in \cref{subsec:FT_sim}, we have the opportunity to assign different weights in the outer decoder for the RHG lattice depending on the expected error at a site. In the most naïve approach, one could simply use the marginal probability that a node undergoes a phase flip to determine the weight; however, this does not take into account correlated phase flips that we expect to see from replacing some nodes with $p$-squeezed states, as we have discussed previously. For instance, in either the case of applying a ring of four $Z$ gates or of applying the identity, neither result in any error in the decoding, but if we simply looked at the marginal probabilities of the sites in the ring, we might pessimistically assume each site has an independent probability of incurring a phase flip, which would result in pessimistic weight assignment.

While a full analysis of correlated errors and weight assignments for general configurations of $p$-squeezed states and GKP states is left to future work, the heuristic choice for weight assignments given by \cref{eq:error_prob} can be found by considering a simple configuration. Amazingly, this choice of weights already provides a significant improvement over the use of marginal probability of error at each site. Here, we detail the motivation for this choice of heuristic weights.

Consider a single node $e_0$ in the RHG lattice surrounded by four neighbors $e_1,e_2,e_3,e_4$, which can be either GKP or $p$-squeezed, so that $e_0$ can have anywhere from 0 to 4 $p$-squeezed states as neighbors. For simplicity, we will assume that the next layers of neighbors of $e_1,e_2,e_3,e_4$ are GKP states. See \cref{fig:pcell,Fig:Raussendorf-Gen} for a visualization of the lattice configuration. Whether $e_0$ is GKP or $p$-squeezed does not impact the following argument. Additionally, we will assume the limit of infinite squeezing ($\delta\rightarrow 0$) for all the sites. Note that these assumptions are used to select a choice of weights, not to run the actual simulation, so discrepancies between the assumptions for choosing weights and the actual parameters of the simulation will nonetheless result in a perfectly usable, albeit suboptimal, set of weights.

In this scenario, for each node that is a momentum eigenstate, it imparts a random displacement -- sampled from the uniform distribution of its $q$-quadrature since we assumed $\delta\rightarrow 0$ -- onto the $p$-quadrature of its neighbors via the action of the CV $CZ$ gates, while each node that is a perfect GKP state does not impart any random displacements onto its neighbors. Let the displacements from $e_1,e_2,e_3,e_4$ be $d_1,d_2,d_3,d_4$, where we specifically mean the excess displacement beyond $n\sqrt{\pi}$~\cite{Noh2020}. Thus, the displacement on $e_0$ is given by $d_1+d_2+d_3+d_4$, assuming the $CZ$ gates are all +1; changing the sign of the $CZ$ gates does not change the argument since $d_1,d_2,d_3,d_4$ are sampled from symmetric distributions. Moreover, let $b_i,\,i=1,2,3,4$ be the binary value returned by performing standard GKP binning on the homodyne output of the neighbors of node $e_i$ other than $e_0$; note that the only shared neighbor of $e_1,e_2,e_3,e_4$ is $e_0$, and since we assumed the nodes beyond $e_1,e_2,e_3,e_4$ were GKP, then we know we will get the same singular outcome $b_i$ on all neighbors of $e_i$ other than $e_0$. Let $b_0$ be the binary value returned by performing standard GKP binning on the homodyne output of $e_0$.

Consider now the scenarios that would cause an error at the level of the qubit decoder. We know that a closed ring of $Z$ gates commutes with the stabilizers of the RHG lattice. If only one of $e_1,e_2,e_3,e_4$ is a momentum eigenstate, then we have already shown that the resulting effect on the binary outcomes $b_i$ is a ring of four $Z$ gates around $e_i$, which we know causes no problem. If two or more of $e_1,e_2,e_3,e_4$ are momentum eigenstates, then we now have potential for error when using standard binning. In particular, for the readout to correspond to a closed ring of $Z$ gates, we require that $(b_0+b_1+b_2+b_3+b_4)\bmod2=0$. This condition will always be true if only one of $d_1,d_2,d_3,d_4$ is sampled from uniform (since we assumed $\delta\rightarrow 0$) while the others are zero, since $d_1+d_2+d_3+d_4$ will then be equal to the only non-zero displacement. We find that if two, three or four of $d_1,d_2,d_3,d_4$ are nonzero, then the condition $(b_0+b_1+b_2+b_3+b_4)\bmod 2=0$ is violated with probabilities of approximately $0.25, 0.33,$ and $0.40$, respectively. This means that even with multiple nodes replaced by $p$-squeezed states, the probability of the resulting readout indicating a series of gates that do not commute with the stabilizer is less than $50\%$, which would be the marginal probability of phase flip at each site. Finally, we use these probabilities of error to assign heuristic, relative weighting in the outer decoder.

\end{document}